\documentclass[12pt]{article}
\usepackage{amsfonts,amsthm,amsmath,amssymb,upgreek,bm}
\usepackage{hyperref}
\usepackage[paper=letterpaper,margin=.7in]{geometry}
\usepackage{graphicx}
\usepackage{units}
\usepackage{float}
\usepackage[export]{adjustbox}
\usepackage{bbold}
\usepackage{xcolor}
\usepackage[shortlabels]{enumitem}
\usepackage{dsfont}

\newcommand{\be}{\begin{equation}}
\newcommand{\ee}{\end{equation}}
\renewcommand{\tilde}{\widetilde}

\newcommand{\stran}{s_{\text{tran}}}
\newcommand{\smax}{s_k}
\newcommand{\stilde}{\tilde{s}_k}

\newcommand{\sprime}{s'}

\numberwithin{equation}{section}

\def\tr{\text{Tr}}
\def\Tr{\text{Tr}}

\begin{document}
\thispagestyle{empty}

\vspace*{.5cm}
\begin{center}

{\bf {\LARGE Replica wormholes and the black hole interior}}

\begin{center}

\vspace{1cm}

 {\bf Geoff Penington, Stephen H. Shenker, Douglas Stanford, and Zhenbin Yang}\\
  \bigskip \rm
  
\bigskip
Stanford Institute for Theoretical Physics,\\Stanford University, Stanford, CA 94305

\rm
  \end{center}

\vspace{2.5cm}
{\bf Abstract}
\end{center}
\begin{quotation}
\noindent

Recent work has shown how to obtain the Page curve of an evaporating black hole from holographic computations of entanglement entropy. We show how these computations can be justified using the replica trick, from geometries with a spacetime wormhole connecting the different replicas. In a simple model, we study the Page transition in detail by summing replica geometries with different topologies. We  compute related quantities  in less detail in more complicated models, including JT gravity coupled to conformal matter and the SYK model. Separately, we give a direct gravitational argument for entanglement wedge reconstruction using an explicit formula known as the Petz map; again, a spacetime wormhole plays an important role. We discuss an interpretation of the wormhole geometries as part of some ensemble average implicit in the gravity description.

\end{quotation}

\setcounter{page}{0}
\setcounter{tocdepth}{2}
\setcounter{footnote}{0}
\newpage

\tableofcontents

\pagebreak

\section{Introduction}
Hawking's calculation \cite{hawking1975} of black hole evaporation by thermal radiation raises deep puzzles about the consistency of black holes and quantum mechanics.   In particular, if the black hole starts in a pure quantum state, how can the apparently thermal radiation at the end of evaporation actually be in a pure state, as quantum mechanics requires?  A quantitative form of this puzzle is encapsulated in the Page curve \cite{Page:1993df}, a plot of the entanglement entropy of the radiation as a function of time.   It increases at short times because of the thermal character of the radiation, but then must decrease to zero at the end of evaporation to be consistent with unitary quantum mechanical evolution.

Holographic duality has supplied a powerful tool for computing entanglement entropy, the Ryu-Takayanagi formula \cite{Ryu:2006bv}.  This formula has been generalized and refined in numerous ways since its original formulation.\footnote{For example \cite{Headrick:2007km,Hubeny:2007xt,Wall:2012uf,Lewkowycz:2013nqa,Faulkner:2013ana,Dong:2016hjy}; for a review see \cite{Rangamani:2016dms}.} In its most general formulation, the Engelhardt-Wall (EW) prescription \cite{Engelhardt:2014gca}, it says that the entropy of a holographic boundary region $B$ is given by the generalized entropy of the minimal quantum extremal surface (QES).

Building on earlier ideas in \cite{Almheiri:2018xdw, Hayden:2018khn}, the authors of \cite{Penington:2019npb, Almheiri:2019psf} considered a black hole in anti-de Sitter space that was allowed to evaporate into an auxiliary system $R$ using absorbing boundary conditions.  They showed that, at exactly the Page time, there was a phase transition in the minimal QES. This caused the boundary entropy to begin decreasing, in accordance with the Page curve. However, as emphasized in \cite{Almheiri:2019psf}, a naive calculation of the entropy of the Hawking radiation showed that it continued to increase, since the semiclassical bulk physics had not changed. 

As advocated in \cite{Penington:2019npb}, and commented on in \cite{Almheiri:2019psf}, the Page curve for the radiation would result from a variant of the usual rules for computing entropy holographically.\footnote{This is a special case of a prescription introduced in \cite{Hayden:2018khn} (near Eqn. 4.14) for the entropy of the \emph{combination} of a boundary region $B$ and auxiliary system $R$. If $B$ is empty, the prescription reduces to \eqref{eq:islandprescription}.} This idea was given an elegant ``doubly holographic'' realization in \cite{Almheiri:2019hni}, where it was called the ``island conjecture.''  Explicitly, the prescription states that the actual entropy $S(R)$ of the Hawking radiation is given by
\begin{align} \label{eq:islandprescription}
S(R) = \min \left\{\underset{I}{\text{ext}} \left[\frac{\text{Area}(\partial I)}{4 G_N} + S_\text{bulk}(I \cup R)\right]\right\},
\end{align}
where $I$ is some ``island'' region of the bulk.\footnote{In \cite{Hayden:2018khn,Penington:2019npb}, \eqref{eq:islandprescription} was justified by imagining throwing the auxiliary, nonholographic system $R$ into an auxiliary \emph{holographic} system. \eqref{eq:islandprescription} then reduces to the usual EW prescription. In \cite{Almheiri:2019hni}, (\ref{eq:islandprescription}) was justified by considering holographic bulk matter and using the Ryu-Takayanagi formula in the ``doubly holographic'' description of the theory.} Here $S_\text{bulk}(I\cup R)$ is the entropy of the island plus the Hawking radiation, computed semiclassically in the original fixed geometry.\footnote{For other recent work on this topic, see \cite{Akers:2019nfi,Almheiri:2019yqk,Rozali:2019day,Chen:2019uhq,Bousso:2019ykv,Almheiri:2019psy}.} 

Since the Ryu-Takayanagi formula can be derived from gravitational path integrals using replicas \cite{Lewkowycz:2013nqa,Faulkner:2013ana,Dong:2016hjy,Dong:2017xht}, the prescription \eqref{eq:islandprescription} should also be derivable directly from the gravitational path integral. Deriving it is one of the main goals of this paper. We will show explicitly how this prescription arises from a replica computation with Euclidean wormholes connecting the different replicas.   While we present our arguments in the context of simple models, their structure is not model dependent and should apply in a general context.

Wormholes of a similar type have recently been used to analyze the late-time behavior of the spectral form factor \cite{Saad:2018bqo,Saad:2019lba} and correlation functions \cite{Saad:2019pqd}, inspired by the puzzle posed in \cite{Maldacena:2001kr}. In both that setting and here, the basic point of the wormholes is to give small but nonzero overlaps between naively orthogonal bulk states. This resonates with the longstanding suspicion that some small corrections to the Hawking calculation (of order $e^{-S_{\rm BH}}$) could be responsible for making black hole evolution unitary.\footnote{A challenge to this idea was provided by \cite{Mathur:2009hf,Almheiri:2012rt}: if the early radiation, the next Hawking quantum, and its interior partner are independent systems, the needed corrections would require a firewall at the horizon. However, if the interior partner is secretly part of the early radiation, then the arguments of \cite{Mathur:2009hf,Almheiri:2012rt} do not apply \cite{Bousso:2012as,Nomura:2012sw,Verlinde:2012cy,Papadodimas:2012aq,Maldacena:2013xja}. The ER=EPR proposal \cite{Maldacena:2013xja} suggested that a geometric connection might play a role in this identification.}

Understanding the Page curve is only part of the black hole information problem. One also wants to understand how information that was thrown into the black hole ends up escaping in the Hawking radiation. As argued in \cite{Penington:2019npb,Almheiri:2019psf}, this can be addressed using \emph{entanglement wedge reconstruction}.\footnote{This was originally conjectured in \cite{Headrick:2014cta,Czech:2012bh,Wall:2012uf}, and established in \cite{Jafferis:2015del,Dong:2016eik,Cotler:2017erl} using the ideas of \cite{Faulkner:2013ana}} In the case of the island conjecture, this idea implies that $R$ contains all of the information in the island $I$, which itself contains much of the black hole interior. Existing derivations of entanglement wedge reconstruction have been indirect,\footnote{A partial exception is \cite{Faulkner:2017vdd}, but it does not immediately apply to the case of an evaporating black hole.} relying on throwing entropy calculations into the meat grinder of modern  quantum information theory. So, the second main goal of this paper is to show using a bulk argument that operators in the radiation can manipulate the interior of the black hole. We show this directly by using gravitational path integrals to evaluate matrix elements of an explicit reconstruction operator defined using the Petz map \cite{petz1986sufficient,petz1988sufficiency} (see \cite{Cotler:2017erl,chen2019entanglement} for a description and recent discussion of the Petz map). Again, Euclidean wormholes play a crucial role. Here they are in some sense connecting the interior of the black hole to the quantum computer acting on the Hawking radiation.

We now give a brief summary of this paper.

In \hyperref[sec:asimplemodel]{{\bf section two}}
we introduce a simple toy model of an evaporating black hole in Jackiw-Teitelboim (JT) gravity \cite{Teitelboim:1983ux,Jackiw:1984je,Almheiri:2014cka}, a two dimensional truncation of near extremal black hole dynamics.   We represent the Hawking radiation by an auxiliary reference system whose states are entangled with interior partner modes represented by ``end of the world branes" (EOW branes) in the black hole interior.   This model is simple enough that we can exactly evaluate the full gravitational path integral for the R\'{e}nyi entropies by summing over all {planar} topologies with the correct boundary conditions.  We find that, before the analog of the Page time, the dominant topology consists of $n$ disconnected copies, one for each replica, of the single replica geometry. In contrast, after the Page time, the dominant topology is connected, with an $n$-boundary Euclidean wormhole connecting all the different replicas. {Our calculation uses a type of Feynman-diagram resummation method applied to spacetime geometry. This is inspired by techniques from the theory of free probability, and it makes it possible to continue the sum over topologies in $n$.}

We also compute the overlap between the different black hole microstates $| \psi_i \rangle$. We find that $\langle \psi_i | \psi_j\rangle = 0$ for $i \neq j$, but that $|\langle \psi_i | \psi_j\rangle |^2 \sim e^{-S}$ due to a wormhole contribution. This can only be consistent if the gravitational path integral represents an ensemble of quantum theories.\footnote{For this particular model this is indeed the case. In Appendix \ref{app:ensemble}, we construct an explicit ensemble of theories dual to our bulk model.}    The R\'{e}nyi entropies are sums of large numbers of these overlaps.  The Page transition comes from the buildup of these small errors in orthogonality.

In \hyperref[sec:petz]{{\bf section three}} we explicitly reconstruct the black hole interior  from the radiation in this simple model.  Our tool is the Petz map which we introduce and explain.   We then evaluate matrix elements of Petz map reconstructions using gravitational path integrals.  We employ  an analytic continuation similar to the replica trick and so consider geometries of different topologies. After analytic calculation, the problem reduces to a bulk field theory calculation in the original spacetime geometry.  

After the Page time, when the connected topology dominates,  the matrix elements of the Petz map reconstruction and original interior operator agree. In contrast, before the Page time, the disconnected topology dominates and the Petz map reconstruction fails to learn anything about the interior. Again we are able to sum these topologies and thereby follow this transition in detail. The error in the reconstruction increases linearly with the dimension of the code subspace. As a result, even long after the Page time, reconstruction is only possible for sufficiently small code subspaces, an example of state dependence. This explains the state dependence required in the Hayden-Preskill decoding process \cite{Hayden:2007cs} and gives a direct derivation of the results about state dependence in entanglement wedge reconstruction from \cite{Hayden:2018khn}.

The toy model calculations have some simplifying features that actual calculations for evaporating black holes do not.  We therefore extend our calculations to two less idealized models.  

 In \hyperref[sec:JT]{{\bf section four}} we calculate an analog of the Page curve using the replica trick in JT gravity coupled to a conformal matter, where the matter sector is allowed to flow freely across the boundaries between two thermofield double black holes.   Perturbative semiclassical gravity would suggest that the entropy of each thermofield double black hole would grow forever.   In contrast, unitarity and more explicitly the island formula  \eqref{eq:islandprescription} implies  that it must saturate at twice the Bekenstein-Hawking entropy.  We analyze this situation using replicas as before.   In this case, we cannot explicitly find the saddle point for integer  $n$ R\'{e}nyi entropies.  But we are able to  calculate the von Neumann entropy by showing that, as  $n \to 1$, the dominant saddle point after the Page time replicates about the minimal QES.  This  reproduces the answer from \eqref{eq:islandprescription}. This argument essentially reproduces the general arguments for the EW prescription from \cite{Dong:2017xht} in this specific context.

In \hyperref[sec:SYK]{{\bf section five}}, we do essentially the same calculations in a UV-complete theory, the SYK model. In this case, we can numerically find a replica-nondiagonal saddle point in the $G,\Sigma$ action that describes the saturation of the $n=2$ R\'{e}nyi entropy. According to the usual intuition in this context  for associating topologies to saddles for the  $G,\Sigma$ action
 \cite{Saad:2018bqo}, this saddle point corresponds to a Euclidean wormhole that connects the two replicas, as in our gravity calculations. 


In \hyperref[sec:desitter]{{\bf section six}} we turn briefly to de Sitter space. We discuss a generalization of the no-boundary proposal for replica computations, and use it to find a Page-like transition in the entropy in a toy model using EOW branes in the JT gravity realization of de Sitter space \cite{Maldacena:2019cbz}.  A notable feature is that the density matrix itself changes character at the analog of the Page transition.

In \hyperref[sec:discussion]{{\bf section seven}} we offer some preliminary remarks about the role of wormholes in theories without ensemble averaging.

We end with  several appendices: In \hyperref[app:zn]{\bf appendix A} we discuss details of the $n$-replica gravitational partition function. In \hyperref[app:dilatongravity]{\bf appendix B} we present analogous calculations in more general theories of two dimensional dilaton gravity, including flat space. In \hyperref[app:ew]{\bf appendix C} we give a direct gravitational path integral derivation of general entanglement wedge reconstruction using the Petz map. In \hyperref[app:ensemble]{\bf appendix D} we present an explicit Hilbert space ensemble precisely dual to the sum over all topologies of the simple model. In \hyperref[app:pagephases]{\bf appendix E} we discuss in some detail the connection between random tensor networks and fixed area states. In \hyperref[app:pagephases]{\bf appendix F} we present a more refined analysis  of the Page transition in the simple model.

\vspace{1em}
Closely related work has been done independently by Ahmed Almheiri, Tom Hartman, Juan Maldacena, Edgar Shaghoulian and Amirhossein Tajdini \cite{Almheiri:2019qdq}.  We have arranged with these authors to coordinate submission of our papers.

\newpage

\section{A simple model}\label{sec:asimplemodel}
\begin{figure}[t]
\begin{center}
\includegraphics[scale = .95]{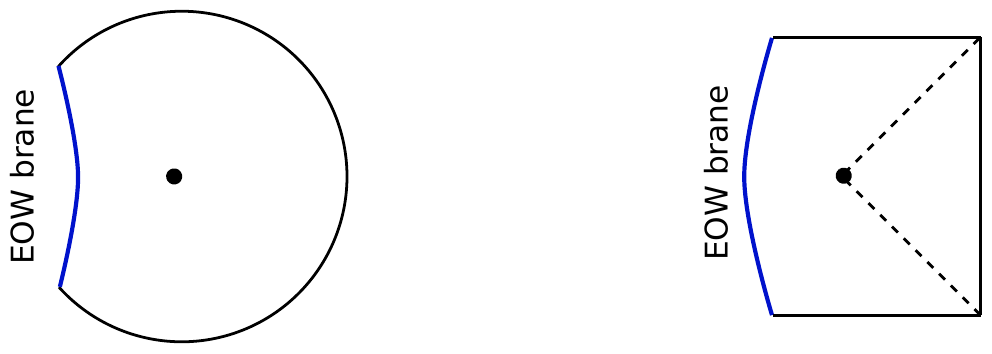}
\caption{{\small Euclidean and Lorentzian geometries for a black hole with an EOW brane behind the horizon.}}\label{fig1}
\end{center}
\end{figure}
\subsection{Setup of the model}
In this section, we will introduce a simple 2d gravity model in which one can derive the ``island'' prescription (and corrections to it) in a very explicit way using the replica trick.\footnote{A similar model has recently been  discussed independently in \cite{Rozali:2019day}.} The starting point is to consider a black hole in JT gravity, with an ``end of the world brane'' (EOW brane) behind the horizon, sketched in figure \ref{fig1}. This can be understood as a $\mathbb{Z}_2$ quotient of an ordinary two-sided black hole, with a particle of mass $\mu$ behind the horizon \cite{Kourkoulou:2017zaj}. The Euclidean action for the system is
\be
 I = I_{\sf{JT}} +  \mu\int_{\sf{brane}}  \mathrm{d}s,
\ee
where the integral is along the worldine of the EOW brane, and the pure JT action is
\be\label{eq:JTaction2}
I_{\sf{JT}} = -\frac{S_0}{2\pi}\left[\frac{1}{2}\int_{\mathcal{M}}\sqrt{g}R + \int_{\partial\mathcal{M}}\sqrt{h}K\right] -\left[ \frac{1}{2}\int_{\mathcal{M}}\sqrt{g}\phi(R+2) +\int_{\partial\mathcal{M}}\sqrt{h}\phi K\right].
\ee
In our discussion of the model, two different types of boundary conditions will be relevant. At the standard asymptotic boundary, we impose
\be
\mathrm{d}s^2|_{\partial\mathcal{M}} = \frac{1}{\epsilon^2}\mathrm{d}\tau^2, \hspace{20pt} \phi = \frac{1}{\epsilon}, \hspace{20pt} \epsilon \rightarrow 0.\label{bc1}
\ee
Here $\tau$ can be interpreted as the imaginary time coordinate of the holographic boundary dual to the JT gravity system. At the EOW brane, we impose the ``dual'' boundary conditions
\be
\partial_n\phi = \mu, \hspace{20pt} K = 0,\label{bc2}
\ee
where $\partial_n$ means the derivative normal to the EOW brane boundary, and $\mu\ge 0$.

We will be interested in the case where the EOW brane has a very large number $k$ of possible internal states, each orthogonal to the others. To model an evaporating black hole, we will think of these states as describing the interior partners of the early Hawking radiation. We will entangle them with an auxiliary system $\sf{R}$, which will model the early radiation of an evaporating black hole. So, all together, the state of the whole system is
\be
|\Psi\rangle = \frac{1}{\sqrt{k}}\sum_{i = 1}^k |\psi_i\rangle_{\sf{B}}|i\rangle_{\sf{R}}.\label{StateWholeSystem}
\ee
Here $|\psi_i\rangle_{\sf{B}}$ is the state of the black hole $\sf{B}$ with the brane in state $i$, and $|i\rangle_{\sf{R}}$ is a state of the auxiliary ``radiation'' system $\sf{R}$.
\begin{figure}[t]
\begin{center}
\includegraphics[scale = .95]{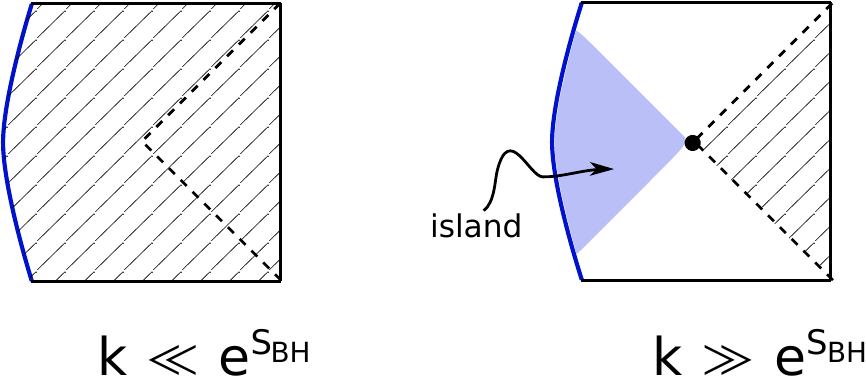}
\caption{{\small We show the expected entanglement wedges for our simple model based on the ``island'' conjecture. In the $k \ll e^{S_{BH}}$ phase, the entanglement wedge of the $\sf{B}$ system (boundary dual of the gravity theory) is the whole spacetime, shown hatched. In the $k \gg e^{S_{BH}}$ phase, an island develops. The entanglement wedge of the $\sf{B}$ system retreats to the exterior of the horizon, and the entanglement wedge of the auxiliary $\sf{R}$ system is the shaded blue island behind the horizon.}}\label{fig2}
\end{center}
\end{figure}

Let's compute the entropy of the $\sf{R}$ system in the state (\ref{StateWholeSystem}), using the island conjecture (\ref{eq:islandprescription}) of \cite{Penington:2019npb,Almheiri:2019psf,Almheiri:2019hni}. If we take the island to be the empty set, then the answer for the entropy is just $\log(k)$, representing the entanglement between the $\sf{R}$ system and the bulk state of the brane. However, we can also consider the case with a nontrivial island, as shown in figure \ref{fig2}. This island contains the EOW brane, so the bulk entropy term $S_{\text{naive}}(\sf{R}\cup I)$ in the island formula (\ref{eq:islandprescription}) will give zero. The remaining area term is interpreted in JT gravity as
\be
\frac{\text{Area}(\partial I)}{4G_N} \rightarrow  S_0 + 2\pi\, \phi(\partial I).
\ee
So the extremization in (\ref{eq:islandprescription}) amounts to putting the boundary of the island at an extremal point of the dilaton $\phi$. The geometry shown in figure \ref{fig1} has an extremal point of the dilaton, represented by the black dot at the bifurcation surface. The extremal value of $S_0 + 2\pi \phi$ is simply the coarse-grained entropy the black hole, $S_{BH}$. So, all together, the island conjecture predicts
\be
S({\sf{R}}) = \text{min}\big\{\log(k),S_{BH}\big\}.\label{islandC}
\ee
The transition between these two answers, as a function of $k$, is analogous to the Page curve.

For the simple model described above, we will derive this answer using the replica trick, along with corrections. We will start by deriving a qualitative first approximation to the full answer, and then gradually progress to calculating the full non-perturbative answer in Section \ref{sec:sumplanar}.

\subsection{Computation of the purity}
We would like to use the 2d gravity path integral to compute the entropy of system $\sf{R}$, in the state $|\Psi\rangle$ given in (\ref{StateWholeSystem}). We can start by discussing the density matrix $\rho_{\sf{R}}$:
\be
\rho_{\sf{R}} = \frac{1}{k}\sum_{i,j = 1}^k|j\rangle\langle i|_{\sf{R}} \ \langle \psi_{i}|\psi_j\rangle_{\sf{B}}.\label{rhoR}
\ee
The matrix elements of $\rho_{\sf{R}}$ are gravity amplitudes $\langle \psi_i|\psi_j\rangle$. These are determined (up to normalization of the states $|\psi_i\rangle$) by a gravity calculation with the following boundary conditions
\be
\langle \psi_i|\psi_j\rangle = \ \includegraphics[scale = .7,valign = c]{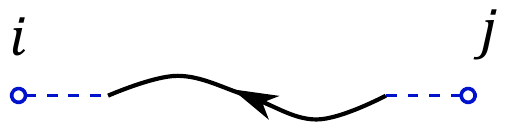}
\ee
Here, the black line is the asymptotic boundary of type (\ref{bc1}), with renormalized length $\beta$. We drew the line with a wiggle to emphasize that the boundary conditions do not constraint its shape. The arrow represents the direction of time evolution, from the ket to the bra.\footnote{In a theory with time-reversal symmetry, we would not draw such an arrow.} At the locations where the dashed blue lines intersect the solid black line, we require that an EOW brane of type $i$ or $j$ should intersect the asymptotic boundary. Here and below, dashed lines carry the index of the EOW brane state.

The leading gravity configuration that satisfies these boundary conditions is the following classical solution
\be
\langle \psi_i|\psi_j\rangle \approx \ \includegraphics[scale = .7,valign = c]{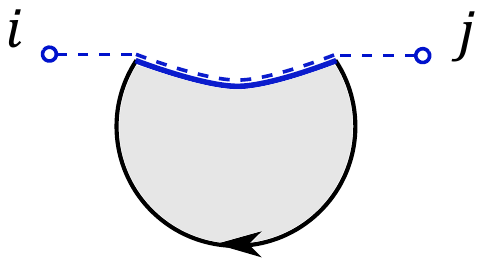}\label{gravCalc1}
\ee
The black asymptotic boundary (with arrow) borders a portion of the hyperbolic disk, and a solid blue EOW brane follows a geodesic between the $i$ and $j$ endpoints. An important feature is that the same EOW brane connects to both the $i$ index and the $j$ index. Because we assume that these correspond to orthogonal internal states of the EOW brane, the result for $\langle \psi_i|\psi_j\rangle$ will be proportional to $\delta_{ij}$. So, based on this computation, it looks like $\rho_{\sf{R}}$ will be maximally mixed, with entropy $\log(k)$:
\be
\rho_{\sf{R}} \stackrel{?}{=} \frac{1}{k}\sum_{i = 1}^k |i\rangle\langle i|.\label{question}
\ee

\begin{figure}[t]
\begin{center}
\includegraphics[scale = .75]{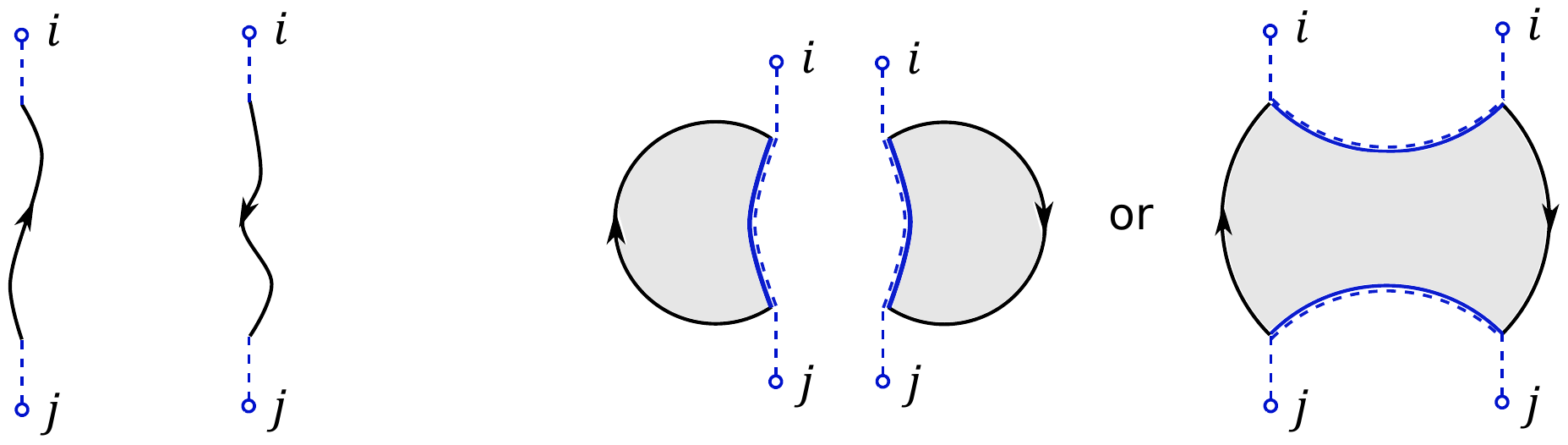}
\caption{{\small The boundary conditions for $|\langle \psi_i|\psi_j\rangle|^2$ are shown at left, and two ways of filling in the geometry are shown at right. To compute the purity, we want to sum over $i,j$ by connecting the dashed lines together. For the disconnected geometry, this will lead to a single $k$ index loop, and for the connected geometry it will lead to two loops.}}\label{fig1f}
\end{center}
\end{figure}

However, we can also try to compute the entropy directly, using the replica trick. Let's start by considering the so-called ``purity'' $\tr(\rho_{\sf{R}}^2)$, which is closely related to the Renyi 2-entropy. From (\ref{rhoR}), one can easily work out that
\be
\tr(\rho_{\sf{R}}^2) = \frac{1}{k^2}\sum_{i,j = 1}^k|\langle \psi_i|\psi_j\rangle|^2.\label{purityFormula}
\ee
The boundary conditions to compute $|\langle \psi_i|\psi_j\rangle|^2$ are shown in the left panel of figure \ref{fig1f}. To compute the purity, we will sum over $i,j$ by connecting the dashed lines in the obvious way.

The crucial point is that there are two different ways of filling these boundary conditions in with 2d geometry, as shown in the right panel of figure \ref{fig1f}. We can either have a disconnected geometry with the topology of two disks, or a connected ``Euclidean wormhole'' geometry with the topology of a single disk.\footnote{Euclidean wormholes are sometimes also referred to as spacetime wormholes, to differentiate them from the more commonly studied {\it spatial} wormholes such as Einstein-Rosen bridges.} To study a two-sided version of this model where we don't take the $\mathbb{Z}_2$ quotient, we would glue two copies of the EOW brane geometries together along the EOW branes. Then the topology would be two disks in the disconnected case, and one cylinder in the connected case.


In both topological classes, there is a classical solution in JT gravity. This is somewhat nontrivial for the connected (wormhole) case, for the following reason. There is a parameter that characterizes the wormhole geometry, which we can take to be the regularized distance from one of the asymptotic boundaries to the other, through the wormhole. By itself, the action of JT gravity provides a pressure that would prefer this distance to be large. But the tension provided by the mass $\mu$ of the particle or EOW brane prefers the length to be short, and provides a counterbalancing force that leads to a stable minimum for the action.\footnote{Note, however, that having a classical solution isn't necessary for the contribution to be meaningful.}

To describe the contributions of these geometries to $\tr(\rho_{\sf{R}}^2)$, we will use the notation $Z_n = Z_n(\beta)$ to represent the gravity path integral on a disk topology with a boundary that consists of alternating segments of $n$ physical boundaries of renormalized lengths $\beta$, and $n$ EOW branes. Using this notation, we can evaluate the sum of the two contributions in figure \ref{fig1f} as
\be
\tr(\rho_{\sf{R}}^2) =  \frac{kZ_1^2 + k^2Z_2}{(k Z_1)^2} = \frac{1}{k} + \frac{Z_2}{Z_1^2}.\label{visible}
\ee
In the numerator, we have the contributions of the two geometries in figure \ref{fig1f}: the disconnected geometry at left has one $k$-index loop, and two copies of the geometry that defines $Z_1$. The connected geometry at right has two $k$-index loops, and a single copy of the geometry that defines $Z_2$. In the denominator, we have divided by the gravity computation that normalizes the density matrix.\footnote{There is a subtlety here. If we normalize the state so that $\tr(\rho_{\sf{R}}) = 1$, we will find that $\tr(\rho_{\sf{R}})^2$ is not exactly one. This can be interpreted as due to the small fluctuation in the normalization of the states. This effect can be taken into account, but in the planar approximation that we will describe below, it is consistent to ignore it.}

We will work out exact formulas for $Z_n$ in JT gravity below, but the basic point can be seen already in a very crude approximation, where we retain only the dependence on the topological $S_0$ term in the JT gravity action (\ref{eq:JTaction2}). This term weights the contribution of a given topology by $e^{S_0\chi}$, where $\chi$ is the Euler characteristic. Since the topology relevant for $Z_n$ is disk-like for any value of $n$, and since $\chi = 1$ for the disk, we will have
\be
Z_n \propto e^{S_0}.\label{s0}
\ee
Using this formula, we see that the second term on the RHS of (\ref{visible}) is proportional to $e^{-S_0}$. So
\be
\tr(\rho_{\sf{R}}^2) = k^{-1} + e^{-S_0} \hspace{20pt} \text{(schematic)}.
\ee

If $k$ is reasonably small the disconnected geometry dominates, and we will find that the purity is $1/k$, which is consistent with (\ref{question}). However, if $k$ gets very big, then the connected geometry dominates and we will find the purity is $e^{-S_0}$, independent of $k$.   This interchange of dominance of the two saddlepoints  is the basic mechanism that prevents the entropy (here the Renyi entropy) of the radiation from growing indefinitely. Note that because the state of the entire system is pure, this implies that the entropy of the excitations in the black hole (EOW brane states) remains finite even when $k$ is very large. The basic mechanism for this is the small nonorthogonality of these states, as we discuss in the next section. When $k$ becomes of order $e^{S_{BH}}$, this nonorthogonality adds up to a large effect.

\subsection{Factorization and averaging}\label{sec:factorization and averaging}

The calculation in the previous section raises some important subtleties. For example, how can this result for $\tr(\rho_{\sf{R}}^2)$ be reconciled with the formula for $\rho_{\sf{R}}$ in (\ref{question})? To address this, let's first consider the amplitude $\langle \psi_i|\psi_j\rangle$. For this quantity, the gravity path integral gives two seemingly-contradictory answers
\be
\langle \psi_i|\psi_j\rangle = \delta_{ij},\hspace{20pt} |\langle \psi_{i}|\psi_j\rangle|^2 = \delta_{ij}+ \frac{Z_2}{Z_1^2}.\label{ansTWO}
\ee
The first of these equations follows from the gravitational calculation in (\ref{gravCalc1}): the answer is proportional to $\delta_{ij}$ because the same EOW brane connects to both the $i$ index and the $j$ index. The two terms in the second equation correspond to the two terms in the RHS of figure \ref{fig1f}. Note that in the second term $i$ does not have to be equal to $j$.

Of course, the two equations in (\ref{ansTWO}) are incompatible in a strict sense. However, suppose that we imagine that the true quantum amplitude is
\be\label{smallfluc}
\langle \psi_{i}|\psi_j\rangle = \delta_{ij} + e^{-S_0/2}R_{ij}
\ee
where $R_{ij}$ is a random variable with mean zero. Then if we interpret the gravity path integral as computing some type of average over the microscopic $R_{ij}$ quantities, then we would interpret the gravity answers as telling us that
\be
\overline{\langle \psi_i|\psi_j\rangle} = \delta_{ij}, \hspace{20pt} \overline{|\langle \psi_{i}|\psi_j\rangle|^2} = \delta_{ij}+ \frac{Z_2}{Z_1^2}.\label{amplitudes}
\ee
Here, the bar indicates an average over $R_{ij}$. Now the equations are compatible, and the extra term in the second equation tells us the variance of $R_{ij}$. Note that the correction to orthogonality in (\ref{smallfluc}) is very small.  But the large number of terms in  (\ref{purityFormula}) enables these small corrections to dominate the final answer for $k \gg e^{S_0}$, giving a qualitatively different result than (\ref{question}).

The conclusion seems to be that the gravity path integral is not literally computing quantum amplitudes, but is instead computing some coarse-grained version, where we average over some microscopic information $R_{ij}$. The lack of factorization of the resulting quantities is familiar from the connection between Euclidean wormholes and disorder.\footnote{A different type of factorization problem in JT gravity has been discussed in \cite{Harlow:2018tqv,Lin:2018xkj,Jafferis:2019wkd}.} This connection goes back to the work of  Coleman \cite{Coleman:1988cy} and of Giddings and Strominger \cite{Giddings:1988cx}, and has been recently discussed in JT gravity in \cite{Saad:2019lba}. It raises important questions in the present context, which we will return to in section \ref{sec:discussion}.

In appendix \ref{app:ensemble} we give a precise description of the random ensemble that is dual to the JT gravity plus EOW branes model.

\subsection{First look at the von Neumann entropy}
\begin{figure}[t]
\begin{center}
\includegraphics[scale = .7]{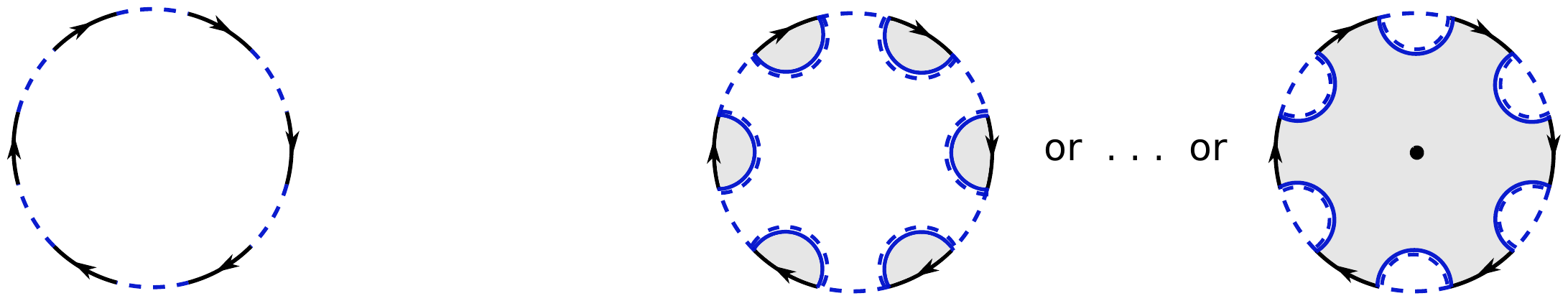}
\caption{{\small The boundary conditions for $\tr(\rho_{\sf{R}}^n)$ with $n = 6$ are shown at left, along with two extreme ways of filling in the geometry, corresponding to the completely connected and completely disconnected options. Note that the geometry at right contains a fixed point of the $\mathbb{Z}_n$ symmetry that rotates the replicas.}}\label{fig1g}
\end{center}
\end{figure}

The standard way to compute the von Neumann entropy using replicas is to use
\be
S_{\sf{R}} = -\tr(\rho_{\sf{R}}\log\rho_{\sf{R}})= -\lim_{n\to 1}\frac{1}{n-1}\log\tr(\rho_{\sf{R}}^n).
\ee
For this to be useful, it is important to be able to compute Renyi entropies in a way that has a good continuation in $n$. This is rather nontrivial, since as $n$ grows large, there are many different geometries that can fill in the boundary conditions for computing $\tr(\rho^n)$ (shown in the left panel of figure \ref{fig1g}). We will show how to handle this problem in the next section.

First, though, we can do a simplified analysis that is appropriate in two extreme limits.
In the case where $k \ll e^{S_{BH}}$, completely disconnected geometries with the topology of $n$ disks dominate. They have a single $k$-index loop, and give the contribution
\be
\tr(\rho_{\sf{R}}^n) \supset \frac{k Z_1^n}{k^n Z_1^n} = \frac{1}{k^{n-1}}.\label{disco}
\ee
Using (\ref{disco}), we find $S_{\sf{R}}= \log(k)$, consistent with (\ref{islandC}). 

In the opposite limit where $k \gg e^{S_{BH}}$, the completely connected geometry with the topology of a single disk dominates. There are $n$ index loops (see figure \ref{fig1g}), and the contribution is
\be
\tr(\rho_{\sf{R}}^n) \supset \frac{k^n Z_n}{k^n Z_1^n} = \frac{Z_n}{Z_1^n}.
\ee
The $k$ factors cancel out, so this answer is purely gravitational. In order to compute the von Neumann entropy, we need to continue $Z_n$ to near $n = 1$. A trick for doing this (see section 3.2 of \cite{Lewkowycz:2013nqa}) is to notice that the geometry associated to $Z_n$ has a $\mathbb{Z}_n$ replica symmetry, and that the $\mathbb{Z}_n$ quotient of the geometry can be continued in $n$. In the limit $n\rightarrow 1$, this becomes the original unreplicated geometry, and the computation of the von Neumann entropy reduces to $S_0 + 2\pi \phi_h$, where $\phi_h$ is the value of the dilaton at the horizon, which is the fixed point of the $\mathbb{Z}_n$ symmetry in the $n\rightarrow 1$ limit. So we conclude that in this phase, the answer is just the thermodynamic entropy of the black hole. Again, this is consistent with (\ref{islandC}).

The main conceptual point that we want to emphasize is that the island extremal surface descends from a Renyi entropy computation that involves replica wormholes.

\subsection{Summation of planar geometries} \label{sec:sumplanar}
So far, we have discussed the completely disconnected and completely connected replica geometries. In the regimes where these dominate, we get the two different answers for the von Neumann entropy, predicted by the island conjecture. In order to make this more precise, and to understand the transition between the two regimes, we need to sum over replica geometries that are intermediate between the completely disconnected and complete connected cases shown in figure \ref{fig1g}.

\begin{figure}[ht]
\centering
	\includegraphics[scale=0.7]{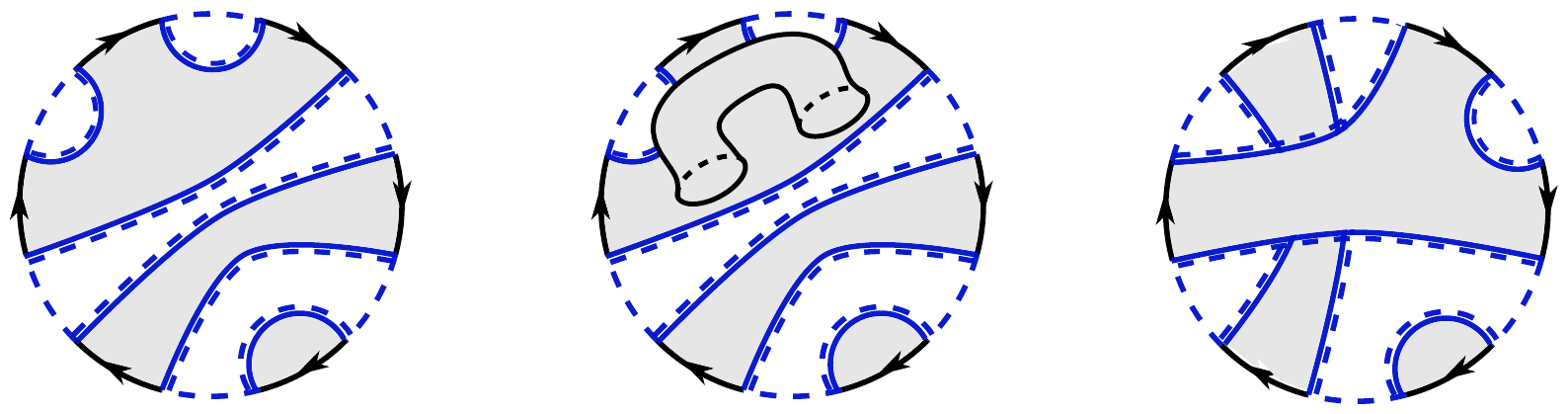}
	\caption{The left figure is an example of planar geometry that we need to include in our analysis. The middle figure has an extra handle and is down by $e^{-2S_0}$. The right figure involves a crossing, and is down by $k^{-2}$ (it has two dashed index loops instead of four).}
	\label{fig:Crossing and Handle}
\end{figure}

The starting point is the boundary condition for $\tr(\rho_{\sf{R}}^n)$, shown in the left panel of figure \ref{fig1g}. In principle, we need to sum over all ways of filling this boundary condition in. To simplify, we will assume that $e^{S_0}$ and $k$ are both large. In this approximation, we only need to sum over geometries that are planar, in the sense explained in figure \ref{fig:Crossing and Handle}. Non-planar geometries are suppressed either by powers of $e^{-2S_0}$ (for adding handles), or by powers of $1/k^2$ (for introducing crossings).

Depending on the relative size of $k$ or $e^{S_0}$, either highly disconnected or highly connected geometries will tend to dominate. We do not assume any particular relationship between $k$ and $e^{S_0}$, so it will be necessary to sum over all planar geometries. In order to do this, it is convenient to define the resolvent matrix $R_{ij}(\lambda)$ of $\rho_R$:
\be
R_{ij}(\lambda)=\left(\frac{1}{\lambda\mathbb{1}-\rho_{\sf{R}}}\right)_{ij} = \frac{1}{\lambda}\delta_{ij}+\sum\limits_{n=1}^{\infty}\frac{1}{\lambda^{n+1}}(\rho_R^n)_{ij}.
\ee
We will write down a Schwinger-Dyson equation for $R_{ij}$ using the planarity property and then use its solution to find the entanglement spectrum of $\rho_R$.\footnote{This analysis was inspired by the ``free probability" results discussed in \cite{speicher2009free} and e.g.~figure 1 of the earlier \cite{Cvitanovic:1980jz}.}

The boundary conditions for $R_{ij}$ correspond to an infinite sum over different numbers of asymptotic AdS$_2$ boundaries, connected by the $k$ index lines associated to the ${\sf{R}}$ system:
\be
\includegraphics[valign = c,scale = .55]{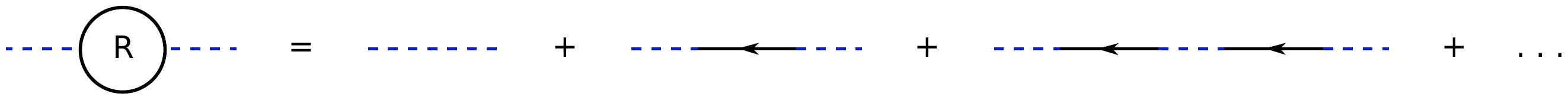}\label{sd1}
\ee
The dashed index lines come with factors of $1/\lambda$, and the solid lines with arrows come with factors $1/(kZ_1)$ that normalize the gravitational path integral. We fill these in with all possible planar geometries:
\be
\includegraphics[valign = c,scale = .55]{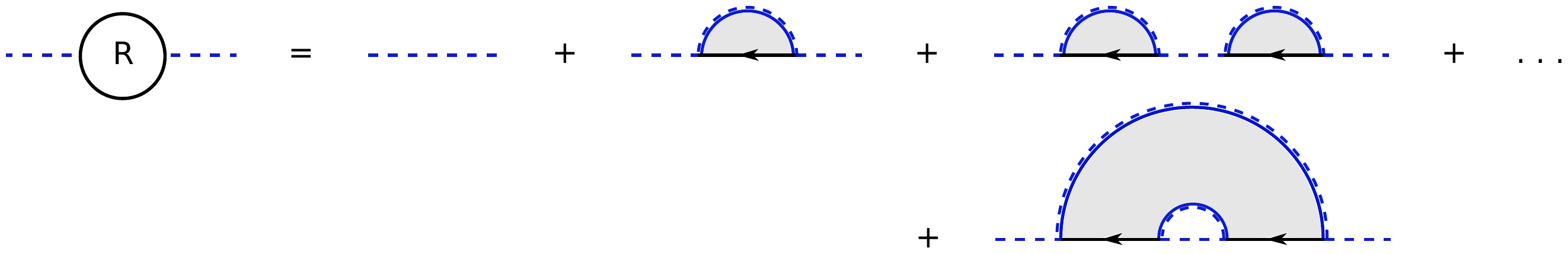}\label{sd2}
\ee
For the one-boundary terms in (\ref{sd1}), there is only a single bulk geometry that can fill it in. For the two-boundary term, there are two possible geometries, and we sum over them in (\ref{sd2}). For the three-boundary term (not shown), there would be five possible geometries.

We can write a Schwinger-Dyson equation that sums these geometries:
\be
\includegraphics[valign = c,scale = .55]{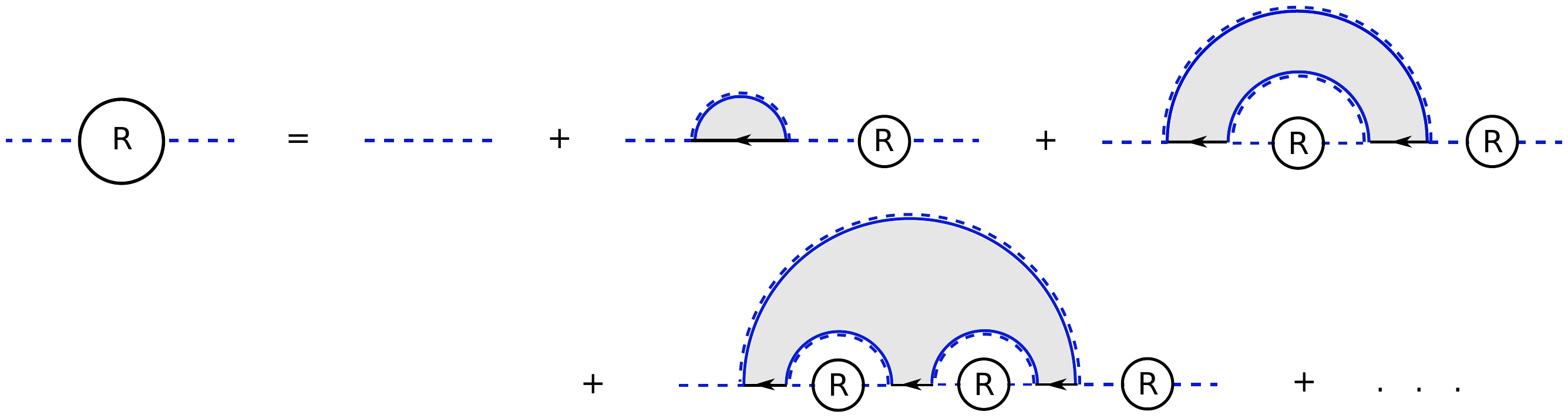}\label{sd3}
\ee
On the RHS, the second term sums all planar geometries in which the first boundary is disconnected from all other boundaries. The third term sums all planar geometries in which the first boundary is connected to one other boundary, and so on.

To write this identity as an equation, it is convenient to define $R$ as the trace of the resolvent matrix,
\be
R(\lambda) = \sum_{i = 1}^kR_{ii}(\lambda).
\ee
Then (\ref{sd3}) is equivalent to
\be
R_{ij}(\lambda)=\frac{1}{\lambda}\delta_{ij}+\frac{1}{\lambda}\sum\limits_{n=1}^{\infty}\frac{Z_n}{(k Z_1)^n}R(\lambda)^{n-1} R_{ij}(\lambda).\label{sd4}
\ee
We can think of the $\delta_{ij}/\lambda$ as a ``bare propagator."  It is the first term in (\ref{sd2}). In the remaining terms, $n$ labels the number of boundaries of the geometry that contains the leftmost boundary. The $Z_n $ factor is the gravitational path integral of an $n$ boundary geometry, which we will analyze below.
For each of the $n$ boundaries, we divided by $ k Z_1$ in order to normalize the density matrix $\rho_R$. Taking the trace of (\ref{sd4}), we find
\be
\lambda R(\lambda)=k+\sum\limits_{n=1}^{\infty}Z_n\frac{R(\lambda)^n}{ k^n Z_1^n} .
\label{eqn:SD-eqn}
\ee

In JT gravity, it is possible to make this equation very explicit.\footnote{For general two dimensional dilaton gravity theories, one can do the summation at the classical level and for large $\mu$. See appendix  \ref{app:dilatongravity}.} The exact formula for $Z_n$ can be worked out using the boundary particle formalism \cite{Yang:2018gdb,Kitaev:2018wpr}. We decompose the path integral for $Z_n$ in the following way
\be
\includegraphics[valign = c,scale=0.25]{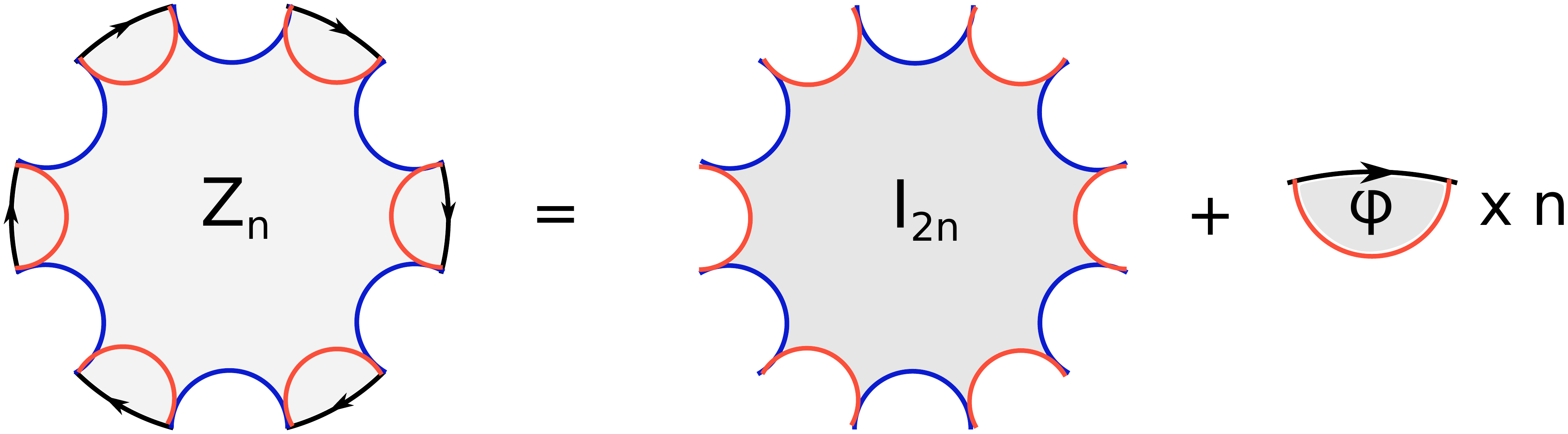}\label{decomp}
\ee
Here, the object $I_{2n}(\ell_1,\dots,\ell_{2n})$ is the JT gravity path integral with $2n$ geodesic boundaries of fixed regularized lengths $\ell$. It has the following expression \cite{Yang:2018gdb}
\be
I_{2n}(\ell_1,\dots,\ell_{2n})=2^{2n}\int_0^\infty \mathrm{d}s\, \rho(s)K_{2is}(4e^{-{\ell_1\over 2}})\dots K_{2is}(4e^{-{\ell_{2n}\over 2}});\hspace{20pt} \rho(s)={s\over 2\pi^2}\sinh(2\pi s),\label{intrepI}
\ee
where $K$ is the modified Bessel function. The object $\varphi$ in (\ref{decomp}) is the Hartle-Hawking state in the geodesic basis \cite{Harlow:2018tqv}. It computes the path integral from the asymptotic boundary (characterized by renormalized length $\beta$) to the geodesic of regularized length $\ell$. It has the expression \cite{Kitaev:2018wpr,Yang:2018gdb}
\be 
\varphi_{\beta}(\ell)=4e^{-{\ell\over 2}}\int_0^{\infty}\mathrm{d}s\, \rho(s)e^{-\beta s^2\over 2}K_{2is}(4 e^{-{\ell\over 2}}).\label{intrepphi}
\ee

The boundary of $I_{2n}$ consists of $n$ geodesics that correspond to EOW branes, and $n$ geodesics that need to be glued to Hartle-Hawking states. In both cases, the procedure is similar: we integrate over the length of the geodesic, with a ``wave function'' that is either the $e^{-\mu\ell}$ weighting for the EOW brane, or the $\varphi_\beta(\ell)$ weighting for the Hartle-Hawking states. Including the correct measure factor for the integral over geodesic lengths (see appendix \ref{app:zn}), the formula for $Z_n$ is
\be
\begin{split}
	Z_n=\,&e^{S_0}\int_0^\infty \mathrm{d}\ell_1 \dots\mathrm{d}\ell_{2n}e^{\ell_{1}+...+\ell_{2n}\over 2} I_{2n}(\ell_1,\dots\ell_{2n})\varphi_{\beta}(\ell_{1})e^{-\mu\ell_2}\dots\varphi_{\beta}(\ell_{2n-1})e^{-\mu \ell_{2n}}\\
	=\,&\int_0^\infty \mathrm{d}s\, \rho(s) y(s)^n;\hspace{20pt} y(s)=e^{-{\beta s^2\over 2}}2^{1-2\mu}|\Gamma(\mu-{1\over 2}+is)|^2.\label{finalZn}
\end{split}
\ee
In going to the second line from the first one, we inserted the integral representations (\ref{intrepI}) and (\ref{intrepphi}), and then did the integrals using formulas discussed in appendix \ref{app:zn}. In $y(s)$, the Boltzmann factor $\exp(-{\beta s^2\over 2})$ and the Gamma function factor $2^{1-2\mu}|\Gamma(\mu-{1\over 2}+is)|^2$ come from the integral with the Hartle-Hawking state and with the brane state, respectively.

Physically, the $s$ parameter can be viewed as giving the energy $s^2/2$ of a particular asymptotic region. The main simplification in the above calculation is the fact that the energy must be the same for all asymptotic boundaries in a single connected geometry.


Mathematically, (\ref{finalZn}) gives an integral representation for $Z_n$, with the property that the $n$ dependence is very simple inside the integral. The sum needed for the Schwinger-Dyson equation (\ref{sd3}) becomes a geometric series. In order to unclutter the equations that follow, we will define rescaled variables
\be
\uprho(s) = e^{S_0}\rho(s), \hspace{20pt} w(s) = \frac{y(s)}{Z_1},\label{rescaledvariables}
\ee 
Then, after summing the geometric series, one finds that the equation for the resolvent is
\be
\lambda R=k+\int_0^\infty \mathrm{d}s\, \uprho(s) {w(s)R\over k- w(s)R}.
\label{eqn: resolvent eqn}
\ee
This equation contains all of the information about the entanglement sepctrum of $\rho_{\sf{R}}$, and it is exact in the planar approximation.\footnote{A similar equation applies to arbitrary dilaton gravity theories, at least in the classical limit and with large $\mu$, see equation \ref{eqn: resolvent general dilaton gravity}.} By solving the equation for $R(\lambda)$, and then taking the discontinuity across the real axis, one can find the density of eigenvalues of the matrix $\rho_{\sf{R}}$:
\be
D(\lambda) = \frac{1}{2\pi i}\left[R(\lambda-i\epsilon) - R(\lambda+i\epsilon)\right].\label{dos}
\ee
The von Neumann entropy can then be computed by evaluating
\be
S({\sf{R}}) = -\int \mathrm{d}\lambda\, D(\lambda)\, \lambda \log(\lambda).\label{vnEnt}
\ee


\subsubsection{Microcanonical ensemble: Page's result}
Let's first use this planar resummation to compute the entanglement spectrum for the case where the black hole is in a microcanonical ensemble. This means that we fix the energy in each asymptotic region, rather than fixing the renormalized length $\beta$. Let $s$ be the chosen value of $s = \sqrt{2E}$, and let $\Delta s$ be the width of a small interval that defines the microcanonical ensemble. It is convenient to define the boldface quantities ${\bf S}$, ${\bf{Z}_1}$ and ${\bf w}$:
\be
e^{\bf S} = \uprho(s) \Delta s, \hspace{20pt} {\bf Z}_n = \uprho(s)y(s)^n\Delta s, \hspace{20pt} {\bf w}(s) = \frac{y(s)}{{\bf Z}_1} = e^{-{\bf S}}.
\ee
Here ${\bf S}$ is the entropy of our microcanonical ensemble, and ${\bf Z}_n$ is the microcanonical version of $Z_n$.

The advantage of this microcanonical ensemble is that the resolvent equation (\ref{eqn:SD-eqn}) simplifies to a quadratic equation for $R$. This equation, and the corresponding solution for the density of eigenvalues (\ref{dos}), are given by
\be
\begin{split}
	&R(\lambda)^2+\left({e^{{\bf S}}-k\over \lambda}-ke^{{\bf S}}\right)R(\lambda)+{k^2e^{{\bf S}}\over \lambda}=0;\\
	&D(\lambda)={k e^{{\bf S}}\over 2\pi \lambda}\sqrt{\left[\lambda-(k^{-{1\over2}}-e^{-{{\bf S}\over 2}})^2\right]\left[(k^{-{1\over 2}}+e^{-{{\bf S}\over 2}})^2-\lambda\right]}+\delta(\lambda) (k-e^{{\bf S}})\,\theta(k-e^{{\bf S}}).\label{microcanonicalAns}
\end{split}
\ee
These equations are precisely the same ones found by Page for the entanglement spectrum of a subsystem of dimension $k$ in a random state of total dimension $k e^{{\bf S}}$, in the planar approximation \cite{Page:1993df}. The fact that we get the random state answer can be understood from the fact that for fixed energy, the random ensemble dual to our JT gravity + EOW branes model (see appendix \ref{app:ensemble}) is the same random state ensemble discussed by Page.

1. \  When $k\ll e^{{\bf S}}$, the range of the spectrum is very narrow and the first term in $D(\lambda)$ is a very narrow semicircle, roughly a delta function. So the spectrum consists of $k$ eigenvalues of size $1/k$. This  describes a maximally mixed density matrix and the von Neumann entropy is $\log k$.

2. \ As $k$ approaches $e^{{\bf S}}$ (the Page time), the distribution develops a $1 / \sqrt{\lambda}$ singularity at the origin. After the Page time, the smooth part of the distribution contains only $e^{\bf{S}}$ eigenvalues. The remaining $k-e^{{\bf S}}$ are exactly zero. The von Neumann entropy has a rather sharp transition from $\log k$ to ${\bf S}$ during the Page transition: $S=\log m-{m\over 2n}$, where $m=\text{min}(k,e^{{\bf S}})$, and $n=\text{max}(k,e^{{\bf S}})$ \cite{Page:1993df}.

3. \ When $k\gg e^{{\bf S}}$, the smooth term can again be approximated as a delta function. The distribution describes a density matrix with $e^{{\bf S}}$ states that are maximally mixed, and the rest unentangled. The von Neumann entropy stays equal to ${\bf S}$.

\subsubsection{Canonical ensemble: smoothing out the transition}
\begin{figure}[t]
\centering
	\includegraphics[scale=0.65]{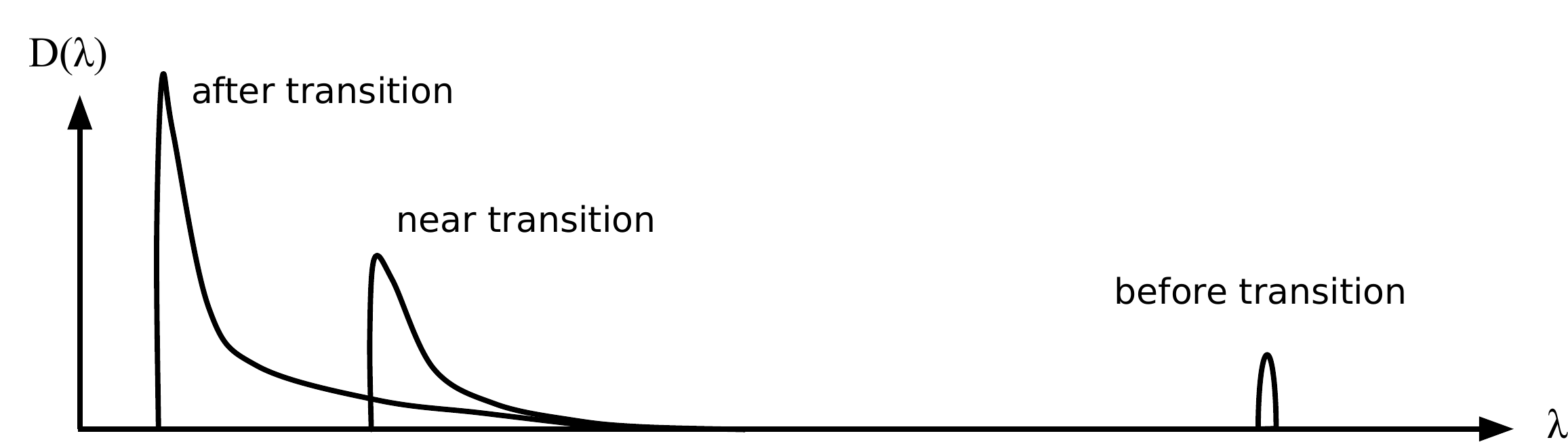}
	\caption{Sketch (not to scale) of the entanglement spectrum in the canonical ensemble. Before the Page transition, the density of states has a very narrow distribution. Near the Page transition, it becomes a shifted thermal spectrum, with a cutoff at some energy. After the Page time, it becomes an ordinary thermal spectrum, again with a cutoff at some energy.}
	\label{fig:densitydistribution}
\end{figure}

In the canonical ensemble, we did not find a way to solve the resolvent equation (\ref{eqn: resolvent eqn}) exactly. However, we can get some intuition by solving the equation numerically. To do this, one can first evaluate $\lambda$ as a function of $R$, find the locus where $\lambda$ is real, and then compute the inverse. The results of this procedure are sketched in figure \ref{fig:densitydistribution}. Long before the Page transition, the distribution of eigenvalues $D(\lambda)$ is very narrow, localized near $\lambda = 1/k$ as in the microcanonical case. Long after the Page transition, the distribution resembles the thermal spectrum of the black hole, with a rather sharp cutoff at $s = s_k$ chosen so that the total number of eigenvalues is $k$:
\be \label{eq:defskmaintext}
k = \int_0^{s_k}\mathrm{d}s\,{\uprho}(s), \hspace{20pt} (\text{definition of $s_k$}).
\ee
In between the two regimes, the curve smoothly interpolates, with no singular feature during the transition.

We will now try to analyze the resolvent equation approximately, with the goal of computing the von Neumann entropy in the semiclassical, small $G_N$ regime. In our formulas for JT gravity, we did not include a $G_N$ parameter explicitly, but it can be restored by taking
\be
\beta \rightarrow  G_N \beta,
\ee
so the semiclassical small $G_N$ limit corresponds to small $\beta$. We give a much more detailed analysis of the same problem in Appendix \ref{app:pagephases}.

As a first step, we can determine the location of the bottom edge in the distribution $D(\lambda)$, i.e.~the smallest eigenvalue $\lambda_0$. This corresponds to a branch point in $R(\lambda)$, which means a location where $d\lambda/dR = 0$. To determine this point, we first write the resolvent equation in the form
\be
\lambda={k\over R}+\int_0^\infty \mathrm{d}s\,\uprho(s){w(s)\over k-w(s)R}.
\label{eqn:lambda(R)}
\ee
In the region near the lower endpoint of the spectrum, $R$ is real and negative, and the above integral can be approximated by dividing the $s$ integral into two intervals where the two different terms in the denominator dominate:
\be
\lambda \approx \frac{k}{R}-\frac{1}{R}\int_0^{s_R}\mathrm{d}s\,\uprho(s) + \frac{1}{k}\int_{s_R}^\infty \mathrm{d}s\,\uprho(s)w(s).\label{approx1}
\ee 
Here $s_R$ is defined by $w(s_R)R = -k$. This approximation is justified in more detail in Appendix \ref{app:pagephases}.

Setting to zero the derivative of (\ref{approx1}) with respect to $R$, we find that the first two terms on the RHS should be equal. Comparison with (\ref{eq:defskmaintext}) then implies that $s_R\approx s_k$. Plugging this back into (\ref{approx1}), we find that the smallest eigenvalue of $\rho_{\sf{R}}$, and the corresponding value of $R$ are
\be
\lambda_0 \approx \frac{1}{k}\int_{s_k}^\infty \mathrm{d}s\,\uprho(s) w(s),  \hspace{20pt} R(\lambda_0) \approx -\frac{k }{w(s_k)}.\label{lambda0}
\ee

The next step is to write the equation for the resolvent (\ref{eqn: resolvent eqn}) and break up the $s$ integral into two parts:
\be
\lambda R\approx k+\int_0^{s_k} \mathrm{d}s\,\uprho(s){w(s)R\over k -w(s) R}+\frac{R}{k}\int_{s_k}^{\infty}\mathrm{d}s\,\uprho(s)w(s)
\label{eqn: approximated revolent equation}
\ee
In the region from $s_k$ to infinity, we replaced the factor of $k-w(s)R$ in the denominator by $k$. This is justified on the following grounds. First, we are going to study the equation in the region $\lambda > \lambda_0$, and in this region $|R| < |R(\lambda_0)|$. Second, $w(s)$ is a decreasing function of $s$, so for $s > s_k$, we have $|w(s)R| \le |w(s_k) R(\lambda_0)| = k$. Intuitively, we have separated the $s$ integral into two terms in (\ref{eqn: approximated revolent equation}) corresponding to the ``pre-Page'' and ``post-Page'' parts of the thermal ensemble. The high energy states with $s > s_k$ are effectively before the Page transition, and the planar resummation (represented by the nontrivial denominator) is not necessary for these states. 

A nice feature of (\ref{eqn: approximated revolent equation}) is that the final term can be recognized as $\lambda_0 R$, so (\ref{eqn: approximated revolent equation}) can be rewritten
\be
(\lambda-\lambda_0)R \approx k+ \int_0^{s_k}\uprho(s) \frac{w(s)R}{k - w(s)R}.
\ee
We can solve this equation in an approximation where the the second term is a small perturbation, and the zero-th order solution is just $k / (\lambda-\lambda_0)$. Iterating the equation once, we find the first-order solution
\be
	R(\lambda)\approx {k\over \lambda-\lambda_0}+{1\over \lambda-\lambda_{0}}\int_0^{s_k} \mathrm{d}s\,\uprho(s){w(s)\over \lambda-\lambda_0-w(s)}.
\label{exp:resolvent1}
\ee 
This approximation is good as long as $\int_0^{s_k}\mathrm{d} s\,\uprho(s){w(s)\over \lambda-\lambda_0-w(s)}\ll k$, which will be true as long as 
\be
\lambda > \lambda_0 +  w(s_k-\delta).\label{condforsuccess}
\ee
with $\delta$ a control parameter. So we conclude that for eigenvalues $\lambda$ satisfying  (\ref{condforsuccess}), we can compute the resolvent to good accuracy. Unfortunately, there is a gap between this value and the actual bottom of the spectrum $\lambda_0$, and we do not have control over the distribution of eigenvalues in this region. Before and during the Page transition, we are saved by the fact that $\lambda_0 \gg w(s_k)$, so for the purposes of computing the von Neumann entropy, this is a narrow region where the density of eigenvalues can be approximated by a delta function with unknown weight. After the Page transition, this is no longer true, but the eigenvalues in the unknown region contribute a small amount to the entropy.

So, using (\ref{eqn: approximated revolent equation}) in the region (\ref{condforsuccess}) and parametrizing our ignorance in the remaining eigenvalues with a delta function, we have
\be
D(\lambda) \approx \# \delta(\lambda-\lambda_0) + \int_0^{s_k-\delta}\mathrm{d}s\,\uprho(s)\delta(\lambda-\lambda_0-w(s)).
\ee
In the region between $s_k-\delta$ and $s_k$, $\lambda_0 \gg w(s)$, so we can approximate $\lambda_0 + w(s) \approx \lambda_0$. This means that we can let the integral run all the way to $s_k$, at the cost of changing the coefficient of the delta function piece. The resulting combined coefficient can be shown to vanish, by the following argument. The distribution of eigenvalues has to satisfy two normalization conditions,
\be
\int \mathrm{d}\lambda\,D(\lambda) = k,\hspace{20pt}\int \mathrm{d}\lambda\,D(\lambda)\lambda = 1.
\ee
The first condition says that $\rho_{\sf{R}}$ has $k$ total eigenvalues, and the second condition says that $\rho_{\sf{R}}$ is normalized. The first of these equations implies that
\be
D(\lambda) = \int_0^{s_k}\mathrm{d}s\,\uprho(s)\delta(\lambda-\lambda_0 - w(s))\label{finalD}
\ee
is correctly normalized without any further delta function piece. One can check that the second condition follows from our result for $\lambda_0$ in (\ref{lambda0}).

This is our final expression for the entanglement spectrum. It can be given a very simple interpretation: we have the spectrum of the first $k$ states of the thermal spectrum of the black hole, shifted so that the total normalization is one. This shift can be understood as the contribution of all of the remaining states in the spectrum of the black hole:
\be
\rho_{\sf{R}}\simeq \mathbb{P} \rho_{BH} \mathbb{P}+{\mathbb{1}\over k}\tr\left(\mathbb{ \bar P} \rho_{BH}\mathbb{\bar P }\right).
\ee
Here $\mathbb{P}$ is the projector into the post-Page subspace with $s < s_k$, and $\mathbb{\bar P}$ is the projector into the pre-Page subspace with $s > s_k$. The operator $\rho_{BH}$ is the density matrix of the thermal ensemble (including the brane) for the black hole.

\begin{figure}[t]
\begin{center}
\includegraphics[width = .45\textwidth]{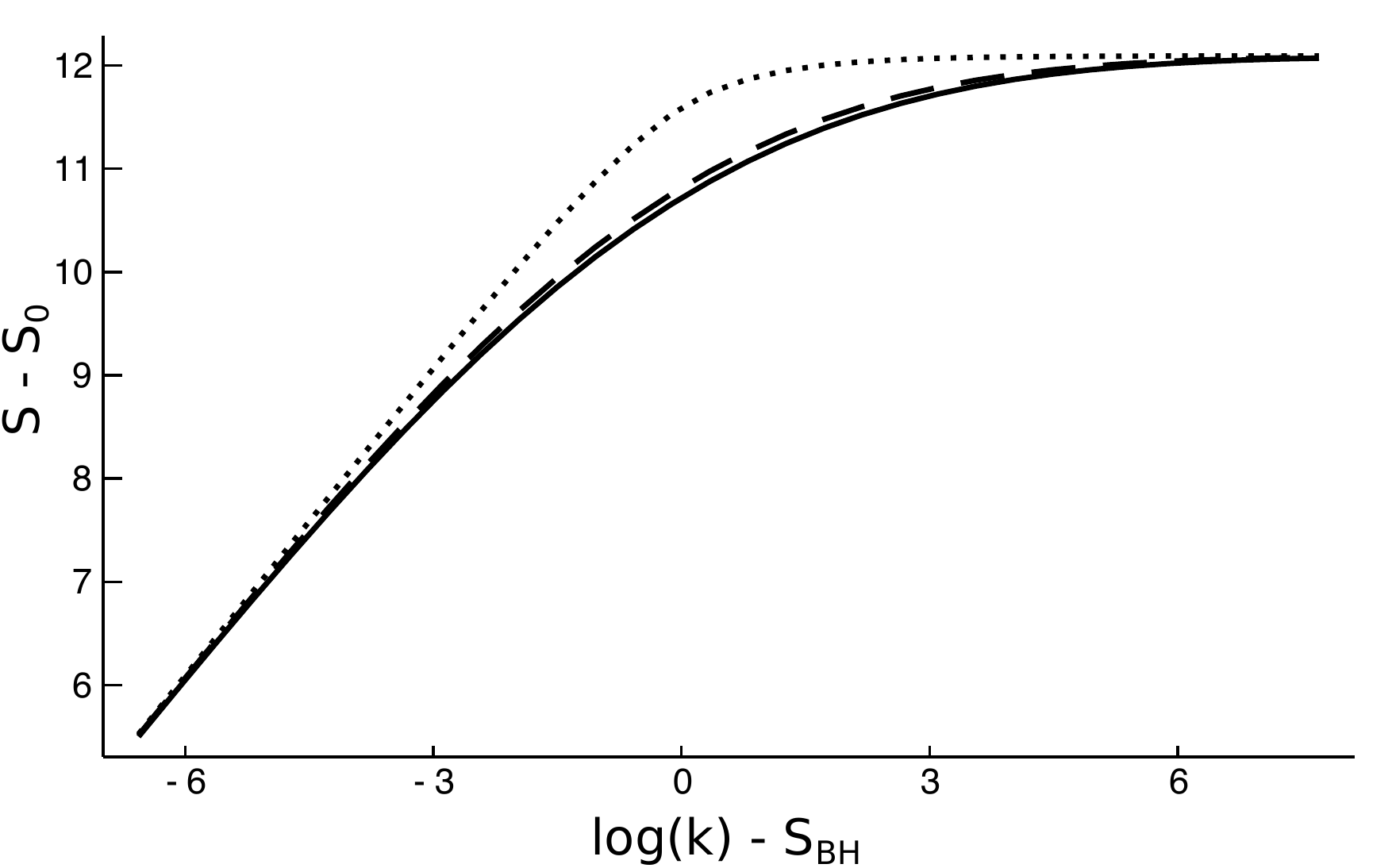}
\caption{{\small We plot the exact entropy in the planar approximation (solid) and the approximation (\ref{approxPage}) (dashed) for the case $\beta = 3$ and large $\mu$. The answer in the microcanonical ensemble with the same entropy is also shown (dotted). The difference between the curves is largest near the Page transition. For small $\beta$, the transition in the canonical curves gets smoothed out over a large region of size $\beta^{-1/2} \sim G_N^{-1/2}$ in $\log(k)$, while the transition in the microcanonical curve always looks the same. The maximum error between the canonical curve and the simple approximation (\ref{approxPage}) will be of order $\beta \sim G_N$.}}\label{fig:entropy}
\end{center}
\end{figure}
Using this result, one can evaluate the von Neumann entropy. Inserting (\ref{finalD}) into (\ref{vnEnt}) and doing the $\lambda$ integral using the delta function, we find
\be
	S\approx -\int_0^{s_k}\mathrm{d}s\,\uprho(s)(\lambda_0+w(s))\log(\lambda_0+w(s)).\label{approxPage}
\ee
A careful analysis shows that the error in this approximation peaks around the Page transition, when it is of order $G_N$ (see appendix \ref{app:pagephases} for details). As a function of $k$, this expression exhibits a smoothed-out version of the Page transition. It differs from the naive answer $\text{min}(k,S_{BH})$ by an amount of order $G_N^{-1/2}$ near the Page transition, but one can show that it agrees to within $O(1)$ precision with the average of the micro-canonical answer over the full thermal ensemble. In figure \ref{fig:entropy} we plot the exact formula for the entropy for $\beta = 3$, obtained by solving the resolvent equation numerically. We also plot the approximation (\ref{approxPage}) and the exact microcanonical answer for comparison.

We would like to emphasize one point about the continuation in $n$. Because they are sensitive to different parts of the thermal spectrum, the different Renyi entropies $\tr(\rho^n)$ experience Page transitions at different values of $k$. In fact, there are values of $k$ for which the Page transition for all of the integer Renyi entropies has already taken place, but the von Neumann entropy is still very close to $\log(k)$. In this regime, a fully connected geometry dominates the computation of the Renyi entropy for every integer $n$, but its continuation to $n = 1$ would give the wrong answer in the von Neumann limit. Nevertheless, the full resummation gives the right answer.



\section{Reconstruction behind the horizon via the Petz map}\label{sec:petz}
In the previous section, we showed (in the simple model) that for large $k$, the entanglement entropy saturates at the thermodynamic entropy of the black hole, and the RT surface is at the horizon. This means that the entanglement wedge of the radiation $\sf{R}$ contains an island behind the horizon.

The location of the entanglement wedge is significant because of the notion of ``entanglement wedge reconstruction,''\footnote{This was originally conjectured in \cite{Headrick:2014cta,Czech:2012bh,Wall:2012uf}, and established in \cite{Jafferis:2015del,Dong:2016eik,Cotler:2017erl} using the ideas of \cite{Faulkner:2013ana}.} which implies that in the large $k$ phase, the radiation system $\sf{R}$ should describe the island. The arguments in favor of entanglement-wedge reconstruction \cite{Faulkner:2013ana,Jafferis:2015del,Dong:2016eik,Cotler:2017erl} are strong but non-constructive. In the next section we will show directly from the gravitational path integral that operators acting in the $\sf{R}$ system can control the region behind the horizon. We will see that the Euclidean wormholes described above play an essential role, connecting the region behind the horizon to the $\sf{R}$ system.

Although we will focus on the simple model in our discussion below, a similar argument applies to the general case of operator reconstruction within the entanglement wedge, e.g.~the famous case of two intervals in the AdS$_3$ vacuum. The argument is essentially the same and we describe this generalization in appendix \ref{app:ew}.

\subsection{Setup}
We would like to slightly enrich the simple model discussed above, so that there is something nontrivial to reconstruct. To do this, we will imagine coupling our JT gravity system to a bulk field theory, containing some propagating degrees of freedom. For simplicity, we will assume that these propagating degrees of freedom cannot feel the difference between different states of the EOW brane.

We will consider a ``code subspace'' of states corresponding to small excitations of the bulk fields propagating on the background described so far (i.e.~by the background corresponding to the state (\ref{StateWholeSystem}), and to the geometry in figure \ref{fig1}). A basis for the code subspace is provided by the states
\be
|\Psi_a\rangle = k^{-1/2}\sum_{j = 1}^k|\psi_{aj}\rangle_{\sf{B}}|j\rangle_{\sf{R}}, \hspace{20pt} a = 1,\dots,d_{\sf{code}}.
\ee
Here, $|\psi_{aj}\rangle$ is a state in which the EOW brane is in state $j$, and we also have a small excitation of the propagating bulk fields, labeled by $a$. 

Because of the variety of geometries that will be important below, we actually have to be somewhat more precise at this point, and define $|\psi_{aj}\rangle$ by a boundary condition rather than a bulk statement. What we really mean is that the $a$ index corresponds to some insertion in the boundary conditions for the Euclidean path integral that defines the bulk state. So, for example, the boundary conditions for an inner product of two such states would be
\be
\langle \psi_{ai}|\psi_{bj}\rangle = \includegraphics[scale = .7,valign = c]{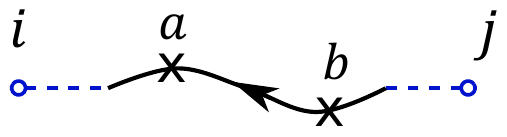}\label{melement3}
\ee
where the labeled ``$\sf{x}$'' marks represent the insertions associated to $a$ and $b$. We assume that e.g.~the $a$ insertion is arranged so that on the leading disk topology, it produces a particular state of the bulk field theory $|a\rangle_{\sf{FT}}$,
\be
\includegraphics[scale = .7,valign = c]{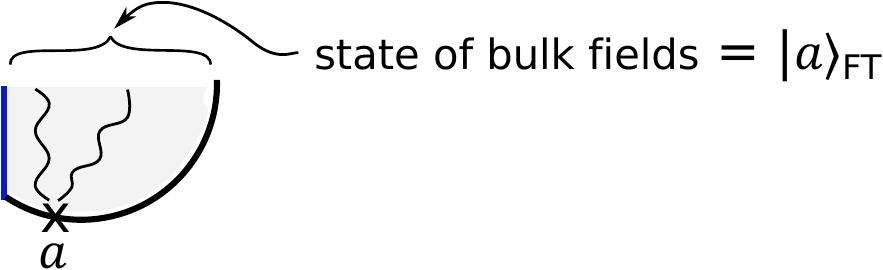}\label{ft}
\ee
Relative to the state with no insertions, $|a\rangle_{\sf{FT}}$ could contain excitations behind the horizon, outside the horizon, or both. To simplify the formulas below, we will assume that the bulk states are indexed and normalized so that to a good approximation $\langle a|b\rangle_{\sf{FT}} = \delta_{ab}$. We will specialize to a more concrete model in Section \ref{sec:nonperPetz}.

We let $\mathcal{O}_{\sf{FT}}$ be some operator that acts within this subspace of bulk states. We will denote its matrix elements in the bulk field theory states as $\mathcal{O}_{ab}$:
\be
\mathcal{O}_{ab}\equiv \langle a|\mathcal{O}_{\sf{FT}}|b\rangle_{\sf{FT}}.
\ee
We define $\mathcal{O}$ to be the representation of the bulk operator $\mathcal{O}_{\sf{FT}}$ in the full boundary system
\be
\mathcal{O} = \sum_{a,b = 1}^{d_{\sf{code}}}|\Psi_a\rangle\langle \Psi_b| \times \mathcal{O}_{ab}.\label{petzOpFirst}
\ee
Note that in general, this ``global'' representation of the operator acts nontrivially on both $\sf{R}$ and $\sf{B}$.

Now we can get to the point. Suppose that we are in the phase with $k \gg e^{S_{BH}}$, so that the entanglement wedge of $\sf{R}$ includes the island behind the horizon. Then if $\mathcal{O}_{\sf{FT}}$ acts within the ``island'' behind the horizon, the general arguments for entanglement-wedge reconstruction suggest that we should be able to find an operator $\mathcal{O}_{\sf{R}}$, acting only on $\sf{R}$, such that 
\be
\langle \Psi_a|\mathcal{O}_{\sf{R}}|\Psi_b\rangle \approx \langle \Psi_a|\mathcal{O}|\Psi_b\rangle.\label{toCheck}
\ee
What we would like to do is show by a direct bulk calculation that this is possible. The tool we will use is the so-called ``Petz map,'' which essentially gives a guess for what operator $\mathcal{O}_{\sf{R}}$ to choose. We will do a bulk calculation to show that it actually works, demonstrating entanglement-wedge reconstruction explicitly.

\subsection{Petz Lite} \label{sec:petzlite}
Before we discuss the Petz map itself, we will go through a simpler (``Petz Lite'') version, which will be good enough for a certain limited class of states. The Petz Lite map works as follows. Given an operator $\mathcal{O}$ on the combined system $\sf{BR}$, we can define a operator on the $\sf{R}$ system using the partial trace
\be
\mathcal{O}_{\sf{R}} = c_0\tr_{\sf{B}}(\mathcal{O}).\label{petzLITE}
\ee
Here, the constant $c_0$ should be chosen so that the identity operator maps to the identity operator. This seems like a very naive guess for an operator mapping, and in fact this Petz Lite version will not work for general states. But it does work in a special class of states, analogous to the ``fixed area'' states of Akers, Rath, Dong, Harlow and Marolf \cite{Akers:2018fow,Dong:2018seb}.

In JT gravity, fixed area is replaced by fixed dilaton $\phi$, and we choose to fix the value of $\phi$ at the horizon to be the value that it has in the classical solution. The advantage of doing this is that when $\phi$ is fixed, we don't impose the equation $R + 2 = 0$ at that point, and instead we allow any value of $R$, including a delta function singularity that corresponds to a conical excess. This freedom makes replica geometries very easy to construct \cite{Akers:2018fow,Dong:2018seb}: we just take $n$ copies of the $n = 1$ geometry, and glue them together in such a way that the horizon is a conical singularity with total angle $2\pi n$.

To proceed, let's define boldface states $|\bm{\psi}_{ai}\rangle$ to be fixed-dilaton versions of the states $|\psi_{ai}\rangle$ defined above. Making this modification in (\ref{petzOpFirst}) and tracing over the $\sf{B}$ system as in (\ref{petzLITE}), we find the candidate operator
\be
\mathcal{O}_{\sf{R}} = \frac{c_0}{k}\sum_{i,j = 1}^k|i\rangle\langle j|_{\sf{R}} \ \sum_{a,b = 1}^{d_{\sf{code}}}\langle\bm{\psi}_{bj}|\bm{\psi}_{ai}\rangle_{\sf{B}} \ \mathcal{O}_{ab}.
\ee
How does this operator manage to affect the region behind the horizon? Note that although $\mathcal{O}_{\sf{R}}$ acts only on $\sf{R}$, its explicit form depends in detail on amplitudes $\langle \bm{\psi}_{ai}|\bm{\psi}_{bj}\rangle$ in the black hole theory $\sf{B}$. Let's imagine that some quantum computer will apply $\mathcal{O}_{\sf{R}}$ to the $\sf{R}$ system. Then in order to compute these amplitudes, the quantum computer will do computations that are equivalent to simulating system $\sf{B}$. This introduces a second copy of $\sf{B}$ into the game.
\begin{figure}[t]
\begin{center}
\includegraphics[width = .7\textwidth]{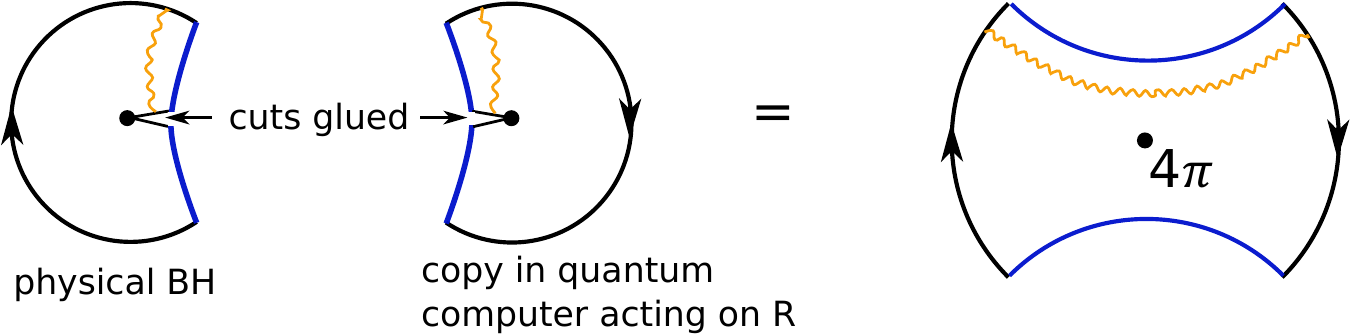}
\caption{{\small The two-replica geometry that contributes to the expectation value (\ref{PetzLiteexp}) of the Petz Lite operator. The wavy line represents the bulk particle. Note that gluing the right geometry to the left geometry has the same effect (from the perspective of the physical left system) as inserting a creation operator behind the horizon. Another sketch of the glued geometry is shown at right. The marked point is the point of fixed dilaton, and it has conical angle $4\pi$. The worldline of the particle created in the quantum computer (wiggly orange line) travels through the wormhole and into the physical black hole. The brane on the top half of each of the figures is a brane of type $j$ and the brane on the top half is of type $i$.}}\label{fig:petzLite}
\end{center}
\end{figure}

The basic point seems to be that the physical copy of the black hole $\sf{B}$ can connect via a Euclidean wormhole to this second copy, providing a geometrical connection between the black hole and $\sf{R}$. This connection is equivalent to inserting an operator behind the horion of the physical black hole.

Let's see how this works in a very simple case, where the code subspace consists of two states: the state $|0\rangle_{\sf{FT}}$ with no bulk particles excited, and $|1\rangle_{\sf{FT}}$ with a single particle behind the horizon. We consider the operator
\be
\mathcal{O}_{\sf{FT}} = |1\rangle\langle 0|\label{example1}_{\sf{FT}}
\ee
which creates a particle. For this case, Petz Lite gives
\be
\mathcal{O}_{\sf{R}} = \frac{c_0}{k}\sum_{i,j=1}^k |i\rangle\langle j|_{\sf{R}}\,\langle\bm{\psi}_{0j}|\bm{\psi}_{1i}\rangle.
\ee
Here $|\bm{\psi}_{0j}\rangle$ is a state with no particle, and $|\bm{\psi}_{1i}\rangle$ is a state with the particle present. If we were to evaluate the amplitudes $\langle\bm{\psi}_{0j}|\bm{\psi}_{1i}\rangle$ using a single replica, we would find zero because the bulk field theory states with and without a particle are orthogonal. So the amplitude would vanish, and we would conclude that $\mathcal{O}_{\sf{R}} = 0$. 

However, as discussed  in section \ref{sec:factorization and averaging}, the gravity path integral seems to require an interpretation as a disorder average, where the amplitude $\langle\bm{\psi}_{0j}|\bm{\psi}_{1i}\rangle$ is only zero ``on average'' in some ensemble implicit in the gravitational description. So we should postpone our concern that the operator vanishes, and see what it does when we compute a matrix element:
\be
\langle \bm{\Psi}_a|\mathcal{O}_{\sf{R}}|\bm{\Psi}_b\rangle = \frac{c_0}{k^2}\sum_{i,j=1}^k\langle \bm{\psi}_{ai}|\bm{\psi}_{bj}\rangle\langle \bm{\psi}_{0j}|\bm{\psi}_{1i}\rangle.\label{PetzLiteexp}
\ee
On the RHS we have a product of two gravitational amplitudes. The first factor corresponds to the physical black hole, and the second factor corresponds to the simulated one in the quantum computer acting on ${\sf{R}}$. In the large $k$ phase, the largest contribution to the expression on the RHS comes from a connected geometry, where the physical black hole is connected to the simulated one by a wormhole. This wormhole is particularly easy to describe for the fixed-dilaton states. We simply glue two copies of the one-replica geometry together along a cut behind the horizon. This corresponds to a geometry with a $4\pi$ angle at the horizon, as sketched in figure \ref{fig:petzLite}. From the perspective of the physical black hole, this wormhole corresponds to an insertion of an operator behind the horizon. It is an operator with a particle present in the ket vector (top half of the cut) and no particle present in the bra vector (bottom half of the cut). So the operator acts as a creation operator behind the horizon.

This is the basic mechanism that underlies entanglement-wedge reconstruction. However, in order to go beyond fixed-dilaton states, we will have to use the full Petz map, rather than the Petz Lite version discussed above.

\subsection{Quantum information discussion of the Petz map}\label{sec:qipetz}
{We now introduce the full Petz map, in its natural environment of quantum channels. The remarks below will not be necessary for our gravity discussion, but we include them for background.}

A reversible quantum channel $\mathcal{N}$ can be defined by the property that it has a type of inverse $\mathcal{R}$, such that $(\mathcal{R}\circ \mathcal{N})(\rho) = \rho$ for all inputs to the channel $\mathcal{N}$. The Petz map \cite{petz1986sufficient,petz1988sufficiency,hayden2004structure} gives a universal formula for $\mathcal{R}$ in terms of $\mathcal{N}$, in cases where exact recovery is possible. We will not write the full formula, since we are only interested in a special case of it, but we refer the reader to  \cite{cotler2019entanglement,chen2019entanglement} for details and previous discussions in the context of entanglement wedge reconstruction.

As a representative example, let's consider a quantum channel that embeds $\mathcal{H}_{\sf code} \subseteq \mathcal{H}_{A} \otimes \mathcal{H}_{\bar A}$, and then traces over $\mathcal{H}_{\bar A}$. The channel is reversible if it is possible to recover $\mathcal{H}_{\text{code}}$ from $\mathcal{H}_A$. In this situation, for any operator $\mathcal{O}_{\sf code}$ acting on $\mathcal{H}_{\sf code}$, the Petz map gives an equivalent operator $\mathcal{O}_A$ acting on $A$:
\begin{align}
\mathcal{O}_A &= \sigma_A^{-1/2} \tr_{\bar A}(V \mathcal{O}_{\sf code}V^\dag)\sigma_A^{-1/2}.\label{petzQI}\\
\sigma_A &= \tr_{\bar A}(\Pi_{\sf code}).
\end{align}
Here $V$ is the encoding map from $H_{\sf code}$ into $\mathcal{H}_A\otimes \mathcal{H}_{\bar A}$, and $\Pi_{\sf code}$ is the projector onto the image of $\mathcal{H}_{\sf code}$ in $\mathcal{H}_{A}\otimes \mathcal{H}_{\bar A}$. By saying that $\mathcal{O}_A$ is equivalent to $\mathcal{O}_{\sf code}$, we mean that for any state in $\mathcal{H}_{\sf code}$, the following two operations produce the same state: (a) acting with $\mathcal{O}_{\sf code}$ and then encoding with $V$ and (b) encoding with $V$ first and then acting with $\mathcal{O}_A$.

To see that (\ref{petzQI}) works, we note that reversiblity of the channel implies that there is an isomorphism $\mathcal{H}_A \cong \mathcal{H}_1 \otimes \mathcal{H}_2 \oplus \mathcal{H}_3$ where $\mathcal{H}_1 \cong \mathcal{H}_\text{code}$ such that any state $|\psi \rangle \in \mathcal{H}_\text{code}$ is mapped by $V$ to the state $|\psi\rangle |\chi\rangle \in \mathcal{H}_1 \otimes \mathcal{H}_2 \otimes \mathcal{H}_{\bar A}$, and where $ |\chi\rangle \in \mathcal{H}_2 \otimes \mathcal{H}_{\bar A}$ is some fixed state.\footnote{$\mathcal{H}_3$ is just there to make sure the dimensions work out. It's unimportant.}

We can now evaluate (\ref{petzQI}). The encoding map acts as
\be
V \mathcal{O}_{\sf code} V^\dagger = \mathcal{O}_{\sf code} \otimes |\chi \rangle\langle \chi|.
\ee
Hence
\be
\text{Tr}_{\bar A} (V \mathcal{O}_{\sf code} V^\dagger) = \mathcal{O}_{\sf code} \otimes \chi_2,\label{trAbar}
\ee
where $\chi_2 = \text{Tr}_{\bar A}  |\chi \rangle\langle \chi|$. Finally
\be
\sigma_A = \tr_{\bar A}(\mathds{1}_1\otimes |\chi\rangle\langle \chi|) = \mathds{1}_1 \otimes \chi_2.\label{sigma}
\ee
Combining (\ref{trAbar}) and (\ref{sigma}), we find $\mathcal{O}_A = \mathcal{O}_{\sf code}\otimes \mathds{1}_2$, as desired.

For channels that are not perfectly reversible, the Petz map will not work perfectly. However, it still achieves very good performance, with an ``average error'' at most twice the optimal average error (using any recovery channel) \cite{barnum2002reversing}. For large code spaces, the performance may be suboptimal for particular``worst-case'' input states, whereas recent work has shown that a slightly more complicated ``twirled Petz map'' has good provable worst-case performance \cite{junge2018universal}. However, for our application, the ordinary Petz map will work well enough.\footnote{In fact, it works better on average than the twirled Petz map.}

For reversible channels, the Petz map reconstruction of a unitary code space operator is itself always a unitary operator (at least when restricted to the image of the code space).\footnote{For approximately reversible channels, it will be approximately unitary.} However, the individual elements that make up the Petz map reconstruction, in particular the factors of $\sigma_A^{-1/2}$, are highly nonunitary.\footnote{For the Petz Lite reconstruction, the problematic nonunitary part is the large constant factor $c_0$.} The Petz map reconstruction may therefore have much higher complexity, as a unitary operator, than the simple description \eqref{petzQI} would suggest. In fact, for evaporating black holes, there is good reason to think that it is exponentially complicated \cite{Harlow:2013tf, PythonsLunch}.

\subsection{Gravitational replica trick for the Petz map}\label{sec:gravPetz}
The above discussion of the Petz map motivates us to amend the naive ``Petz Lite'' formula to
\be
\mathcal{O}_{\sf{R}} = \sigma_{\sf{R}}^{-1/2}\tr_{\sf{B}}(\mathcal{O})\sigma_{\sf{R}}^{-1/2}.\label{petzOp}
\ee
Here $\sigma_{\sf{R}}$ is the trace over $\sf{B}$ of the projector onto the code subspace,
\be
\sigma_{\sf{R}} = \sum_{a = 1}^{d_{\sf{code}}}\tr_{\sf{B}} |\Psi_a\rangle\langle \Psi_a| = \sum_{i,j = 1}^k|i\rangle\langle j|_{\sf{R}} \ \sum_{a = 1}^{d_{\sf{code}}}\frac{\langle\psi_{aj}|\psi_{ai}\rangle_{\sf{B}}}{k}.
\ee
To check how well (\ref{petzOp}) works, we will compute the LHS of (\ref{toCheck}) in gravity and compare it with the RHS. The main complication in this calculation is the fact that it involves the $-1/2$ power of an operator. To deal with this, we will use a version of the replica trick, defining
\be
\mathcal{O}^{(n)}_{\sf{R}} = \sigma_{\sf{R}}^{n}\,\tr_{\sf{B}}(\mathcal{O})\,\sigma_{\sf{R}}^{n}.\label{sigmaR}
\ee
For positive integer $n$, matrix elements of this operator can be evaluated using replicas. In a first pass, we will take a naive approach and analytically continue the dominant term in the bulk calculation from integer $n$ to $n = -1/2$. Later, we will show how to do this analytic continuation more precisely, using a summation of planar geometries.

We will now begin the calculation. After writing out the integer $n$ expression and evaluating the inner products in the $\sf{R}$ system explicitly, one gets an expression that depends only on inner products of the $|\psi_{ai}\rangle$ states in the $\sf{B}$ system:
\begin{align}
\langle \Psi_a|\mathcal{O}^{(n)}_{\sf{R}}|\Psi_b\rangle &= \frac{1}{k^{2n+2}}\langle \psi_{a i_0}|\psi_{b j_0}\rangle_{\sf{B}} \times \langle \psi_{a_1 i_1}|\psi_{a_1 i_0}\rangle_{\sf{B}}\dots \langle \psi_{a_n i_n}|\psi_{a_n i_{n-1}}\rangle_{\sf{B}}\times \langle \psi_{b' j_n}|\psi_{a' i_n}\rangle_{\sf{B}} \notag\\ &\hspace{30pt}\times \langle \psi_{b_n j_{n-1}}|\psi_{b_n j_{n}}\rangle_{\sf{B}}\dots \langle \psi_{b_1 j_0}|\psi_{b_1 j_{1}}\rangle_{\sf{B}} \times \mathcal{O}_{a'b'}.\label{cancelled}
\end{align}
In this expression, all indices except for $a$ and $b$ are summed over. The pattern of index contractions is best understood by writing a diagram for the boundary conditions for the corresponding bulk path integral. Using the translation of inner products to boundary conditions shown in (\ref{melement3}), and dropping the ``$\sf{x}$'' symbols for clarity, we find
\begin{align}
\langle \Psi_a|\mathcal{O}^{(n)}_{\sf{R}}|\Psi_b\rangle &= \frac{1}{(k Z_1)^{2n+2}} \times \includegraphics[valign = c, width = .20\textwidth]{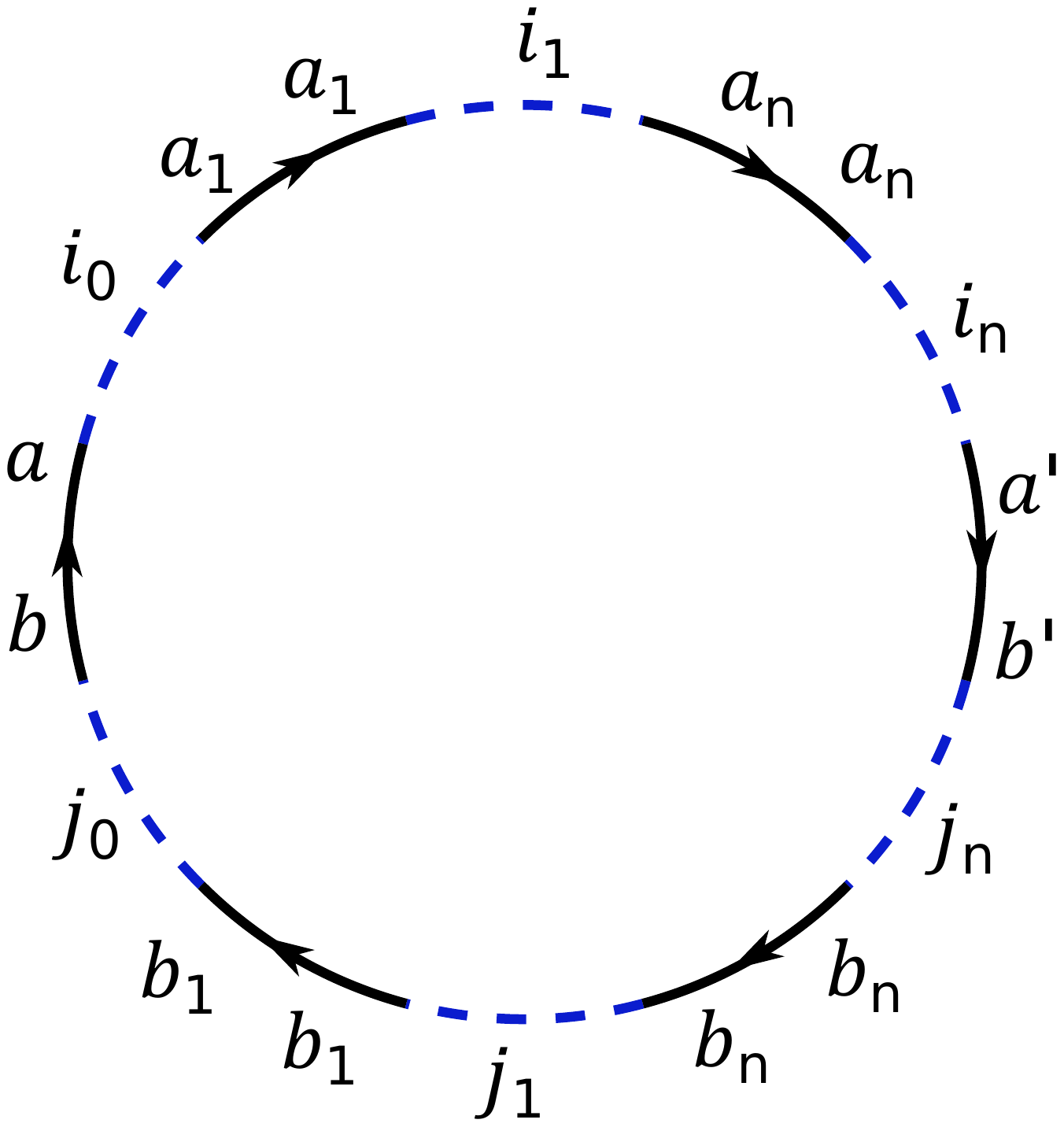}\times  \mathcal{O}_{a'b'}.\label{picture}
\end{align}
In this diagram, and in similar ones below, we draw the case $n = 2$. In general there will be $2n+2$ black segments corresponding to asymptotic boundaries.

In the phase $k \gg e^{S_{BH}}$ where the connected geometries dominate, the boundary conditions will be filled in by a completely connected geometry
\be
\includegraphics[valign = c, width = .21\textwidth]{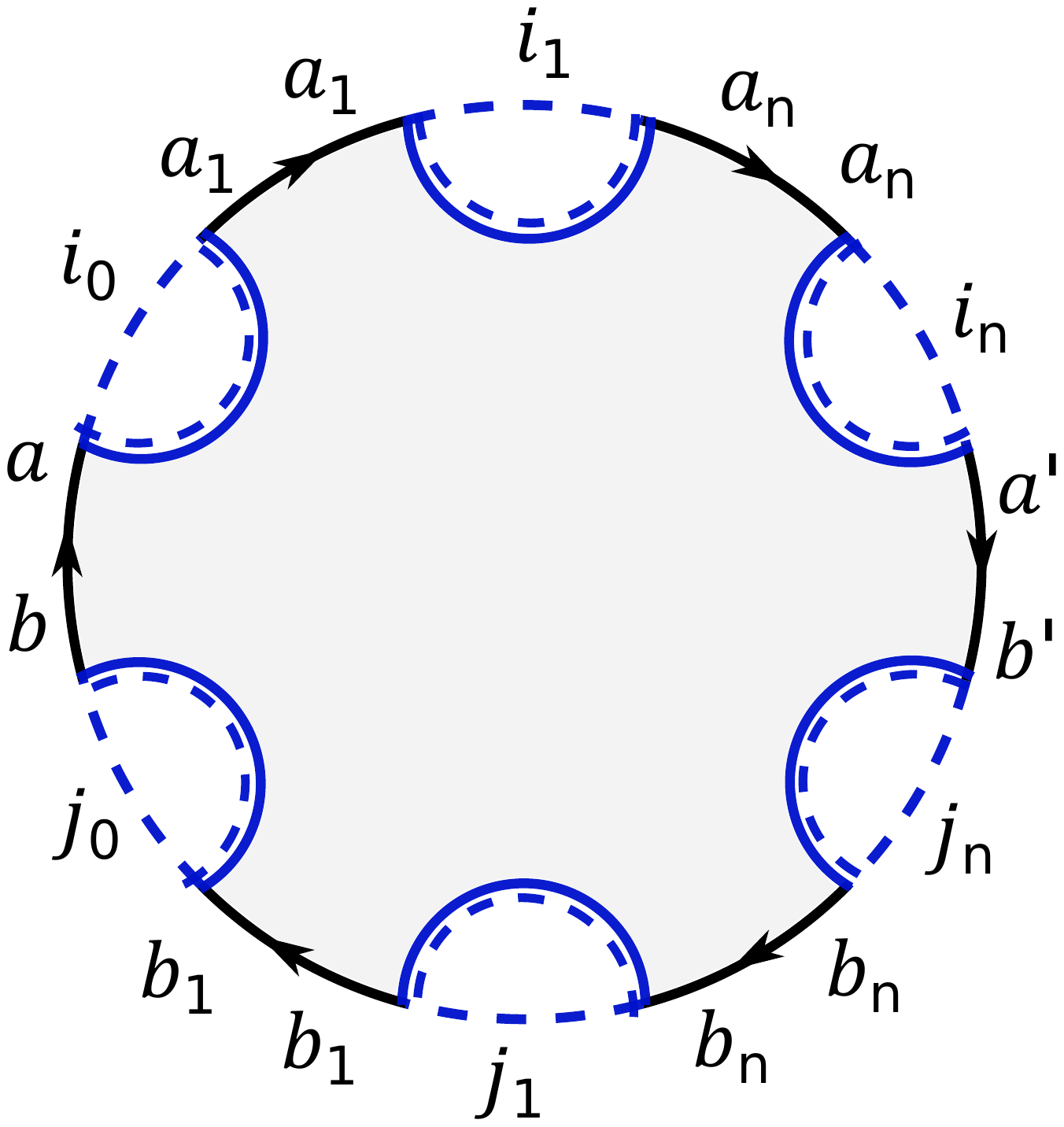}
\ee
We now need to evaluate this bulk computation. There are $2n+2$ index loops for the $k$ states of the EOW brane, which gives a factor of $k^{2n+2}$ that cancels against the $k^{2n+2}$ in the denominator of (\ref{picture}). What remains is a purely gravitational calculation, involving a path integral over gravity and over the propagating bulk fields. 

As a first approximation (order $G_N^{-1}$), we can ignore the matter fields altogether and just work out the answer from the gravity computation. In this approximation, the geometry has a $\mathbb{Z}_{2n+2}$ symmetry and we can analytically continue the $\mathbb{Z}_{2n+2}$ quotient of the geometry in $n$. In the limit $n \rightarrow -1/2$, it becomes a disk with a single boundary, for which the gravitational path integral gives $Z_1$. This cancels against the $Z_1^{2n+2}$ in the denominator, so the gravitational answer is simply one.

\subsection{Reducing to a bulk field theory calculation}\label{sec:ftpetz}
Now that we have evaluated the leading-order gravitational contribution, we need to compute the contribution from the propagating bulk fields. In the leading approximation, this means a field theory calculation on the fixed background, with the $\mathbb{Z}_{2n+2}$ symmetry. We can view the geometry as built out of $2n+2$ pieces, which we separate with dotted lines:
\be
\includegraphics[valign = c, width = .21\textwidth]{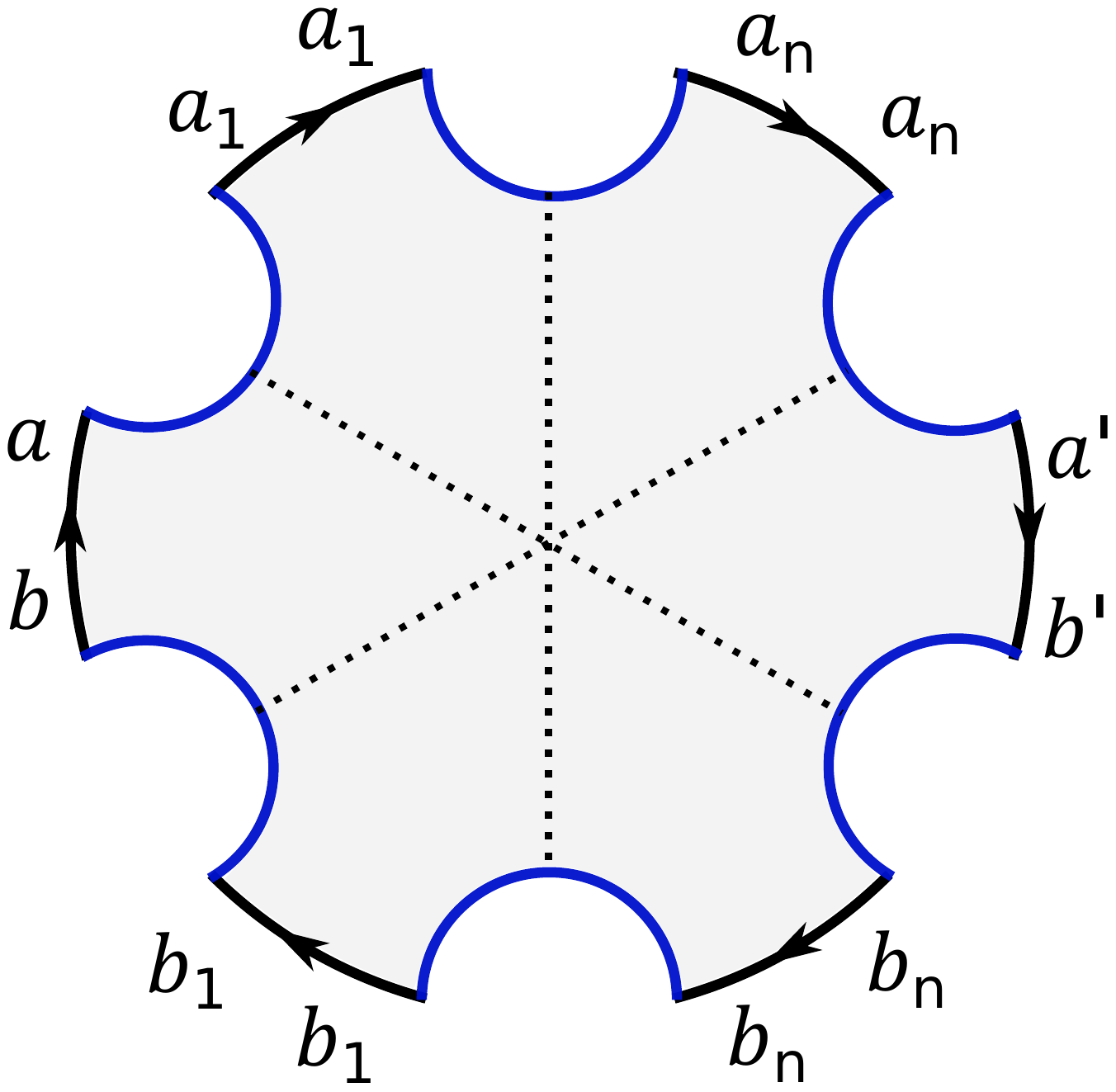}\label{pieces}
\ee
The field theory path integral on each of these pieces gives an operator with bra and ket corresponding to the dotted lines that border each piece. The particular operator we get in each case depends on the angle $\theta$ between the dotted lines, and it also depends on the field theory boundary conditions at the asymptotic boundaries, in particular, it depends on the $a$ and $b$ indices in (\ref{pieces}). We will refer to this operator as $M$, and a certain sum over such operators as $\widehat{M}$:
\be
M(a,b;\theta) = \includegraphics[scale = .3,valign = c]{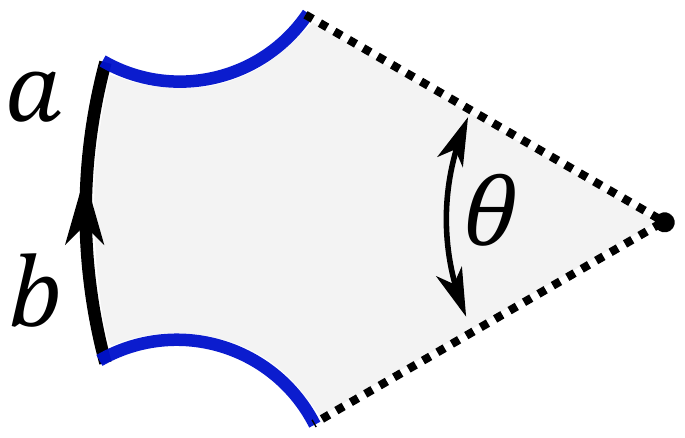}, \hspace{40pt} \widehat{M}(\theta) = \sum_{a = 1}^{d_{\sf{code}}}M(a,a;\theta).
\ee
In this notation, the field theory computation we are interested in is
\be
\langle \Psi_a|\mathcal{O}_{\sf{R}}^{(n)}|\Psi_b\rangle = \sum_{a'b' = 1}^{d_{\sf{code}}}\tr\Big[M(a,b,\theta)\widehat{M}(\theta)^n M(b',a',\theta)\widehat{M}(\theta)^n\Big]\mathcal{O}_{a'b'} + O(G_N).\label{intin}
\ee
The trace in this expression is over the Hilbert space of the bulk fields on one of the dotted lines. Since $\mathcal{O}^{(n)}_{\sf{R}}$ involves $2n+2$ replicas, we should set $\theta = 2\pi/(2n+2)$.

There are two sources of $n$-dependence in (\ref{intin}): the number of powers of $\widehat{M}$, and the value of  $\theta$. We would like to continue to $n = -1/2$, and it is convenient to do this in stages. In the first stage, we set $\theta  = 2\pi$, leaving the number of powers of $\widehat{M}$ general. When $\theta = 2\pi$, the $M$ operator becomes a path integral on the standard unbackreacted geometry,
\be
M(a,b,2\pi) \hspace{5pt} = \hspace{5pt}\includegraphics[scale = .3,valign = c]{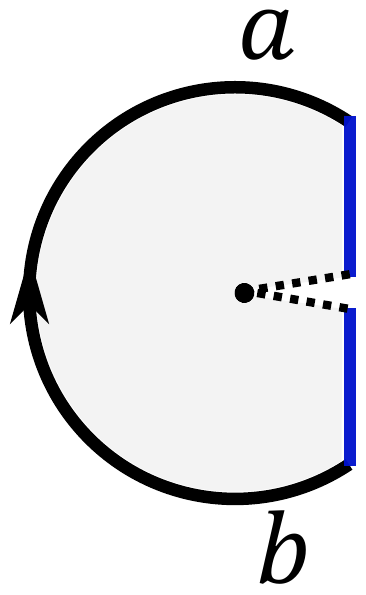} \hspace{5pt}=\hspace{5pt} \tr_{\sf{outside}}|b\rangle\langle a|_{\sf{FT}}.\label{hwuwu} 
\ee
In deriving the second equality here, we used (\ref{ft}). The notation $\tr_{\sf{outside}}$ means a trace over the region outside the horizon, in the Hilbert space of the bulk field theory. Substituting this in, we find that with $\theta = 2\pi$, the RHS of (\ref{intin}) can be written as $
\langle a|\mathcal{O}_{\sf{inside}}^{(n)}|b\rangle_{\sf{FT}}$
where
\begin{align}
\mathcal{O}_{\sf{inside}}^{(n)} &=\sigma_{\sf{FT}}^{n}\tr_{\sf{outside}}(\mathcal{O}_{\sf{FT}})\sigma_{\sf{FT}}^{n}\\
\sigma_{\sf{inside}} &= \tr_{\sf{outside}}\sum_{a} |a\rangle\langle a|_{\sf{FT}}.
\end{align}

As a second stage, we now take the remaining dependence on $n$ to $-1/2$, and we note that the answer takes the form of a Petz map, but now for a simpler problem defined entirely within the Hilbert space of the bulk field theory on the fixed background. Specifically, it is the Petz map $\mathcal{O}_{\sf{FT}}\rightarrow \mathcal{O}_{\sf{inside}}$ for reconstructing the bulk field theory operator $\mathcal{O}_{\sf{FT}}$ using only the region inside the horizon. If $\mathcal{O}_{\sf{FT}}$ acts behind the horizon, then this reconstruction is possible, so this auxiliary Petz map will succeed, meaning that
\be
\langle a|\mathcal{O}_{\sf{inside}}|b\rangle_{\sf{FT}} \approx  \mathcal{O}_{ab}.\label{poil}
\ee
This follows from the properties of the Petz map discussed in section \ref{sec:qipetz}, but it can also be shown directly as follows. Suppose that $\mathcal{O}_{\sf{FT}} = \Pi_{\sf{code}}A_{\sf{inside}}\otimes \mathds{1}_{\sf{outside}}\Pi_{\sf{code}}$, where $A_{\sf{inside}}$ commutes with $\Pi_{\sf{code}}$. Then one can check that $\mathcal{O}_{\sf{inside}} = A_{\sf{inside}}$, and the reconstruction works perfectly.

Let's now summarize. We were originally interested in a reconstruction problem where the goal is to reconstruct a bulk field theory operator $\mathcal{O}_{\sf{FT}}$ using an operator $\mathcal{O}_{\sf{R}}$ acting only on the ``radiation'' system $\sf{R}$. We started with a candidate for $\mathcal{O}_{\sf{R}}$ suggested by the Petz map. Using a bulk argument, we showed that
\be
\langle \Psi_a|\mathcal{O}_{\sf{R}}|\Psi_b\rangle = \langle a|\mathcal{O}_{\sf{inside}}|b\rangle_{\sf{FT}} + O(G_N).
\ee
where the matrix element in the RHS is defined purely in the bulk field theory on a fixed background. The operator $\mathcal{O}_{\sf{inside}}$ can be interpreted as the result of an auxiliary Petz map problem defined entirely in the field theory on the fixed background. This auxiliary Petz map attempts to reconstruct $\mathcal{O}_{\sf{FT}}$ using only the region behind the horizon. Such reconstruction will obviously be possible if $\mathcal{O}_{\sf{FT}}$ acts only behind the horizon, so (\ref{poil}) will hold and therefore (\ref{toCheck}) will hold also. So we conclude that in the island phase, reconstruction of operators behind the horizon is possible using only $\sf{R}$.

\subsection{The disconnected phase}
To understand this result a little better, we can ask why reconstruction fails if we are in the disconnected phase with $k \ll e^{S_{BH}}$. In this phase, the geometry that dominates (\ref{cancelled}) is the completely disconnected one, similar to the geometry shown in the center of figure \ref{fig1g}. At leading order, the gravitational part of this computation reduces to $Z_1^{2n+2}$, with an additional factor of $k$ for the single $k$-index loop. These factors cancel the prefactors in (\ref{picture}) in the limit $n\rightarrow -1/2$. So the gravitational and $k$-index part of the computation gives one, just as in the connected phase.\footnote{One could ask the following question: if in the limit $n\rightarrow -1/2$, the purely gravitational part of the answer is one for both the connected and the disconnected geometries, then should we sum both? Briefly, the answer is no. A full analysis including the transition is in section \ref{sec:nonperPetz} below.} The computation therefore reduces again to a purely field-theory calculation on a fixed background.

Formally, this field theory calculation can be interpreted as an auxiliary Petz-map calculation, but of a rather trivial kind. The analog of the geometries in (\ref{hwuwu}) is the one shown on the RHS of (\ref{hwuwu}), but with no cut at all. As a field theory operator, this corresponds to 
\be
\tr_{\sf{entire \ bulk}}|b\rangle\langle a|_{\sf{FT}}.
\ee
So we replace $\tr_{\sf{outside}}$ by a trace over the entire bulk Hilbert space. The associated Petz map is one that attempts to reconstruct the operator $\mathcal{O}_{\sf{FT}}$ from the trivial subsystem of the bulk theory, after tracing out everything. This is obviously not possible, so reconstruction fails.

\subsection{Planar resummation for the Petz map}\label{sec:nonperPetz}
In the above analysis of the Petz map, for the most part we assumed that the connected geometry dominates. In this section, we will do the full sum over planar geometries, using a method similar to the one we used to compute the entanglement spectrum of $\rho_{\sf{R}}$. 

In this section, we will remove the bulk fields from the theory, and go back to considering the theory with only the EOW brane. In order to have a nontrivial code subspace of states
\be
|\Psi_a\rangle = k^{-1/2}\sum_{j = 1}^k|\psi_{aj}\rangle_{\sf{B}}|j\rangle_{\sf{R}}, \hspace{20pt} a = 1,\dots,d_{\sf{code}},
\ee
we will interpret both the $a$ and $j$ indices as describing the state of the EOW brane. The $j$ index is entangled with the radiation, and the $a$ index labels the state within the code subspace. To reiterate: in our previous discussion, this $a$ index labeled some state of the bulk fields, but to make the following analysis simple we will now take $a$ to describe another aspect of the EOW brane (e.g.~the state of some qubit that propagates right alongside the EOW brane).

The main challenge in the computation is the existence of the operators $\sigma^{-1/2}$ in the formula for the Petz map. It will be convenient to write a version of this formula for integer powers of $\sigma$ as
\be
\mathcal{O}_{\sf{R}}^{n_1,n_2}=\frac{1}{d_{\sf{code}}}\upsigma_R^{n_1}\tr_B (\mathcal{O}) \upsigma_R^{n_2}, \hspace{20pt} \upsigma_{\sf{R}} = \frac{1}{d_{\sf{code}}}\sigma_{\sf{R}}.
\ee
In the rest of this section, we will use the rescaled $\upsigma$ defined here, which is normalized as a density matrix. What we would like to do is make sense of the ``analytic continuation to $n_1 = n_2 = -1/2$'' of this expression. To make this precise, it will be helpful to introduce a generating function of these operators for all values of $n_1,n_2$ as $\mathcal{O}_{\sf{R}}(\lambda_1,\lambda_2)$:
\be
\mathcal{O}_{\sf{R}}(\lambda_1,\lambda_2)=\sum_{n_1,n_2 = 0}^\infty \frac{\mathcal{O}_{\sf{R}}^{n_1,n_2}}{\lambda_1^{n_1}\lambda_2^{n_2}}, \hspace{30pt} \mathcal{O}_{\sf{R}}^{n_1,n_2} = \oint_{\infty} \frac{\mathrm{d}\lambda_1}{2\pi i}\frac{\mathrm{d}\lambda_2}{2\pi i}\lambda_1^{n_1-1}\lambda_2^{n_2-1}\mathcal{O}_{\sf{R}}(\lambda_1,\lambda_2).\label{transform}
\ee
Later we will see how to deform the integration contour to get an answer that can be continued to $n_1 = n_2 = -\frac{1}{2}$, which gives the actual Petz map.

We would like to compute the sum over planar bulk geometries of the matrix elements of $\mathcal{O}_{\sf{R}}(\lambda_1,\lambda_2)$. The geometries can be divided into two classes. In the first class, the asymptotic boundary corresponding to the physical black hole (with $a,b$ indices in (\ref{picture})) is connected to the asymptotic boundary on the other side of the circle, where the operator $\mathcal{O}_{a'b'}$ acts. In the second class, these two asymptotic boundaries are not in the same connected component of the geometry. The contribution of the first class of geometries to the matrix element is proportional to $O_{ab}$, since the index lines associated to the EOW branes require $a = a'$ and $b = b'$. In the second class, the contribution is proportional to $ \delta _{ab}\tr\, \mathcal{O}$. In other words, we have the matrix elements:
\be
\langle\Psi_a|\mathcal{O}_{\sf{R}}(\lambda_1,\lambda_2)|\Psi_b\rangle=c_1(\lambda_1,\lambda_2) \mathcal{O}_{ab}+c_2(\lambda_1,\lambda_2)\delta_{ab} \tr \, \mathcal{O}.
\ee
Ideal reconstruction would mean that $c_1 = 1$ and $c_2 = 0$. Since the Petz map is normalized so that the identity operator always reconstructs perfectly, we must have $c_1+c_2 d_{\sf{code}}=1$. We can compute $c_1$ by summing all geometries from the first class.

\begin{figure}[t]
\centering
	\includegraphics[scale=0.6]{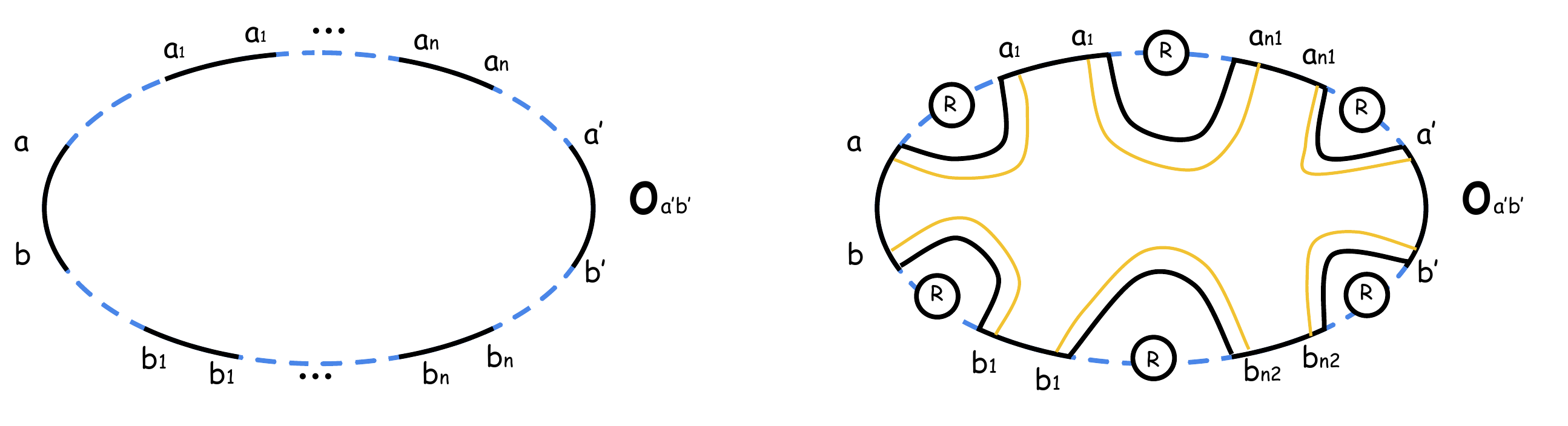}
	\caption{Boundary conditions (left) and a planar contribution to $\langle\Psi_a|\mathcal{O}_{\sf{R}}(\lambda_1,\lambda_2)|\Psi_b\rangle$. This contribution is an element of the first class, since the leftmost boundary is in the same connected geometry as the rightmost one.}
	\label{fig:petzcorr}
\end{figure}

Geometries belonging to the first class can be further grouped as follows. We are guaranteed that there is a geometry that connects the boundary of the physical black hole to the ``opposite side'' boundary. In addition, this geometry is connected to some number $m_1$ of boundaries in the ``top half'' of the circle, and $m_2$ boundaries in the ``bottom half.'' This defines one connected component with $2+m_1+m_2$ boundaries, but in between each of these boundaries we can have a full summation of planar geometries. These can be summed over using the resolvent for $\upsigma_{\sf{R}}$:
\be
\widehat{R}(\lambda) = \tr\frac{1}{\lambda - \upsigma_{\sf{R}}} \hspace{20pt} (\text{this equals $R(\lambda) = \tr\frac{1}{\lambda - \rho_{\sf{R}}}$ but with $e^{S_0}\to d_{\sf{code}}e^{S_0}$}).\label{ressub}
\ee
See figure \ref{fig:petzcorr} for a graphical description of this. The resulting expression for $c_1(\lambda_1,\lambda_2)$ in the planar approximation is
\be
c_1(\lambda_1,\lambda_2)=\sum_{m_1, m_2=0}^{\infty}\frac{d_{\sf{code}}Z_{m_1+m_2+2}}{(d_{\sf{code}}kZ_1)^{m_1+m_2+2}}\widehat{R}(\lambda_1)^{m_1+1}\widehat{R}(\lambda_2)^{m_2+1}\lambda_1\lambda_2.\label{c1}
\ee

In this formula, $d_{\sf{code}}$ and $e^{S_0}$ only appear in the combination $d_{\sf{code}}e^{S_0}$. This is clear for the explicit factors of $d_{\sf{code}}$, since they multiply either $Z_1$ or $Z_{m_1+m_2+2}$, both of which are proportional to $e^{S_0}$. But it is also true for the implicit factors in $\widehat{R}$, see (\ref{ressub}). This will have an important implication: in order for the connected geometries to dominate the Petz map computation, we will need to have $k \gtrsim d_{\sf{code}}e^{S_{BH}}$, rather than the Page time condition $k \gtrsim e^{S_{BH}}$. This is in keeping with expectations based on \cite{Hayden:2007cs, Hayden:2018khn, Penington:2019npb}. In analyzing the equations below, we will simplify by setting $d_{\sf{code}} = 1$, and remembering that it can be restored by rescaling $e^{S_0} \to d_{\sf{code}}e^{S_0}$. 

Using the integral representation (\ref{finalZn}), and the notation (\ref{rescaledvariables}), we can rewrite (\ref{c1}) as
\be
\begin{split}\label{c1first}
	c_1(\lambda_1,\lambda_2)&=\lambda_1\lambda_2\int_0^\infty \mathrm{d}s\,\uprho(s){w(s)R(\lambda_1) \over k- w(s)R(\lambda_1)}{w(s)R(\lambda_2)\over k-w(s)R(\lambda_2) }.
\end{split}
\ee
To compute the coefficient $c_1^{n_1,n_2}$, we can use the contour integral in (\ref{transform}) and interchange the $s$ and $\lambda$ integrals to write
\be
c_1^{n_1,n_2} = \int_0^\infty \mathrm{d}s\,\uprho(s)\oint_{\infty}\frac{\mathrm{d}\lambda_1}{2\pi i}\frac{\mathrm{d}\lambda_2}{2\pi i}\lambda_1^{n_1}\lambda_2^{n_2}{w(s)R(\lambda_1) \over k- w(s)R(\lambda_1)}{w(s)R(\lambda_2)\over k-w(s)R(\lambda_2) }.
\ee
For non-integer $n$, the function $\lambda^n$ has a branchpoint at infinity, so we can't continue the integral as written. However, we can first deform the contour for the $\lambda_1,\lambda_2$ integrals to a new region where continuation will be possible. In particular, for each of the $\lambda_1,\lambda_2$ integrals, we separately deform the integration contour to surround the cut in the resolvent along the positive real axis.\footnote{The denominators $k - w(s)R(\lambda)$ can be shown not to vanish on the principal sheet.} After doing this there is no problem in continuing in $n$, and for $n_1 = n_2 = -1/2$, we find
\be
c_1 = \int_0^\infty \mathrm{d}s\,\uprho(s)\left(\oint_{\mathcal{C}}\frac{\mathrm{d}\lambda}{2\pi i}{\lambda^{-\frac{1}{2}}w(s)R(\lambda) \over k- w(s)R(\lambda)}\right)^2.\label{c1final}
\ee

\subsubsection{Microcanonical ensemble}
Let's first analyze this equation in the microcanonical ensemble with fixed $s$. Then the Schwinger-Dyson equation for the resolvent (\ref{eqn: resolvent eqn}) implies
\be
\lambda R = k + e^{\bf{S}}\frac{w(s)R}{k - w(s)R}.
\ee
Substituting this into the formula for $c_1$ (\ref{c1final}), one finds
\be
c_1 = e^{-{\bf S}}\left(\oint_{\mathcal{C}}\frac{\mathrm{d}\lambda}{2\pi i}\lambda^{1/2}R(\lambda)\right)^2 = e^{-{\bf S}}\left(\int_{\mathcal{C}}\mathrm{d}\lambda\lambda^{1/2}D(\lambda)\right)^2.\label{c11}
\ee
We can do the integral using the result for the resolvent or the density of states in (\ref{microcanonicalAns}), and one finds (restoring $d_{\sf{code}}$ by taking $e^{\bf{S}}\to e^{\bf{S}}d_{\sf{code}}$)
\be
c_1 = \begin{cases}\frac{k}{d_{\sf{code}}e^{{\bf S}}} + \dots  & k \ll d_{\sf{code}}e^{\bf{S}} \\ 1 - \frac{d_{\sf{code}}e^{\bf S}}{4k} +\dots & k \gg d_{\sf{code}}e^{\bf{S}}\end{cases}.\label{c1ans}
\ee
So the reconstruction works well in the region $k \gg d_{\sf{code}}e^{\bf{S}}$ where connected geometries dominate. In this region, (\ref{c1ans}) gives the leading small correction due to disconnected geometries.

\subsubsection{Canonical ensemble}
To analyze $c_1$ in the canonical ensemble, we can go back to (\ref{c1final}) and insert the leading expression $R(\lambda) \approx k/(\lambda-\lambda_0)$ from (\ref{exp:resolvent1}). The $\lambda$ integral is
\be
\oint_{\mathcal{C}} \frac{\mathrm{d}\lambda}{2\pi i} {\lambda^{-1/2}w(s)R(\lambda)\over k -w(s)R(\lambda)}\approx \oint_{\mathcal{C}} \frac{\mathrm{d}\lambda}{2\pi i}\frac{\lambda^{-1/2}w(s)}{\lambda - \lambda_0 - w(s)}=-\frac{w(s)}{(\lambda_0 + w(s))^{1/2}},\label{LHScorr}
\ee
so we find the expression
\begin{align}
c_1&\approx \int_{0}^{\infty}\mathrm{d}s\,\uprho(s){w(s)^2\over \lambda_{0}+w(s)}\label{rhorho}.
\end{align}
Near the Page transition, the integral is dominated by the region $s < s_k$, where the $w(s)$ term in the denominator is larger than $\lambda_0$. So we can approximate the integral as 
\be
c_1 \approx \int_0^{s_k} \mathrm{d}s\,\uprho(s) w(s) = 1 - \int_{s_k}^\infty \mathrm{d}s\,\uprho(s)w(s).\label{corrone}
\ee
This has a simple interpretation: part of the thermal ensemble is ``pre-Page,'' and part of the thermal ensemble is ``post-Page.'' The above integral gives the fraction of the ensemble that is Post-page. Note that this notion of post- and pre-Page refers to the Page time with $e^{S_0}$ replaced by $d_{\sf{code}}e^{S_0}$.

\begin{figure}[t]
\begin{center}
\includegraphics[width = .45\textwidth]{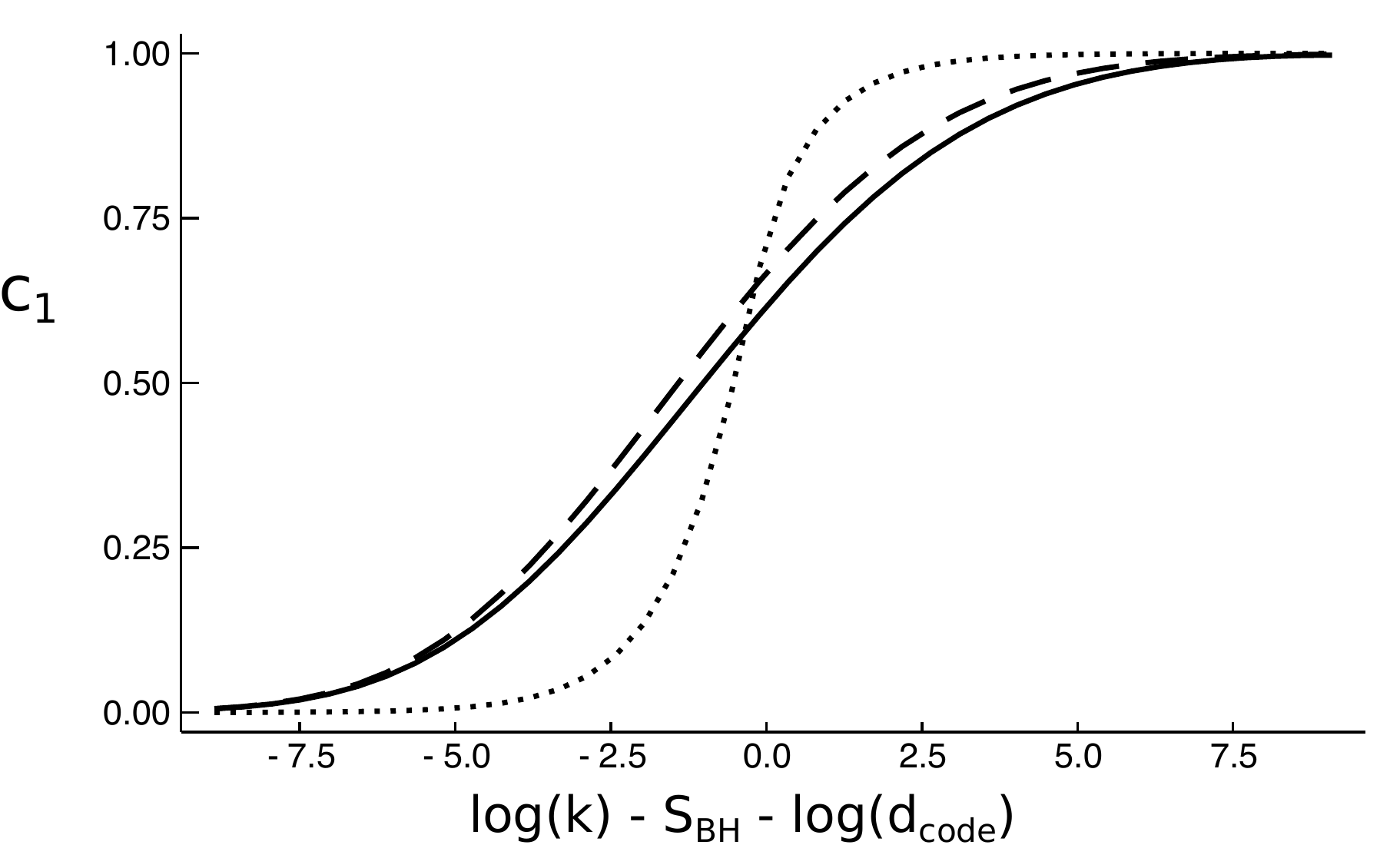}
\caption{{\small For $\beta = 3$ and large $\mu$, we plot the exact $c_1$ in the planar approximation (solid) and the simple approximation (\ref{rhorho}) (dashed). The answer in the microcanonical ensemble is also shown (dotted). For smaller $\beta$, the transition in the canonical ensemble takes place over a longer interval in $\log(k)$, of order $\sim\beta^{-1/2}$.}}\label{fig:petzerror}
\end{center}
\end{figure}

After the Page transition, the correction to one in (\ref{corrone}) becomes very small, and the leading correction comes from a new place: the correction to the assumed form $k/(\lambda-\lambda_0)$ of the resolvent. This correction is computed in the final term in (\ref{exp:resolvent1}), and working out the first-order change in the LHS of (\ref{LHScorr}), one finds
\begin{align}
c_1 \approx 1 - \frac{1}{k}\int_0^\infty\hspace{-7pt}\mathrm{d}s_1\int_0^{s_k}\hspace{-7pt}\mathrm{d}s_2\,\frac{\uprho(s_1)\uprho(s_2)w(s_1)w(s_2)}{(w(s_1)^{1/2}+w(s_2)^{1/2})^2}.
\end{align}
In order to simplify this answer, we set $\lambda_0 = 0$, which is a good approximation after the Page transition. This integral is dominated by the region where $s_1\approx s_2$. Making this approximation, it then evaluates to the weighted average of the error in the microcanonical ensemble (\ref{c1ans}), with the weighted average taken over the Post-page portion of the thermal ensemble. This is consistent with the lower bounds on the reconstruction error derived in \cite{Hayden:2018khn, Penington:2019npb}.

In figure \ref{fig:petzerror}, we give a plot of the exact answer for $c_1$ in the microcanonical ensemble, and in the canonical ensemble with $\beta = 3$.

\section{Radiating black holes in equilibrium}\label{sec:JT}
In the simple model discussed above, the Page transition results from a competition between the entropy of the black hole and the number of states $k$ of an EOW brane. The interesting regime is a rather artificial one, where $k$ is extremely large, of order $e^{S_{BH}}$. However, for a physical radiating black hole, we expect that a similar effect is accomplished by going to late time, where the large number of states is provided automatically by the large amount of Hawking radiation.

The analysis for a radiating black hole (even in a simple theory like JT gravity) is more complicated than for the simple model: it involves intrinsically Lorentzian physics, and the backreaction of the Hawking radiation is essential for finding the wormhole solutions. So we will not be able to go as far as for the simple model. However, in this section we will make some preliminary comments. First, we will explain the relationship between the wormhole topologies and the island extremal surface. Second, we will analyze the continuation near $n = 1$ of the replica-symmetric wormhole explicitly, and make contact with the quantum extremal surface. Finally, we will explain qualitatively why we expect solutions to exist for integer values of $n > 1$.

The setup we will discuss is the following. We consider two thermodynamically stable black holes (e.g.~large black holes in AdS) that are weakly coupled together in such a way that they can exchange Hawking radiation. We refer to the two black holes as 1 and 2. As the Hawking radiation of 1 falls into 2, and vice versa, the entropy of each black hole grows. In thermal equilibrium, the two black holes will exchange Hawking radiation at a constant rate. A naive semiclassical computation will show that the entanglement between the black holes grows linearly with time forever. This eventually exceeds the coarse-grained entropy by an arbitrarily large amount. The Page curve in this context would be an entanglement entropy that follows the linear growth for a while, before saturating at a value that corresponds to the coarse-grained entropy of one of the two black holes.

\subsection{JT gravity setup}
We now analyze this situation in JT gravity. (2d gravity is convenient for drawing pictures, but the topological argument relating replica wormholes to the island extremal surface is similar in any spacetime dimension: one just replaces each point in the discussion below by a sphere.) We will consider two black holes in JT gravity, coupled together by a bulk matter CFT in a way that will be described below. Before getting into replicas and wormholes, let's begin by discussing the computation that gives the ordinary thermal partition function $Z(\beta)$ for this combined system.

For a single black hole, the boundary conditions (\ref{bc1}) for computing $Z(\beta)$ would be to specify that the boundary is a loop of length $\beta/\epsilon$ and to set the dilaton to $1/\epsilon$ at this boundary. Here, $\epsilon \rightarrow 0$ is a holographic renormalization parameter. We would then look for gravity configurations that can ``fill in'' these boundary conditions. Up to symmetries, the unique classical solution is the hyperbolic disk
\be
\mathrm{d}s^2 = \mathrm{d}\rho^2 + \sinh^2(\rho)\mathrm{d}\theta^2 = 4\frac{\mathrm{d}r^2 + r^2\mathrm{d}\theta^2}{(1-r^2)^2},\label{rotinv}
\ee
where we use the portion of the geometry inside the radius
\be
r = 1 - \frac{2\pi}{\beta}\epsilon.
\ee
This cutoff radius has been chosen so the length of the boundary is $\beta / \epsilon$. The  dilaton profile is
\be
\phi = \phi_h \cosh(\rho) = \phi_h \frac{1+r^2}{1-r^2}, \hspace{20pt} \phi_h = \frac{2\pi}{\beta}.
\ee
The value of $\phi_h$ was chosen so that $\phi = 1/\epsilon$ at the boundary.

Let's now discuss the computation of $Z(\beta)$ with two coupled black holes. Neglecting the matter fields for the moment, the boundary conditions will be the same in the two copies. We can fill these in with the same solution (\ref{rotinv}), but where we now take the two boundaries to be at radii
\be
r_{\pm} = 1 \pm \frac{2\pi}{\beta}\epsilon.\label{radii}
\ee
The region $r < r_-$ is the first black hole, and the region $r > r_+$ is the second black hole. (Note that this is another copy of the hyperbolic disk, as one can check by taking $r\rightarrow 1/r$.) A sketch of the full configuration is here:
\begin{equation}
\includegraphics[valign = c,scale = .55]{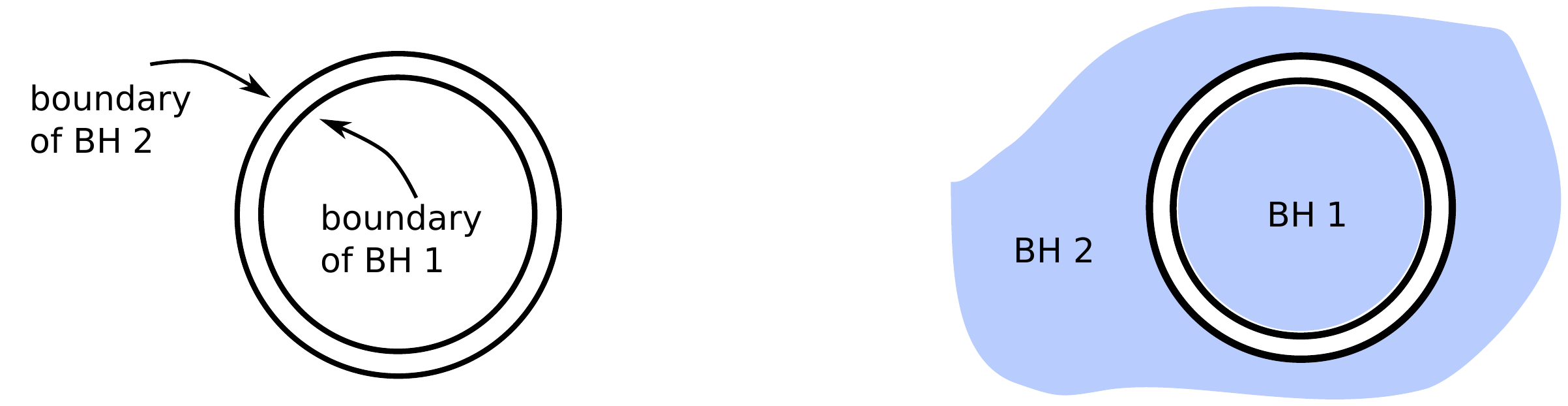}\label{notwist}
\end{equation}
Next we should discuss the bulk matter fields and their coupling between the two black holes. A convenient choice is to include a matter CFT propagating in the bulk theory, and then allow these matter fields to pass freely from one system to the other, with ``transparent'' boundary conditions. So, for example, let's consider the geometry described above. For small $\epsilon$, the geometry of the two black holes is conformal to the entire plane, and (up to a Weyl transformation that affects some observables), the bulk matter fields are simply propagating on this plane without seeing any unusual feature at $r = 1$ where we transition from one black hole to the other.

So far, we have described a situation in Euclidean signature, appropriate for computing $Z(\beta)$. But we can also continue it into Lorentzian signature. The setup corresponds to two copies of the thermofield double black hole. In other words, each of the two black holes 1 and 2 have both a left asymptotic boundary $L$ and a right asymptotic boundary $R$. The boundary conditions for the matter fields allow particles to pass between the two $L$ asymptotic boundaries, and similarly for the two $R$ boundaries. Up to a Weyl rescaling, the bulk matter fields feel like they are propagating on a Lorentzian cylinder, where half of the spatial $S^1$ is in black hole 1, and the other half of the spatial $S^1$ is in black hole 2.

In this situation, we would like to compute the entropy of the two-sided black hole 1, namely the entropy of the subsystems $1L\cup 1R$. Let's start by considering the boundary conditions to compute the Renyi $n$-entropy in the $t = 0$ state prepared by cutting the Euclidean path integral in half. These boundary conditions correspond to $n$ replicas of (\ref{notwist}), with boundary twist operators inserted, shown as zigzag cuts here:
\begin{equation}
\includegraphics[valign = c,scale = .55]{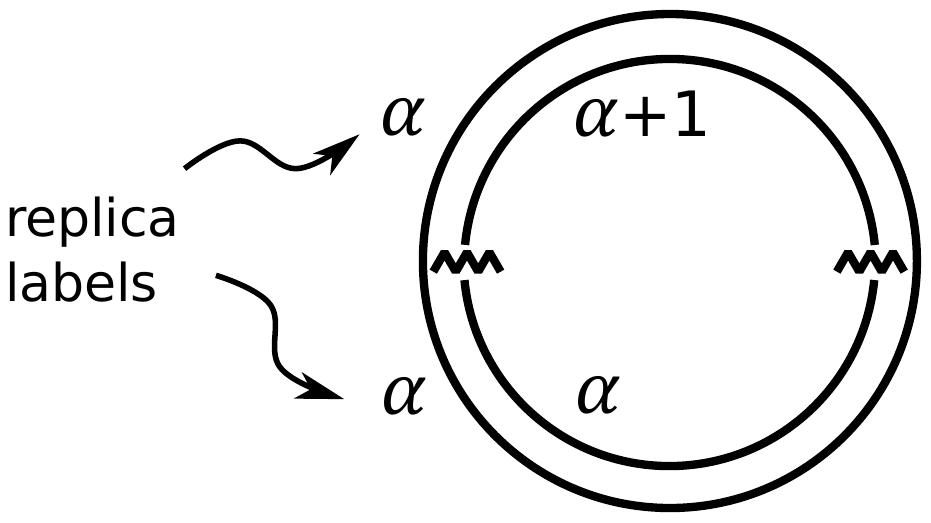}\label{twist}
\end{equation}
As we pass through these twist operators from below to above, we transition from replica $\alpha$ of the inner circle to replica $\alpha+1$. If the systems were not interacting with each other, this would be a trivial relabeling. But with interactions it becomes nontrivial: below the twist operators replica $\alpha$ of the system 2 (outer circle) interacts with replica $\alpha$ of system 1 (inner circle), and above them replica $\alpha$ of system 2 interacts with replica $\alpha+1$ of system 1.

The boundary conditions described above will compute the Renyi $n$-entropy at time zero of the two-sided BH 1. By moving the location of the twist operators, we can compute the entropy at other times. In principle, we can move the two twist operators independently. From the rotational symmetry of (\ref{twist}), it is clear that if we rotate the twist operators around the circle by the same angle, we will not change the answer. In Lorentzian signature, this corresponds to invariance of the answer under forwards evolution on the $R$ system, and backwards evolution on the $L$ system. However, if we move \emph{both} forwards in Lorentzian time, there is nontrivial time-dependence.

\subsection{Renyi topologies for the  empty set and for the island}
Having specified the boundary conditions, we can now ask what bulk geometries can fill them in. Our goal is to understand what topologies for the Renyi entropy computation lead to the ``island'' and ``empty set'' extremal surfaces in the von Neumann limit.

Let's first discuss the empty set. This arises from a Renyi entropy computation with the topology of $2n$ disk geometries: $n$ for system 1, and $n$ for system 2. Each replica of the boundary conditions is associated to a distinct replica in the bulk, and the different geometries are only connected by their boundary interactions. The geometry can be represented in a convenient way by filling in the interior and exterior of (\ref{twist}), and extending the cuts corresponding to the twist operators into the bulk, from one side of system 1 to the other:
\begin{equation}
\includegraphics[valign = c,scale = .55]{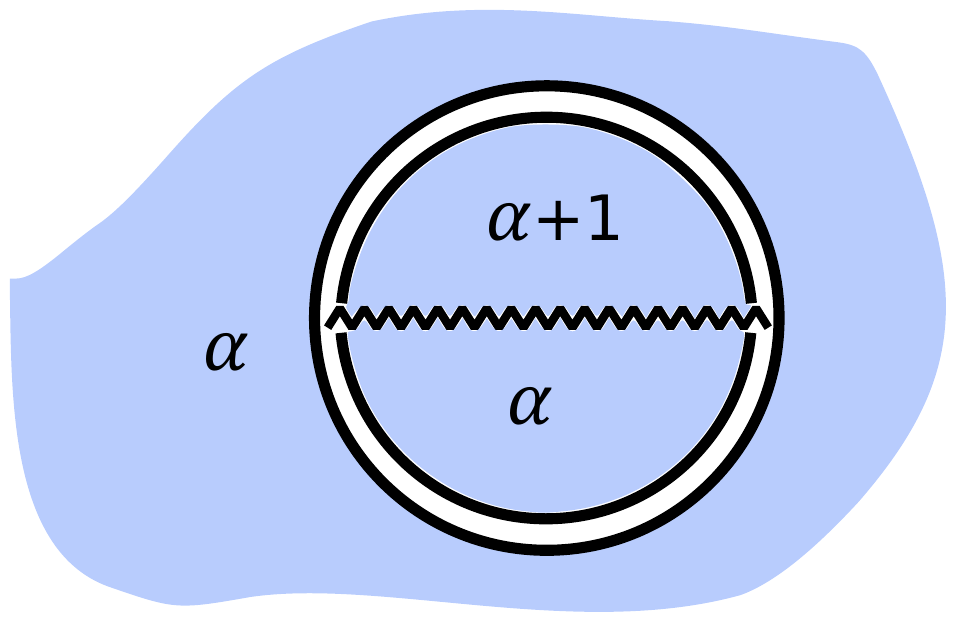}\label{bulk1}
\end{equation}
The path integral of the bulk matter fields on this geometry computes the Renyi $n$-entropy of the bulk matter fields on a particular bulk slice. This slice corresponds to the location of the cut in (\ref{bulk1}), and it can be understood as passing across the full Einstein-Rosen bridge of system one.\footnote{This is equal to the same entropy for system two since the global state of the bulk fields is pure.} In other words, we are computing the total entropy of the matter fields in system one. As we move the twist operators forwards in time, this entropy will grow, due to the fact that the Hawking radiation of one black hole falls into the other and vice versa. In general, the Renyi entropy computation will backreact on the gravity computation, so the geometry of the disks will be modified. But in the von Neumann limit $n\rightarrow 1$ where the entropy computation does not backreact on the geometry, we find simply
\be
S_{\text{vN}} = (\text{entropy on full slice across ERB of black hole 1}).\label{b1a}
\ee
In writing this expression, we used the fact that the gravitational contribution to the path integral gives one in non-backreacting limit $n\rightarrow 1$. As we move the twist operators forwards in time, the RHS of (\ref{b1a}) will grow linearly in time forever.

Let's now discuss the island. For the Renyi $n$-entropy, we leave the $n$ disks of system 2 in place, but we replace the $n$ disks of system 1 by a completely connected $\mathbb{Z}_n$-symmetric geometry that we will refer to as a ``pinwheel.'' A sketch of the pinwheel for $n = 6$ (with its $\mathbb{Z}_n$ quotient shaded) looks like this:
\be
\includegraphics[valign =c, scale = .35]{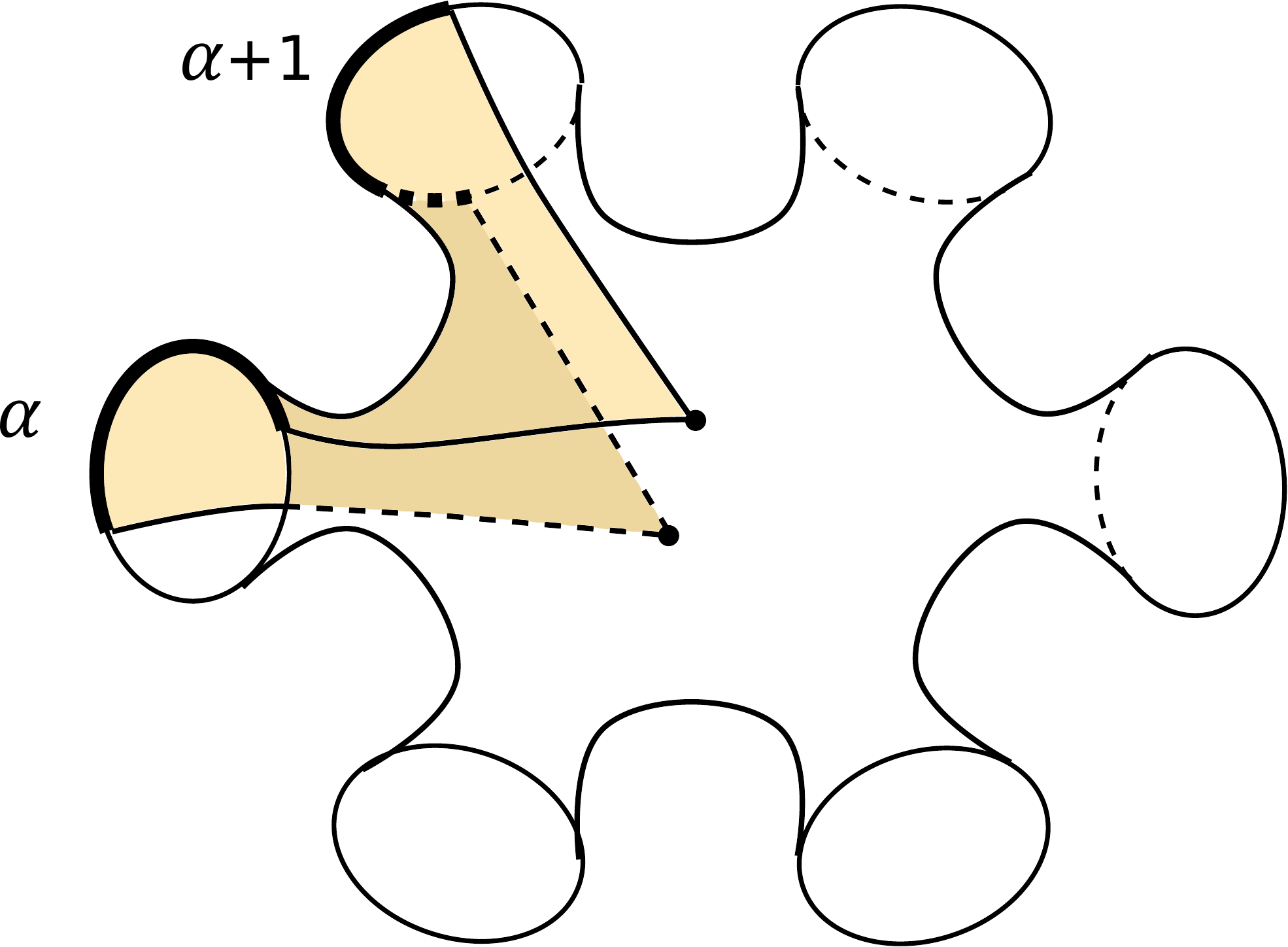}
\label{fig:pinwheel}
\ee
For the moment, let's not worry about whether the geometry in (\ref{fig:pinwheel}) is a solution or not (we will come back to this). What we would like to do is understand what it looks like in the limit $n\rightarrow 1$. Strictly speaking, this geometry does not have a continuation in $n$, but its $\mathbb{Z}_n$ quotient does. The quotient (shaded orange in the above sketch) is characterized by the requirement that the conical angle around the two points marked by black dots should be $2\pi/n$. 

The shaded region intersects the asymptotic boundaries in two semicircles, shown with heavy black curves. These two semicircles are the two inner semicircles of (\ref{twist}), so in particular, they are joined by bulk field theory interactions to the same replica of system 2, which will be filled in by a disk geometry. For $n\approx 1$, the quotient of the full geometry including system 2 therefore looks like the following:
\be
\includegraphics[valign =c, scale = .55]{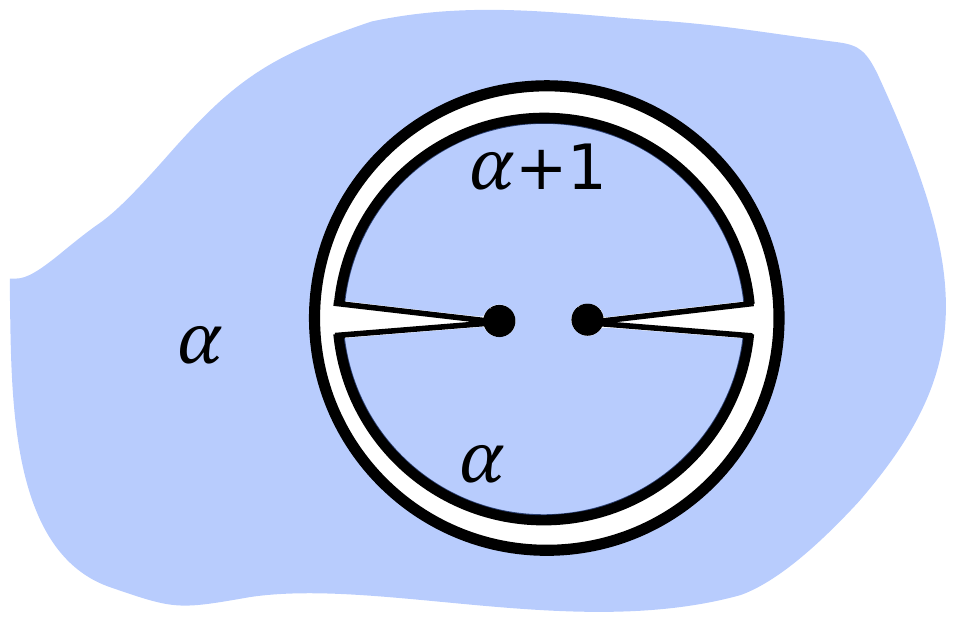}
\label{bulk3}
\ee
The inner region is the continuation to $n$ near one of the shaded portion of (\ref{fig:pinwheel}). Note that the conical angle around the black dots is close to $2\pi/n$, which is close to $2\pi$.

One can evaluate the contribution of this topology to the gravity path integral, in a very crude approximation where we keep only the topological $S_0$ term in the JT action. In order to compute the von Neumann entropy, we need to consider
\be
S_{\text{vN}} = \lim_{n\to 1}\left[\frac{1}{n-1}\log\frac{\mathcal{Z}_n}{\mathcal{Z}_1^n}\right].\label{vvn}
\ee
Here $\mathcal{Z}_n$ is the bulk computation with the boundary conditions for the Renyi $n$ entropy, and $\mathcal{Z}_1$ is the bulk computation of the partition function of the combined system. At the level of the topological action, these partition functions are simply $e^{\chi S_0}$, where $\chi$ is the total Euler characteristic:
\begin{align}
\mathcal{Z}_n &\sim Z_{\text{pinwheel}}\times Z_{\text{disk}}^n \sim e^{(2-n)S_0}\times e^{nS_0}\\
\mathcal{Z}_1^n &\sim Z_{\text{disk}}^{2n} \sim e^{2nS_0}
\end{align}
Here, the $\sim$ symbol means equivalence at the level of the topological part of the action.
Substituting these expressions in (\ref{vvn}), we find that the topological part of the gravitational path integral contributes $S_{\text{vN}} \sim 2S_0$. 

Finally, let's discuss the contribution from the bulk field theory. It is helpful to compare the island geometry in (\ref{bulk3}) to the empty-set one in (\ref{bulk1}). In the empty-set geometry (\ref{bulk1}), the cut associated to the twist operators on the boundary extended all the way across the bulk. In the limit $n\rightarrow 1$, the corresponding field theory computation was the entropy of the bulk fields on that slice. In the island geometry (\ref{bulk3}), the cut ends on the two black dots, which are the fixed points of the $\mathbb{Z}_n$ symmetry. In the limit $n\rightarrow 1$, the corresponding field theory computation is the entropy of bulk fields in the portion of system 1 outside the black dots. Adding this together with the gravity contribution described above, we have very roughly
\be
S_{\text{vN}} =  2S_0 + (\text{bulk entropy in BH 1 outside black dots}).
\ee
At the level of our topological analysis, this is consistent with the island conjecture (\ref{islandC}).

In order to go beyond this crude analysis, we would like to find actual on-shell solutions in JT gravity coupled to a bulk CFT, which will require going to late times. As in many discussions of JT gravity, it is convenient to think about doing the path integral in two stages. First, we integrate over the dilaton. This imposes a delta function constraint that the metric should be hyperbolic, with $R = -2$. Next, we integrate over the moduli space of hyperbolic metrics, and over the space of ``boundary wiggles,'' which represent the shape of the cutoff within the parent hyperbolic manifold \cite{Maldacena:2019cbz}. Finding true solutions appears to involve two complications: first, in order to find on-shell solutions, we need to continue the twist operators forwards in Lorentzian time a certain amount. Second, for integer Renyi index $n$, the solution seems to involve an interesting but somewhat nontrivial configuration of these boundary wiggles. This problem can be avoided by continuing the off-shell action near $n = 1$, and observing that in this limit, only a finite dimensional space of wiggles can get excited. We study this problem next.

\subsection{The pinwheel geometry for \texorpdfstring{$n \approx 1$}{n near 1}}
Let's begin by analyzing the pinwheel geometry in more detail, and working out its off-shell action near $n = 1$. The simple final answer could be derived quickly from the methods in \cite{Lewkowycz:2013nqa,Dong:2017xht}, but we believe the following derivation may generalize better to integer $n >1$.


We are interested in the special case of a pinwheel geometry with $\mathbb{Z}_n$ symmetry, and with a hyperbolic metric $R = -2$. Such geometries can be parametrized by the length $b$ of the geodesic that separates off any one of the asymptotic regions. Other geometrical quantities can be computed in terms of this $b$ parameter. For example, consider the two fixed points of $\mathbb{Z}_n$ symmetry, shown with black dots in (\ref{fig:pinwheel}). If we set $2\rho$ to be the geodesic distance between the two points, then one can show using hyperbolic geometry that
\be
\cosh\Big(\frac{b}{4}\Big) = \sin\Big(\frac{\pi}{n}\Big)\cosh(\rho).
\ee
An important special case of this formula is the limit $n\rightarrow 1$, where
\be
b = 2\pi \mathrm{i} - 4\pi \mathrm{i} \cosh(\rho)\, (n{-}1) + O\big((n{-}1)^2\big).\label{nnearone}
\ee
The pinwheel has $n$ of these geodesics of length $b$, separating off the $n$ asymptotic regions from the rest of the geometry. The asymptotic region outside each of these geodesics is a ``trumpet'' geometry
\be
\mathrm{d}s^2 = \mathrm{d}\sigma^2 + \cosh^2(\sigma) \frac{b^2}{(2\pi)^2}\mathrm{d}\theta^2\label{trumpet}
\ee
with $\sigma \ge 0$. The locus $\sigma = 0$ is the geodesic of length $b$, and the asymptotic boundary is at $\sigma \rightarrow \infty$. 

In addition to the parameter $b$, the pinwheel geometry is characterized by more subtle ``boundary wiggles,'' which represent the shape of the cutoff surface at large $\sigma$. These can be parametrized by giving the angular coordinate $\theta$ in (\ref{trumpet}) as a function of the proper length along the cutoff surface. It is convenient to use a rescaled proper-length coordinate $\tau$, which runs from zero to $\beta$. The full off-shell action for the $\mathbb{Z}_n$-symmetric pinwheel geometry reduces \cite{Maldacena:2019cbz} to the Schwarzian action for the trumpet geometry \cite{Saad:2019lba}
\be
nI_{\text{Sch}} = -n\int_0^\beta\mathrm{d}\tau\, \text{Sch}(e^{-\frac{\theta b}{2\pi}},\tau) = \frac{n}{2}\int_0^\beta\mathrm{d}\tau\left[\frac{\theta''^2}{\theta'^2} + \frac{b^2}{(2\pi)^2}\theta'^2\right].
\ee
Putting the boundary wiggles on shell means no wiggles at all, $\theta(\tau) = \frac{2\pi}{\beta}\tau$. However, the result is still not on shell with respect to $b$, because one finds $I \propto b^2$.


So far, we have taken $n$ to be general. But for $n$ very close to one, the $\mathbb{Z}_n$ quotient of the pinwheel geometry becomes the hyperbolic disk, with two marked points corresponding to the fixed points of the $\mathbb{Z}_n$ symmetry, see (\ref{bulk3}). The hyperbolic disk has exact SL(2,R) symmetry, and for $n$ close to one, we expect an approximate SL(2,R) symmetry. This leads to a separation of scales in the action: four special modes become parametrically soft. Three of these are parametrized by an SL(2,R) transformation, and one is parametrized by the distance $2\rho$ between the $\mathbb{Z}_n$ fixed points. We would like to work out a ``slightly off-shell'' action with arbitrary values of these modes, but with everything else on-shell.

A finite SL(2,R) transformation that preserves the unit circle is
\be\label{preserves}
x \rightarrow \frac{zx  +y}{\bar{y}x + \bar{z}}.
\ee
The corresponding configuration of the Schwarzian variable $\theta(\tau)$ is determined by
\be
e^{i\theta} = \frac{z e^{2\pi \mathrm{i}\tau/\beta} + y}{\bar{z} + \bar{y}e^{2\pi \mathrm{i}\tau/\beta}}.\label{expitheta}
\ee
The Schwarzian derivative $\text{Sch}(e^{-\frac{\theta b}{2\pi}},\tau)$ for this function $\theta(\tau)$ can be computed using the composition rule $\text{Sch}(f\circ g,\tau) = \text{Sch}(f,g)g'^2 + \text{Sch}(g,\tau)$, where $f(x) = x^{\frac{\mathrm{i}b}{2\pi}}$ and where $g(\tau) = e^{i\theta}$ given in (\ref{expitheta}). The integral over $\tau$ can be converted to an integral over $e^{i\theta}$, and can then be done by contour integration. The result for the action is
\be
I_{\text{Sch}} = -\frac{2\pi^2}{\beta}\left[1 + \left(1 + \frac{b^2}{(2\pi)^2}\right)\frac{y \bar{y} + z\bar{z}}{y\bar{y} - z\bar{z}}\right].\label{schactio}
\ee
Now taking the limit $n\rightarrow 1$, we can use (\ref{nnearone}) to expand this off-shell action near $n = 1$ as
\begin{align}
I_{\text{Sch}} &= -\frac{2\pi^2}{\beta} -\frac{8\pi^2}{\beta}\cosh(\rho)\frac{y \bar{y} + z\bar{z}}{y\bar{y} - z\bar{z}}(n-1) + O\big((n{-}1)^2\big).
\end{align}

It will be convenient to think about this action in a slightly different way. We can act with an SL(2,R) transformation on the entire geometry (\ref{bulk3}) to ``straighten out'' the boundary mode (\ref{expitheta}). This transformation will leave the metric invariant, since the hyperbolic disk metric (\ref{rotinv}) is invariant under (\ref{preserves}), but it will move the two marked points. A straightforward calculation shows that
\begin{align}
I_{\text{Sch}}=-\frac{2\pi^2}{\beta} - 2\pi \big[\phi(\rho_1) +\phi(\rho_2)\big](n-1)+ O\big((n{-}1)^2\big),\label{offshell}
\end{align}
where $\rho_1,\rho_2$ are the $\rho$ coordinates of the new marked points, and 
\be
\phi = \frac{2\pi}{\beta}\cosh(\rho) = \frac{2\pi}{\beta}\frac{1+r^2}{1-r^2}
\ee
is the dilaton profile in the $n = 1$ geometry. This simple form of the final answer would follow more directly from the general results in \cite{Lewkowycz:2013nqa,Dong:2017xht}.

Let's briefly summarize. For $n$ near one, the pinwheel geometry has four special modes with action of order $(n-1)$. These can be parametrized by the locations of the two $\mathbb{Z}_n$ fixed points, considered as marked points in the hyperbolic disk. The off-shell action, as a function of the locations of these points, is related to the value of the dilaton at these points, (\ref{offshell}).

\subsection{The bulk field theory calculation}
So far we have just studied the gravitational action of the pinwheel geometry. We also need to analyze the contribution from the bulk matter CFT. As explained in \cite{Faulkner:2013ana}, the matter path integral can be understood as computing 
\be
\tr(\widehat{\rho}_n^n)
\ee
where $\widehat{\rho}_n$ is the un-normalized density matrix prepared by the path integral on the $\mathbb{Z}_n$ quotient geometry. For $n\approx 1$, this approaches (\ref{bulk3}). In principle, we need to consider two types of corrections to $n = 1$, 
\be
\widehat{\rho}_n^n = \widehat{\rho}_1 + (n-1)\big[\partial_n \widehat{\rho}_n|_{n = 1}+\widehat{\rho}_1\log(\widehat{\rho}_1)\big] + O\big((n-1)^2\big).
\ee
The first $O(n-1)$ term can be shown to correspond to a small shift in the background value of the dilaton at the fixed points, due to quantum fluctuations of the propagating matter fields \cite{Faulkner:2013ana,Dong:2017xht}. This term will be subleading in the discussion below, and we will neglect it. However, the second $O(n-1)$ term leads to an important contribution for $n\approx 1$:
\be
\log\mathcal{Z}_n \supset (n-1)\times (\text{bulk entropy in cut region of BH 1}).
\ee
We need to evaluate this entropy. This sounds potentially difficult, because the cut region in (\ref{bulk3}) consists of two intervals. But in the late-time limit of interest, it can be well-approximated \cite{Almheiri:2019yqk} using the formula for the entropy of a single interval, which is simple in 2d CFT.

Let's review the CFT formula for the entropy of an interval. Suppose we are interested in a metric of the form
\be
\mathrm{d}s^2 = \Omega(x)^2\mathrm{d}x^i \mathrm{d}x^i, \hspace{20pt} i = 1,2
\ee
We consider the state to be the vacuum in the $x$ coordinates. For a conformal field theory, one would naively expect the entropy to be Weyl-invariant. But this is not quite true because of the need for a cutoff. If we assume that the appropriate cutoff is one in the physical distance, then the regularized entropy of an interval is (see e.g.~\cite{Calabrese:2009qy})
\be
S = \frac{c}{3}\log\left(|x - x'|\sqrt{\Omega(x)\Omega(x')}\right) + (\text{UV divergence ind. of $x,x'$}).\label{entform}
\ee
In the twist-operator formalism for the entropy, this $\Omega$ dependence comes from the Weyl-transformation properties of the twist operators.

As a warm-up, we would like to apply this to work out the regularized entropy of the interval between $r = r_1$ and $r = 1$ (at fixed $\theta$) in the metric
\be
\mathrm{d}s^2 = 4\frac{\mathrm{d}r^2 + r^2\mathrm{d}\theta^2}{(1-r^2)^2}, \hspace{20pt} 0\le r \le \infty.
\ee
This metric is conformal to the plane, and we consider the state that is the ordinary vacuum in the plane. From (\ref{entform}), the entropy of an interval $[r_1,r_2]$ at fixed $\theta$ would be
\be
S_{\text{bulk}} = \frac{c}{3}\log\left(\frac{r_2-r_1}{\sqrt{(1-r_1^2)(1-r_2^2)}}\right) + (\text{UV divergence ind. of $r_1,r_2$}).
\ee
In the present case we are interested in taking $r_2 = 1$. As $r_2$ approaches 1, the expression diverges. This reflects the divergent entanglement between the two black holes due to the coupling at the boundary. This is an IR divergence in the bulk and a UV divergence in the boundary. In principle, this should be cut off by smearing out the coupling between the two boundaries so that very high-energy modes are not coupled. Although we will not discuss the details of this regularization, it will lead to an expression of the form
\be
S_{\text{bulk}} = \frac{c}{3}\log\frac{1-r_1}{\sqrt{1-r_1^2}} + \text{const} = -\frac{c}{6}\rho_1 + \text{const}
\ee
where the constant depends on the regularization.

\begin{figure}[t]
\begin{center}
\includegraphics[scale = 1.2]{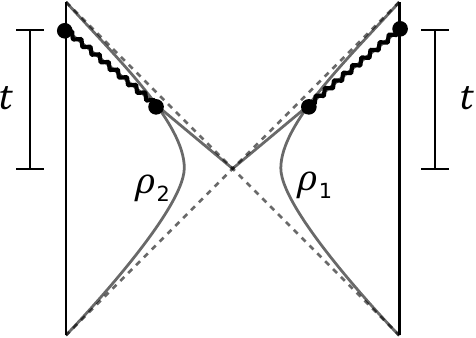}
\caption{{\small The black dots on the boundaries correspond to the boundary twist operators, and the black dots in the center are the marked points corresponding to the $\mathbb{Z}_n$ fixed points. For large $t$, the entropy on the two intervals marked with wavy lines is approximately the same as the sum of the entropies on the two intervals taken separately.}}\label{fig:lorentz}
\end{center}
\end{figure}
Now, the entropy that we actually need to compute is the one for two intervals, where each interval stretches between one of the $\mathbb{Z}_n$ fixed points in (\ref{bulk3}) and one of the boundary twist operators. This problem simplifies in a late time Lorentzian configuration like the one shown in figure \ref{fig:lorentz}. As we increase $t$, the two intervals become far apart and uncorrelated with each other, and the entropy approaches the sum of two separate single-interval computations,
\be
S_{\text{bulk}} \approx \left[-\frac{c}{6}\rho_1 + \text{const}\right] + \left[-\frac{c}{6}\rho_2 + \text{const}\right].
\ee
In this formula, we assumed that the bulk and boundary twist operators were at the same Killing times. We won't work out the formula for different times, but we note that the action is stationary under first-order changes of these Killing times.

\subsection{Saddle point}
We can now put the gravity computation together with the bulk CFT computation and find a saddle point. For $n$ near one, we have the off-shell contribution 
\be
\mathcal{Z}_n \approx \int \mathrm{d}^2x_1\mathrm{d}^2x_2 \exp\Big\{ (2-n)S_0 - n I_{\text{Sch}} + (n-1)S_{\text{bulk}}\Big\}
\ee
where the integral is over the locations of the two marked points in the disk geometry. In the case where the boundary twist operators have been continued to late time, this integral has a saddle point for a Lorentzian configuration of the two marked points, corresponding to the one shown in figure \ref{fig:lorentz}. In principle, we should extremize the action over both coordinates. But for late time, the extremum will be such that the points $x_1$ and $x_2$ are at the same Killing time as the corresponding boundary twist operators.\footnote{This is because the calculation is invariant under time-reversal symmetry, once we ignore correlations between the two intervals.} We then have to extremize over the radii. The action is
\be
-nI_{\text{Sch}} + (n-1)S_{\text{bulk}} = n\frac{2\pi^2}{\beta} + (n-1)\left[\frac{4\pi^2}{\beta}\big(\cosh(\rho_1) + \cosh(\rho_2)\big) - \frac{c}{6}(\rho_1+\rho_2)\right] + O\big((n-1)^2\big).
\ee
This action has an extremum at $\sinh(\rho_1) = \sinh(\rho_2) =  \beta c/24\pi^2$. This is analogous to the quantum extremal surface of \cite{Penington:2019npb,Almheiri:2019psf,Almheiri:2019hni,Almheiri:2019yqk}. Here we have shown how it arises from a $n\to 1$ limit of a Euclidean wormhole calculation.\footnote{In the setup described here, both entangled systems are gravitational, and we could consider another saddle point where BH 1 remains disk-like and BH 2 forms a pinwheel. By making the $S_0$ parameters of the two theories different, we can make one of these configurations dominate over the other.}

\subsection{Comments on integer \texorpdfstring{$n$}{n}}\label{sec:commentsoninteger n}
Finding saddle points explicitly for integer $n > 1$ seems to be be significantly more difficult than in the $n\approx 1$ limit. One reason is that it is no longer possible to restrict to the SL(2,R) modes of the Schwarzian theory: a more general $\theta(\tau)$ is necessary. Gluing the two black holes together then involves a ``conformal welding'' problem. Without getting into the details of this, we would like to explain qualitatively why the Renyi entropy should be finite in the limit $t\rightarrow \infty$, using an argument similar to one in \cite{Saad:2018bqo}. We caution the reader that the discussion in this section and the next one \ref{sec:factorizationProblem} is preliminary.

As a first step, consider the auxiliary computation 
\be
\tr\left(e^{-(\beta+\beta')H_{1+2}}\right) = \tr\left(e^{-\beta H_{1+2}}e^{iH_{1+2}t} e^{-\beta' H_{1+2}}e^{-iH_{1+2}t}\right) = \includegraphics[valign = c, width = .25\textwidth]{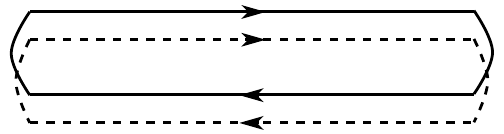}.\label{contourcomp}
\ee
Here $H_{1+2}$ represents the Hamiltonian of the combined boundary dual of BH 1 and BH 2, including the interaction. In the first equality, we inserted some cancelling factors of $e^{\pm i Ht}$, and in the figure at right we sketched a path integral contour that would compute this quantity. The solid and dashed lines represent the boundary duals of the two BH systems. The ``caps'' at the left and right end represent the Euclidean evolution by $\beta$ and $\beta'$, and the long horizontal portions represent the forwards and backwards Lorentzian evolution.

This quantity is exactly independent of time, and it is helpful to imagine taking $t$ to be very large. Then the contour has a time-translation invariance that is broken only near the endpoints. The gravity solution that fills this in will also have this property, and in order for the answer to be independent of $t$, this time-translation-invariant gravity geometry must have zero action per unit time. In fact, the relevant geometry is easy to identify; it is a piece of a thermofield-double-like geometry, which is invariant under forwards evolution on one side and backwards evolution on the other. The action is zero per unit time because of a cancellation between the two sides.\footnote{We say TFD-like, because if $\beta\neq\beta'$, the two boundaries will not be on opposite sides, but at some general angle. Also, this TFD-like geometry actually consists of two TFD-like geometries, one for each of BH 1 and BH 2. However, because of the interactions between the quantum fields, it is not exactly the same as two separate TFDs; in particular, the quantum state of the matter fields will be somewhat entangled.}
\begin{figure}[t]
\begin{center}
\includegraphics[width = .9\textwidth]{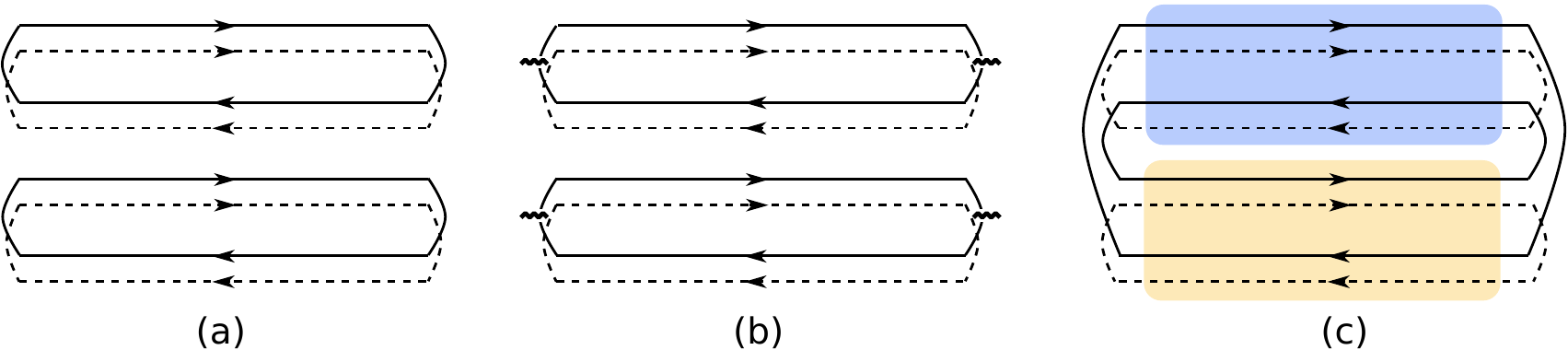}
\caption{{\small In (a) we sketch the boundary time contours for two copies of (\ref{contourcomp}), where the long horizontal part is the Lorentzian portion and the ``caps'' at the ends are Euclidean. In (b) we insert boundary twist operators (wiggly lines) in order to compute the purity $\tr(\rho_1^2)$. In (c) we represent these twist operators explicitly by changing the pattern of connection of the contours. We can make a configuration with bounded action as $t\to\infty$ by pasting the solution from the corresponding region of (a) into the shaded regions of (c). The configuration is thermofield-double-like in these two regions.}}\label{fig:contours}
\end{center}
\end{figure}

Now let's use this auxiliary computation. The contour for the Renyi 2-entropy is shown in figure \ref{fig:contours}(b) and \ref{fig:contours}(c). The long Lorentzian part of this contour is exactly the same as for two copies of (\ref{contourcomp}), shown in figure \ref{fig:contours}(a). The difference has to do with the way the contours are connected together at the ends (and also the amount of Euclidean evolution at the ends). To find a configuration whose action does not grow with time, we can paste the solution from part (a) into the shaded region of (c), and then fill in the remainder in some way that is consistent with the new boundary conditions. Since this modification happens near the ends of the contours, it will cost an amount of action that does not depend on the length of the time interval, at least in the limit of large $t$. In two dimensions, the topology of the resulting gravity configuration will consist of a cylinder for BH 1 and two separate disks for BH 2.

This explains why there are gravity configurations for the Renyi entropy that have bounded action as $t\rightarrow \infty$, but it doesn't show that there are classical solutions. A potential subtlety is the following. In (\ref{contourcomp}) there are two parameters $\beta,\beta'$. The action along the Lorentzian part of the contours will be zero for any values of these parameters. We expect that these are stabilized to saddle point values when we take into account the action penalty from gluing in the caps at the ends. We have not shown this reliably in gravity, but we will verify it in a related context by finding numerical solutions in the SYK model in the next section.

\subsection{A factorization problem}\label{sec:factorizationProblem}
In the context of the simple model, we argued in section \ref{sec:factorization and averaging} that the answers from the gravity path integral must be interpreted in terms of some implicit ensemble average. In order to make this argument, we showed that the gravity path integral predicted $\langle \psi_i|\psi_j\rangle = 0$ but $|\langle \psi_i|\psi_j\rangle|^2 \neq 0$. Can one say something similar for a more physical black hole?

In the simple model, $|\psi_i\rangle$ was the state of the black hole after projecting the radiation $\sf{R}$ onto a definite state $|i\rangle$. For a radiating black hole, an analog is as follows. We consider a case with only one black hole system, which starts out in a microcanonical version of the TFD state, called $|E\rangle$\footnote{For concreteness, we assume a width $\Delta$ of order the thermal scale associated to energy $E$.}
\be
|E\rangle \propto \sum_{|E_i-E|\le \Delta}|E_i\rangle_L|E_i\rangle_{R}.
\ee
To project onto a state in which the black hole radiates a particular sequence of Hawking quanta, we could act on this initial state with a sequence of annihilation operators
\be
a_{i_m}(t_m)\dots a_{i_1}(t_1)
\ee
where $i_j$ refer to a particular sequence of Hawking radiation modes, and $t_j$ are the times at which these modes are extracted. Acting with a sequence of this type will lower the energy of the black hole, and eventually lead to evaporation. However, in order to make our arguments as sharp as possible, we would like the ``evaporation'' process to continue forever, so we will intersperse creation operators between the annihilation operators in such a way that the energy of the black hole remains constant.

Concretely, we will evolve forwards in time on the $R$ boundary of the two-sided state $|E\rangle$, applying a sequence of $m$ creation and annihilation operators to various bulk field theory modes, roughly once per unit time,
\be
\mathcal{O}(i_1\dots i_m;t_1\dots t_m) = a_{i_m}^\dagger(t_k) a_{i_{m-1}}(t_{m-1}) \dots a_{i_2}^\dagger(t_2) a_{i_1}(t_1).\label{Adef}
\ee
Here, for simplicity, we are taking the bulk matter theory to be a product of free field theories. It will be convenient to constrain $\mathcal{O}$ so that it has approximately zero conserved charges in the boundary theory. So, in particular, we balance the number of creation and annihilation operators so that when we act with $\mathcal{O}$ on a state, we do not change the energy significantly.

Let's choose two different operators of the type (\ref{Adef}), and call them $A$ and $B$. Then we can consider states obtained by acting with these operators on the $R$ system of the two-sided state $|E\rangle$:
\be
|\psi_A\rangle = A^{(R)}|E\rangle, \hspace{20pt} |\psi_B\rangle = B^{(R)}|E\rangle.
\ee
We expect that in JT gravity coupled to a bulk free theory, the inner product between these states will either be zero or will approach zero as we increase the number of operator insertions $m$,
\be
\frac{\langle \psi_A|\psi_B\rangle}{\sqrt{\langle \psi_A|\psi_A\rangle\langle \psi_B|\psi_B\rangle}} \to 0.
\ee
We can think of this inner product as the one-point function of $A^\dagger B$ in the state $|E\rangle$. As described in section 6.2 of \cite{Saad:2019pqd}, we can get a nonzero answer for the square of this one point function, by considering a Euclidean wormhole with two boundaries. What we would like to do now is argue that the gravity answer for
\be
\frac{|\langle \psi_A|\psi_B\rangle|^2}{\langle \psi_A|\psi_A\rangle\langle \psi_B|\psi_B\rangle}\label{numdenom}
\ee
approaches a constant as the number of insertions $m$ goes to infinity.

\begin{figure}[t]
\begin{center}
\includegraphics[width = \textwidth]{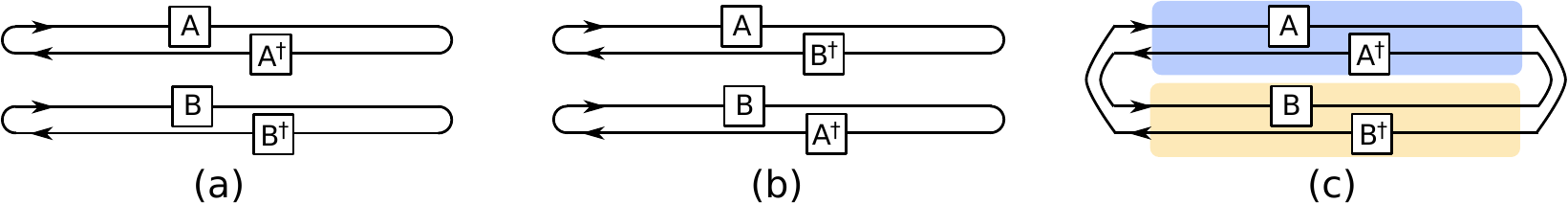}
\caption{{\small In (a) we sketch the boundary time contours for computing $\langle \psi_A|\psi_A\rangle\langle \psi_B|\psi_B\rangle$. In (b) we sketch the contours for computing $|\langle \psi_A|\psi_B\rangle|^2$, and in (c) we rearrange them. In the shaded regions, we glue the relevant part of the solution from (a).}}\label{fig:contours4}
\end{center}
\end{figure}
The argument is very similar to the one we used in the last section to argue that Renyi entropies approach a finite answer as $t\rightarrow \infty$. The basic point is that the long Lorentzian portions of the contour (where the operators act) is the same for the numerator and denominator of (\ref{numdenom}), see figure \ref{fig:contours4}. By gluing the gravity saddle point for the computation of the denominator into the long Lorentzian parts of the computation of the numerator, we will find a finite answer, which is suppressed only by the action involved in gluing on the different set of ``caps'' at the ends. The resulting topology is that of a cylinder connecting the two asymptotic boundary circles, as described in \cite{Saad:2019pqd}.\footnote{In the microcanonical setting described here, we expect to find a stabilized solution based on this approach. In the canonical ensemble we do not expect one, based on arguments from \cite{Saad:2018bqo,Saad:2019pqd}.}

\section{SYK computations with two replicas}\label{sec:SYK}
JT gravity coupled to matter fields is not a UV complete theory, due to a divergence in the moduli space integral for long thin tubes. There is no reason to suspect that this is a  problem for the current calculations, but to build confidence we undertake a related calculation in the UV complete SYK model \cite{Sachdev:1992fk,KitaevTalks,Kitaev:2017awl}. We will numerically find a solution that gives a non-decaying contribution to the Renyi 2-entropy, which provides a check of the argument given in section \ref{sec:commentsoninteger n}.

We study the same physical arrangement as described in section \ref{sec:JT}, but with the black hole systems 1 and 2 replaced by SYK systems 1 and 2, interacting with each other in a way that will be described below. The goal is to compute the purity $\tr (\rho^2)$, for system 1, as a function of time. Using the replica-diagonal SYK saddle point, this purity will decrease exponentially in time forever (in a microcanonical ensemble). But in the exact theory, it must saturate at an exponentially small floor value. This is a good target for a nontrivial ``wormhole'' saddle point, and the interchange of dominance is the Renyi 2-entropy version of the Page transition.

This SYK setup, and the replica-diagonal saddle point, was considered previously by Gu, Lucas, and Qi in \cite{Gu:2017njx}. In order to find the ``wormhole'' saddle point, we will essentially import an appropriate modification of the ``double cone'' solution found in \cite{Saad:2018bqo}. 

To start, define two separate SYK models, called 1 and 2, with $N_1$ and $N_2$ fermions, respectively. In order to compute the Renyi 2-entropy, we will need to consider two replicas of this system. So we will write a general fermion with three different indices, $\psi_{i,a}^\alpha$
\be
\psi_{i,a}^\alpha: \hspace{15pt} i \in \{1,\dots,N_a\} = \text{flavor}, \hspace{15pt} a \in \{1,2\} = \text{physical system}, \hspace{15pt} \alpha \in \{1,2\} = \text{replica}.
\ee
For each value of the replica and subsystem indices $\alpha$ and $a$, we can define an SYK Hamiltonian\footnote{We take $q$ to be a multiple of four to avoid some factors below.}
\be
H_a^\alpha = \hspace{-5pt}\sum_{1\le i_1<\dots <i_q\le N_a}\hspace{-20pt}J_{i_1\dots i_q;a} \ \psi_{i_1,a}^\alpha\dots \psi_{i_q,a}^\alpha.
\ee
Note that the number of fermions in the two physical systems $a = 1,2$ are in general different, and the couplings $J_{i_1\dots i_q;1}$ and $J_{i_1\dots i_q;2}$ are independently drawn. However, the couplings do not depend on the replica index $\alpha$.

So far, the two physical systems are independent. We would like to introduce a coupling between them. In the computation of the Renyi entropy, different replicas will need to interact with each other, so we define a general operator with two replica indices $\alpha$ and $\alpha'$:
\be
V^{\alpha\alpha'} = \hspace{7pt}\hspace{-20pt}\sum_{\substack{1\le i_1<\dots<i_{\hat{q}}\le N_1 \\ 1\le j_1<\dots<j_{\hat{q}}\le N_2}}\hspace{-20pt}\hat{J}_{i_1\dots i_{\hat{q}},j_1\dots j_{\hat{q}}} \ \psi_{i_1,1}^\alpha\dots \psi_{i_{\hat{q}},1}^\alpha \ \psi_{j_1,2}^{\alpha'}\dots \psi_{j_{\hat{q}},2}^{\alpha'}.
\ee
Regardless of the replica indices, this operator always couples the two physical systems 1 and 2 to each other, and never to themselves.\footnote{We will take $\hat{q}$ to be even, so that $V$ preserves separate fermion parity symmetries on the two subsystems.}

Let's now specify the physical setup more precisely. The system starts out in a thermofield double state for the combined interacting $1+2$ system, with four subsystems $1L,1R,2L,2R$. For weak interactions, this thermofield double is a state with relatively little entanglement between the 1 system and the 2 system. (Most of the entanglement is between e.g.~1$L$ and 1$R$, the two sides of the TFD.) We then evolve the systems forward in time, computing the entropy of system $1L\cup 1R$. The state of the combined system is invariant under forwards evolution on the $R$ systems and backwards evolution on the $L$ systems. We will evolve forwards on $R$ only, leaving $L$ alone.

After evolving for time $t$, we would like to compute the entropy of the $1L\cup 1R$ system. This measures the entanglement between the total 1 system and the total 2 system. We expect this entropy to grow linearly in time, before saturating at a late time value determined by a nontrivial saddle point.
\begin{figure}[t]
\begin{center}
\includegraphics[width = .27\textwidth]{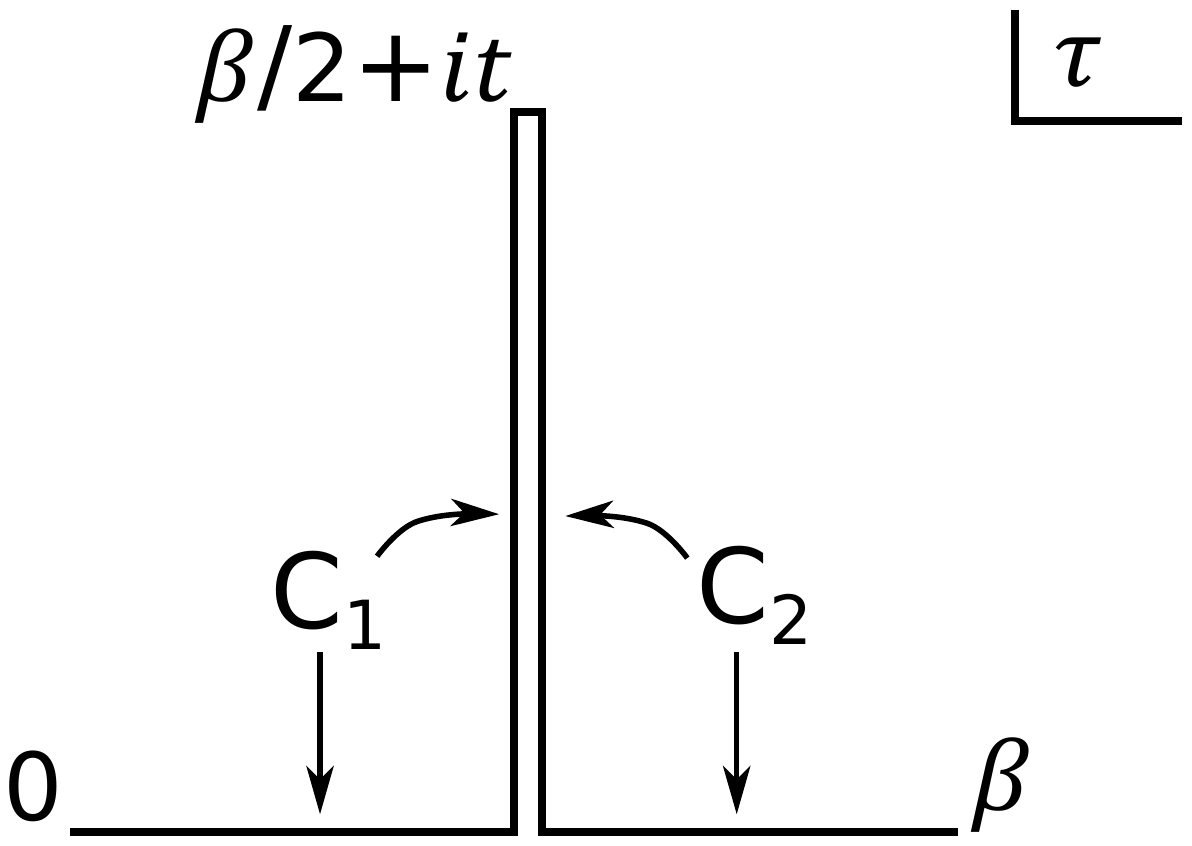}
\caption{{\small The time contour $C$, divided into $C_1$ and $C_2$. Note that $\tau = \beta$ is identified with $\tau = 0$, so $C$ is closed. One swap operator is inserted at $\tau = 0$ and another is inserted at $\tau = \beta/2+it$.}}\label{timeContour}
\end{center}
\end{figure}
More precisely, at time $t$, we will study the purity $\tr\left(\rho_1(t)^2\right)$ of the combined $1L \cup 1R$ subsystem. This quantity can be computed by a path integral with two replicas, and with insertions of a swap operator that swaps the replica index of the $1L$ and $1R$ subsystems. We expect it to decay exponentially until saturating at a floor value.

Including the effect of the swap operators, the path integral for the purity can be written as 
\be
\tr\left(\rho_1(t)^2\right)= \frac{1}{Z(\beta)^2}\int \mathcal{D}\psi_{i,a}^\alpha e^{-I}\label{toAVG}
\ee
where the action is \cite{Gu:2017njx}\footnote{In reading this equation, remember that the interaction terms $V$ always couple system 1 to system 2: the superscripts refer to the replica indices.}
\begin{align}
I &= \int_C \mathrm{d}\tau \ \sum_{\substack{a = 1,2 \\ \alpha = 1,2}}\Bigg[\sum_{i = 1}^{N_a}\psi_{i,a}^\alpha \partial_\tau \psi_{i,a}^\alpha + H_{a}^\alpha \Bigg]+ \int_{C_1}\mathrm{d}\tau \left[V^{11} + V^{22}\right] + \int_{C_2}\mathrm{d}\tau \left[V^{12} + V^{21} \right].
\end{align}
Each system has two replicas, and we can either couple a particular replica of system 1 to the same replica on system 2, or to the opposite one. These two possibilities are realized on the two components of the contour $C = C_1 \cup C_2$. Switching this coupling at the transition between the two contours is equivalent to inserting swap operators at those transition points. It will be helpful below to write the full action with a condensed notation
\be
\int_C \mathrm{d}\tau \sum_{\alpha\gamma}V^{\alpha\gamma}g^{\alpha\gamma}(\tau)
\ee
where $g(\tau)$ is the identity matrix for $\tau$ in $C_1$, and $g(\tau)$ is the swap operator for $\tau$ in $C_2$.

We can now compute the disorder average (over couplings $J$) of (\ref{toAVG}), by taking the couplings to be Gaussian random variables with mean zero and with
\begin{align}
\langle J_{i_1\dots i_q;a} J_{i_1'\dots i_q';a'}\rangle &=J^2 \delta_{i_1 i_1'}\dots \delta_{i_q i_q'}\delta_{aa'} \frac{ (q-1)!}{N_a^q}\\
\langle \hat{J}_{i_1\dots i_{\hat{q}};j_1\dots j_{\hat{q}}} \hat{J}_{i_1'\dots i_{\hat{q}'};j_1'\dots j_{\hat{q}}'}\rangle &= \hat{J}^2\delta_{i_1 i_1'}\dots \delta_{i_q i_q'}\delta_{j_1j_1'}\dots \delta_{j_{\hat{q}}j_{\hat{q}}'}\frac{(\hat{q}!)^2}{q(N_1 N_2)^{\hat{q}}}.
\end{align}
As usual in SYK calculations, the result for the disorder average can be written in terms of $G,\Sigma$ collective fields\footnote{Here we are intentionally making a small mistake and neglecting the fluctuations in $Z(\beta)$ in the ensemble of couplings. Such fluctuations are small $\sim N^{-q/2}$ and not significant for our analysis.}
\be
\langle \tr\left(\rho_1(t)^2\right)\rangle = \frac{1}{Z(\beta)^2}\int\mathcal{D}\Sigma_{a}^{\alpha\alpha'}\mathcal{D}G_a^{\alpha\alpha'}e^{-I}.
\ee
Note that one only needs collective fields that are diagonal in the physical system index $a$, although in general we need off-diagonal fields in the replica indices $\alpha,\alpha'$. The action is explicitly
\begin{align}
I &= \sum_a N_a\left\{-\log\text{Pf}\left(\partial_\tau \delta^{\alpha\alpha'}-\Sigma_a^{\alpha\alpha'}\right) + \frac{1}{2}\int_C\mathrm{d}\tau_1\mathrm{d}\tau_2\sum_{\alpha,\alpha'}\left[\Sigma_a^{\alpha\alpha'}(\tau_1,\tau_2)G_a^{\alpha\alpha'}(\tau_1,\tau_2) - \frac{J^2}{q}G_{a}^{\alpha\alpha'}(\tau_1,\tau_2)^q\right]\right\}\notag\\&\hspace{20pt}- \sqrt{N_1N_2}\frac{\hat{J}^2}{2q}\int_C\mathrm{d}\tau_1\mathrm{d}\tau_2 \sum_{\alpha\alpha'\gamma\gamma'}G_1^{\alpha\alpha'}(\tau_1,\tau_2)^{\hat{q}} \ g^{\alpha\gamma}(\tau_1)g^{\alpha'\gamma'}(\tau_2) \ G_2^{\gamma\gamma'}(\tau_1,\tau_2)^{\hat{q}}.\label{SYKACTION}
\end{align}
The saddle point equations that stationarize this action are
\begin{align}
G_a &= (\partial_\tau - \Sigma_a)^{-1}\label{saddlePt}\\
\Sigma_a^{\alpha\alpha'}(\tau_1,\tau_2) &= J^2G_a^{\alpha\alpha'}(\tau_1,\tau_2)^{q-1} + \sqrt{\frac{N_{\hat{a}}}{N_a}}\hat{J}^2G_a^{\alpha\alpha'}(\tau_1,\tau_2)^{\hat{q}-1}\sum_{\gamma\gamma'}g^{\alpha\gamma}(\tau_1)g^{\alpha'\gamma'}(\tau_2)G_{\hat{a}}^{\gamma\gamma'}(\tau_1,\tau_2)^{\hat{q}}.\notag
\end{align}
In the second line, we are using the notation $\hat{a}$ to mean the opposite physical system, so if $a =1$ then $\hat{a} = 2$. As usual in SYK equations, the first equation is more complicated than it seems: for fixed $a$, the quantities $G_a$ and $\Sigma_a$ are viewed as matrices, acting on the vector space parametrized by $\tau,\alpha$. The $\partial_\tau$ operator is also viewed in these terms, but is taken to be diagonal.

These equations can be discretized on the contour (\ref{timeContour}), and solved numerically, using the standard iterative approach that is common in the SYK literature. There are two different types of solutions that will be important for us, and we will discuss them in turn.

	\begin{figure}[t]
\begin{center}
\includegraphics[width = \textwidth]{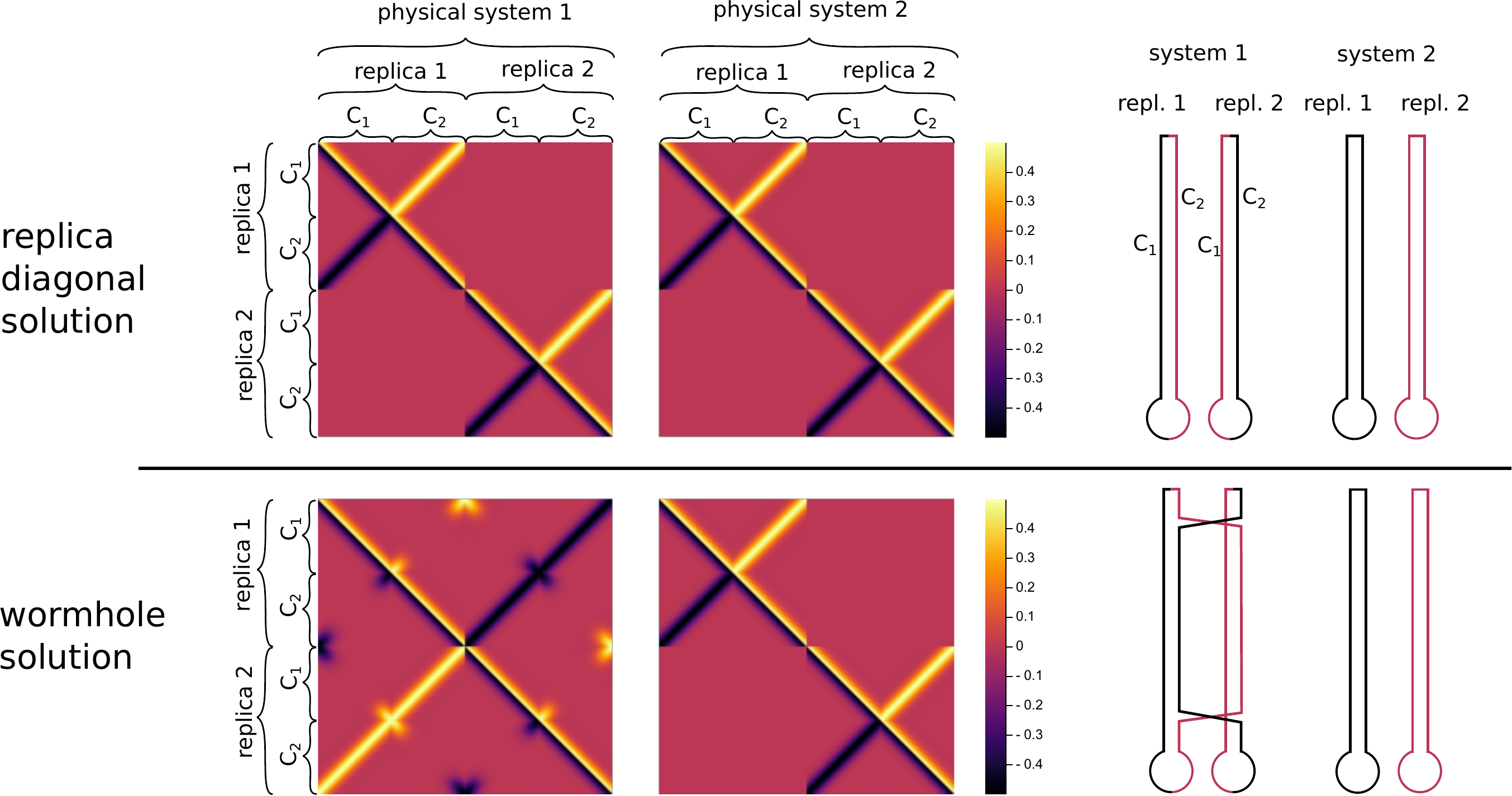}
\caption{{\small Numerical solutions to (\ref{saddlePt}). The heatmaps at left represent $G_1(t,t')$ and $G_2(t,t')$ for two different types of solution discussed in the main text. We plot at infinite temperature so the solutions are real. The stripe along the main diagonal is somewhat trivial, representing large correlation between adjacent points along the contour. The other features in the heatmap represent nontrivial correlations. Contours of system 1 are interacting with the contours of system 2 with the same color. The contours are drawn so that nearby points have large correlation in the solution. For this plot, $Jt = 20, \beta J = 0, \hat{J} = J/2, q = 4, \hat{q} = 2$.}}\label{figheatmap}
\end{center}
\end{figure}

\subsection{The replica-diagonal solution}
We start with the replica-diagonal solution shown in the top row of figure \ref{figheatmap}. This is the solution that was considered in \cite{Gu:2017njx}. As shown there, it has an action that initially grows linearly with time, implying an exponential decrease of the purity. 

The growth in the action can be understood as follows. First consider the case with no interaction between system 1 and system 2, so $\widehat{J} = 0$. In this case, the twist operators have no effect, and action evaluated on the saddle point $G,\Sigma$ configuration is exactly independent of time. This saddle point is simply the standard thermal solution $G(\tau)$ of the SYK model for each of the four copies (two noninteracting physical systems, with two replicas each), analytically continued along the contour $C$:
	\be
G^{\alpha,\alpha'}_1(\tau,\tau') = G^{\alpha,\alpha'}_2(\tau,\tau') = \delta^{\alpha,\alpha'}G(\tau-\tau').\label{leading}
	\ee
On this solution, if we evaluate the action with $\widehat{J} = 0$, the answer must be exactly independent of $t$.

Now, let's start with this solution and treat the $\hat{J}^2$ term on the second line of (\ref{SYKACTION}) as a perturbation. Evaluating the action by plugging in the unperturbed solution (\ref{leading}), the second line of (\ref{SYKACTION}) works out to
	\begin{align}
I &\supset -\sqrt{N_1N_2}\frac{\hat{J}^2}{q}\left[\int_{C_1}\mathrm{d}\tau_1\mathrm{d}\tau_2  \ G(\tau_1-\tau_2)^{2\hat{q}} + \int_{C_2}\mathrm{d}\tau_1\mathrm{d}\tau_2  \ G(\tau_1-\tau_2)^{2\hat{q}}\right] \\
& = \bigg[\sqrt{N_1N_2}\frac{2\hat{J}^2}{q}\int_{-\infty}^\infty \mathrm{d}t' G(it')^{2\hat{q}}\bigg] \times t + O(1).\label{INITIAL}
	\end{align}
	Here, the $O(1)$ term represents a non-growing contribution as $t$ becomes large. So the action will grow linearly in time, at least for while. This corresponds to an exponential decrease in the purity.

\begin{figure}[t]
\begin{center}
\includegraphics[width = .48\textwidth]{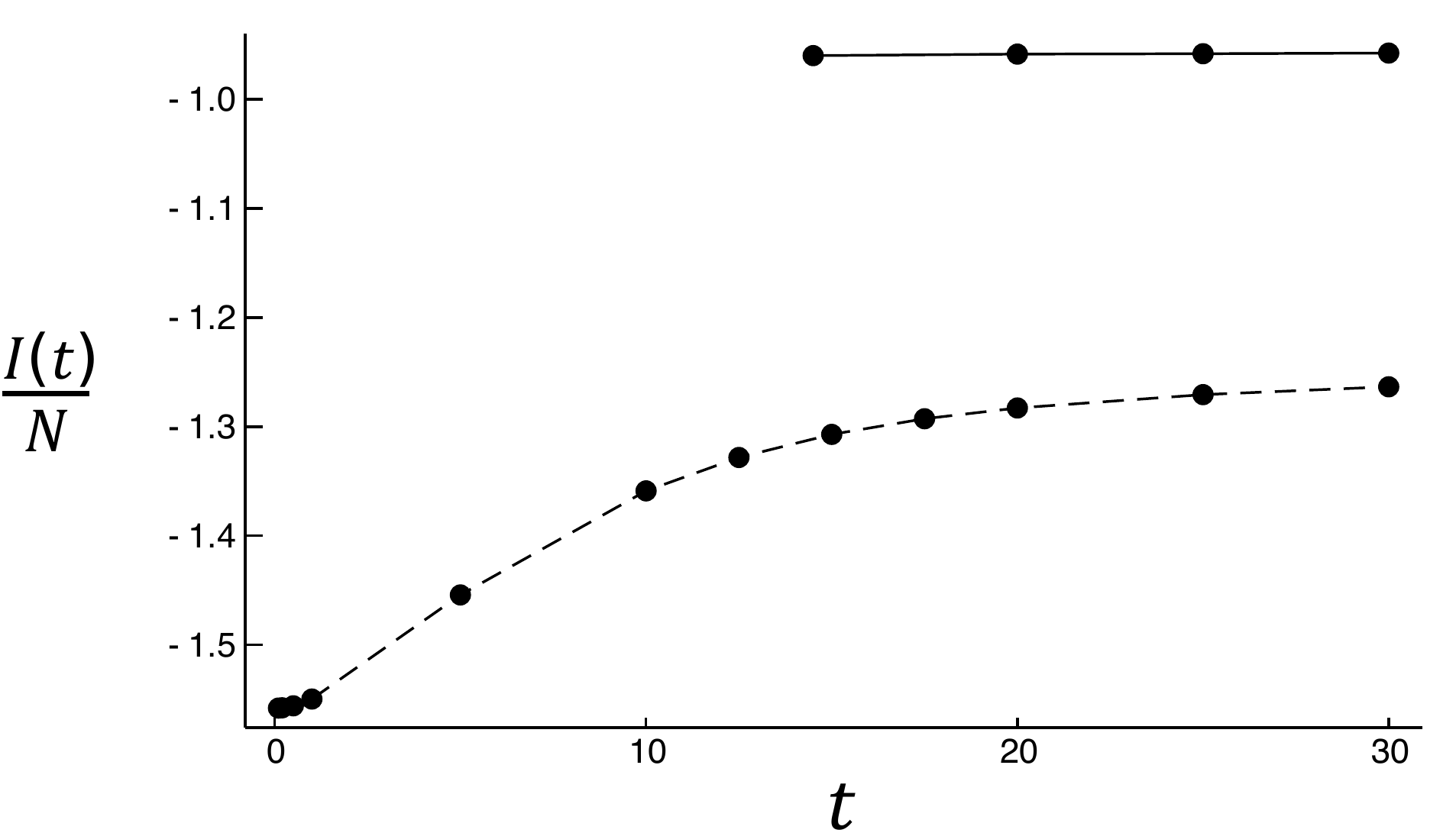}
\includegraphics[width = .49\textwidth]{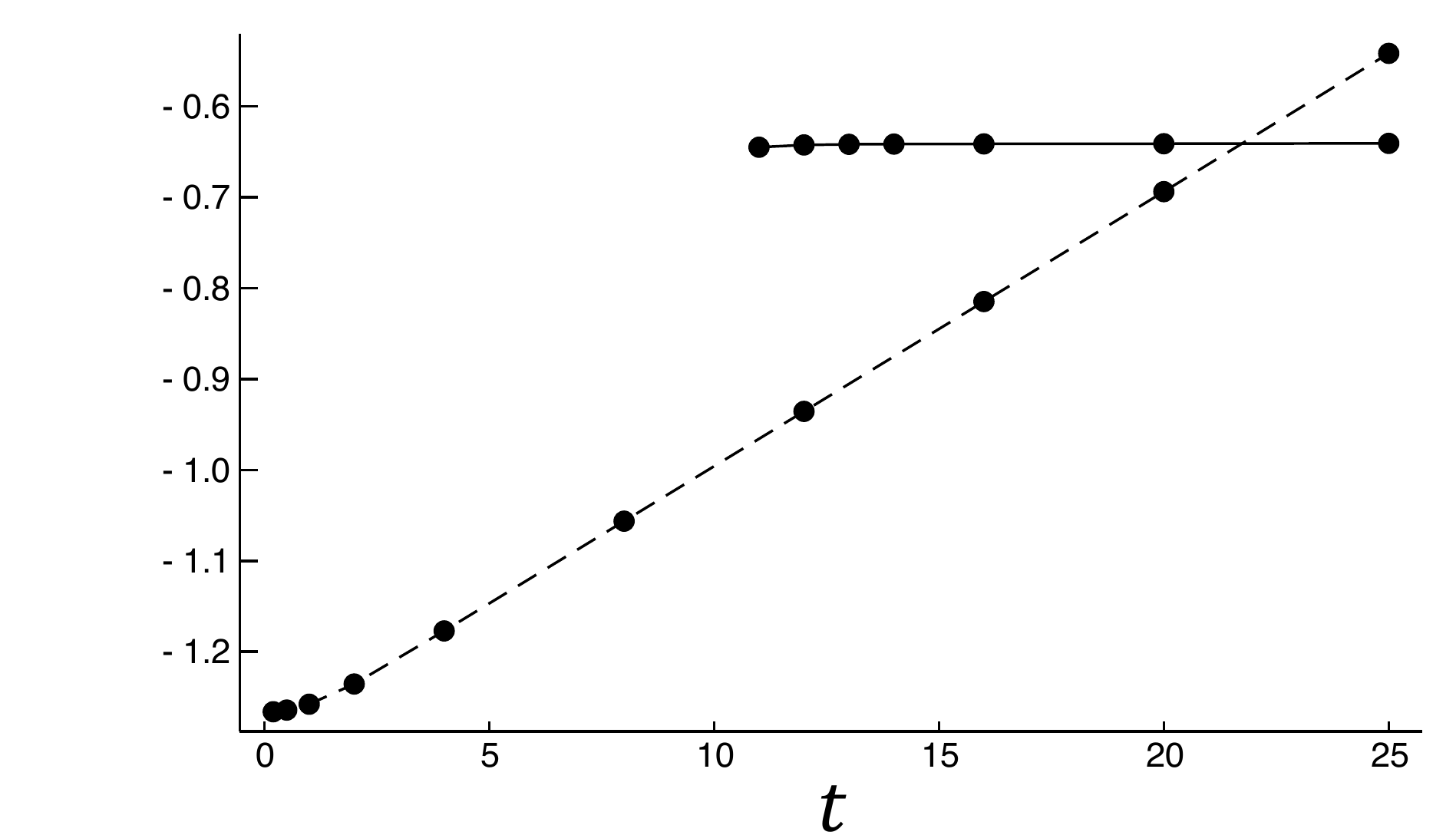}
\caption{{\small The action of the replica-diagonal solution (dashed) and the wormhole solution (solid) as a function of $t$. The left plot is in the canonical ensemble at $\beta J = 2.4$. The right plot is in the microcanonical ensemble at the corresponding energy, $E/(NJ) = -0.03$ per system. The black dots are data, and the curves are to guide the eye. The parameters were $J = 1, \hat{J} = J/2$, $q = 4$ and $\hat{q} = 2$.}}\label{figAction}
\end{center}
\end{figure}

This derivation is only valid for early times $\widehat{J}^2t/J \ll 1$, so that the interaction can be treated perturbatively. A subtle detail \cite{Gu:2017njx} is that in the canonical ensemble, this linear growth does not continue forever; instead, the action saturates at some finite value (see the left panel of figure \ref{figAction}). One might be tempted to interpret this saturation as the Page transition. However, what this saturation actually represents is the contribution of very low energies in the tail of the canonical ensemble, where the dynamics are slow enough that the two physical systems essentially do not evolve and entangle.\footnote{At the quantum mechanical level, the contribution of such states will still decrease with time (a power-law decrease in the purity), but we do not see this at the level of the classical action.}

This is a real effect in the Renyi 2-entropy, but such ``tail'' effects cannot significantly effect the von Neumann entropy (which isn't affected by tail effects the way Renyi entropies are), so we view it as a distraction. In order to avoid this subtlety altogether, we can work in the microcanonical ensemble. Then one finds that the action continues to increase linearly, as shown in the right panel of \ref{figAction}. So we conclude that in the microcanonical ensemble, the replica-diagonal saddle point gives an answer for the purity that exponentially decays forever.

\subsection{The wormhole solution}
The other class of solutions can be motivated by the discussion in section \ref{sec:commentsoninteger n}, which suggests that there should be a solution with an approximately time-translation invariant portion corresponding to a TFD-like correlation between different replicas, glued in some way to a configuration with the correct boundary conditions at the ends of the contour. We refer to this as a wormhole solution because its pattern of correlation is the same as that of the wormhole geometry from section \ref{sec:commentsoninteger n}.

In order to find such a  solution numerically, the procedure we followed was as follows. For the first few iterations of the Schwinger-Dyson equations, we included an explicit source in the SYK equations that encourages replica-off-diagonal correlations like the ones expected based on section \ref{sec:commentsoninteger n}. After a few iterations, we then set the source to zero and continued iterating until convergence. If the time $t$ is larger than a critical value (discussed below), we found that the iterations converged to a nontrivial solution like the one shown in figure \ref{figheatmap}. The pattern of correlations in this solution is precisely the one expected based on the argument in section \ref{sec:commentsoninteger n}. 

The numerical value of the action is independent of time to a good approximation, see figure \ref{figAction}.\footnote{To compute the action accurately, we used extrapolation in the size of the matrices that represent the discretized $G,\Sigma$ variables. The maximum size we used was $1600\times 1600$ each for system 1 and system 2. To speed up convergence, one can start with a refinement of the converged solution for smaller grid size.} To reiterate the discussion from section \ref{sec:commentsoninteger n}, this can be understood as follows: as we make the time larger, the only aspect of the solution that changes is the TFD-like portion in the middle of the time contours gets extended. Since this TFD-like configuration has exactly zero action, the action does not change.\footnote{Note that the SYK action is not local in time, but for solutions that decay exponentially in time like these, it is local enough for this argument to apply.}

Since the action of the replica-diagonal solution increases linearly (in the microcanonical ensemble), there will eventually be a transition between the two solutions. For the setup that we have described here, the transition is at a time that is independent of $N$, but proportional to $1/\hat{J}^2$.

It would be desirable to understand this solution better, since it seems to involve several interesting aspects of thermalization and chaos. One example of this has to do with the critical time at which the solution first starts to exist. Empirically, this is based on the following: near the endpoints of the contour, the solution needs to have a ``normal'' pattern of correlation, in which contours $C_1$ and $C_2$ of each replica are highly correlated with each other. However, as we move away from the endpoints, this pattern of correlation is replaced by correlation between $C_1$ and $C_2$ of opposite replicas. The decay of the first type of correlation appears to be due to a scrambling effect, sourced by the perturbation due to the coupling between the systems, $\hat{J}$. Based on this, one would predict that the critical time at which the ``wormhole'' solutions start to exist is logarithmic in the coupling $\hat{J}$
\be
t_{\text{first exist}} \propto \log(J/\widehat{J}).
\ee
This appears to be consistent with numerics (not shown).

\section{Entropy and replica wormholes in de Sitter}\label{sec:desitter}
In this section, we will tentatively discuss some entropy computations using replica wormholes in de Sitter space.\footnote{We thank Jorrit Kruthoff, Juan Maldacena, Mehrdad Mirbabayi and Eva Silverstein for discussions about de Sitter space.} Our starting point is the no-boundary proposal \cite{PhysRevD.28.2960} for the wave function of de Sitter space, or more precisely its generalization in \cite{PhysRevD.34.2267,barvinsky2008density,Maldacena:2019cbz,maldacenaStrings2019} to a no-boundary proposal for the density matrix. In this version, we compute a density matrix for the universe by summing over all geometries that end on the boundary conditions for the bra and ket vectors of the density matrix. In this sum, one can have separate disconnected geometries attached to the bra and ket (these terms would also be included in the original no-boundary proposal) but also connected geometries in which the bra and ket are distinct boundaries of a single connected spacetime. 

One can also generalize this further to a no-boundary proposal for the tensor product of $n$  copies of the density matrix $\rho$. In this case, we have $n$ bra-type boundaries and $n$ ket-type boundaries, and we sum over all spacetimes (connected or otherwise) that end on these $2n$ boundary conditions. 

This immediately leads to a somewhat surprising conclusion. Naively, it would appear that connected geometries will lead to a mixed density matrix. However, to check this, let's compare $\tr(\rho^n)$ and $\tr(\rho)^n$. In both cases, the boundary conditions consist of $2n$ boundary components: $n$ ket-type boundaries and $n$ bra-type boundaries. To compute either quantity, we identify these boundaries in bra-ket pairs, and integrate over the mutual boundary conditions for each pair. $\tr(\rho^n)$ and $\tr(\rho)^n$ correspond to two different ways of pairing up the $2n$ boundaries. However, the no-boundary rules described above are invariant under arbitrary permutations of the ket-type boundaries and the bra-type boundaries. So $\tr(\rho^n) = \tr(\rho)^n$ and the state is pure.

This seems at odds with the fact that the no-boundary answer for a single copy of the density matrix will be mixed. As in the discussion of black holes in section \ref{sec:factorization and averaging}, a possible interpretation is that the gravity path integral is describing an ensemble of theories in which the state of the universe is pure. The average of the density matrix is the mixed no-boundary density matrix. But the average of the entropy is zero. (Another possible interpretation is that our no-boundary prescription for $\rho^{\otimes n}$ is too aggressive. But we will keep it for the moment and see where else it leads.)

So far, we have only discussed the entropy of the whole universe. The entropy of subsystems is more interesting, but more difficult to compute. So we will study a simple model inspired by the end-of-the-world brane model for near-extremal black holes in section \ref{sec:asimplemodel}.
This model is a two dimensional nearly de Sitter space  described by  Jackiw-Teitelboim gravity with positive cosmological constant \cite{Maldacena:2019cbz,Cotler:2019nbi}. 
We take the asymptotic time slice to be a segment that starts and ends on two EOW branes with a large number of orthogonal internal states. The boundary conditions for an unnormalized ket vector look like the following
\be\label{eqn:dSstate}
\langle i j|\Phi\rangle= \ \includegraphics[scale = .7,valign = c]{images/melement.pdf}
\ee
Here, we are imagining that we measure the length of the spatial interval to be some renormalized value $\ell$, which is held fixed in the rest of the discussion. The density matrix $\rho$ consists of two copies of this boundary condition, which can be filled in as follows:
\begin{align}
\langle ij|\rho|i'j'\rangle \ &= \  \ \includegraphics[scale=0.6,valign = c]{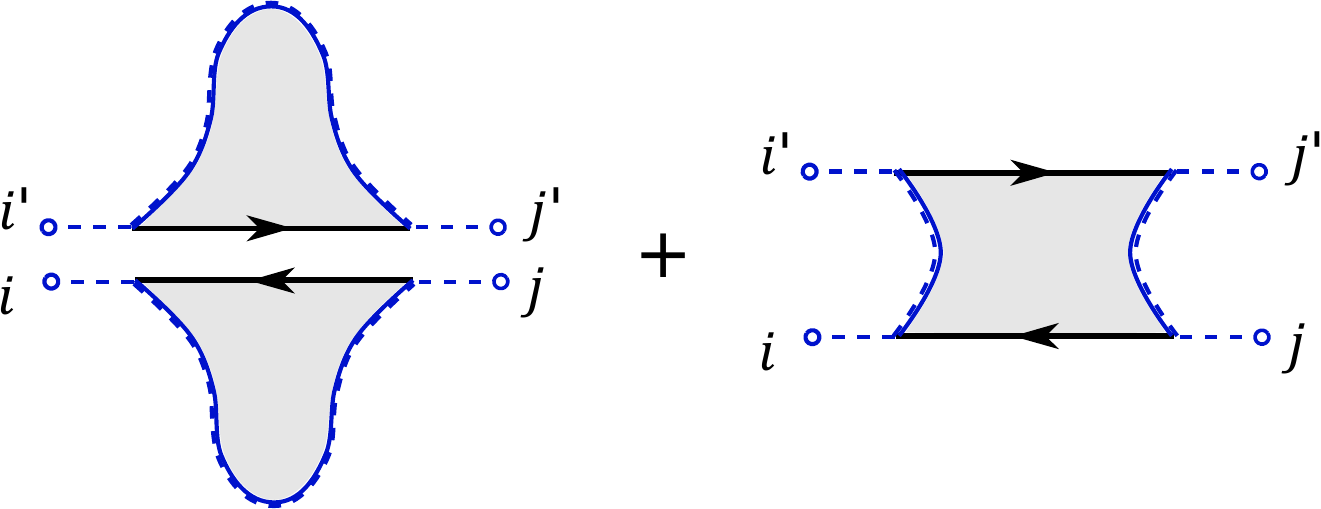} \ \ = \ \delta_{ij}\delta_{i'j'}|\tilde{Z}_1|^2 \ + \ \delta_{ii'}\delta_{jj'}\tilde{Z}_2.
\end{align}
The fact that there are multiple topologies already for the density matrix makes the story in de Sitter space richer than for the black hole. Note that due to the second term, this is a mixed state. Here, in nearly de-Sitter gravity, the path integrals are given by analytic continuation of JT gravity path integrals, see \cite{Maldacena:2019cbz,Cotler:2019nbi} (or more generally \cite{Maldacena_2003,Anninos:2011ui}) for details of this analytic continuation. For the simple case where the brane is massless, we have
\be
\tilde Z_{1}= 2e^{S_0}\int ds\rho(s)e^{i\ell s^2/2}|\Gamma(-\tfrac{1}{2}+is)|^2, \hspace{20pt} \tilde Z_{2}= 4e^{S_0}\int ds\rho(s)|\Gamma(-\tfrac{1}{2}+is)|^4.
\ee
In order to compute e.g.~$\tr(\rho)$, the primed and unprimed indices will be contracted. The first diagram will contribute $e^{2S_0}k$, and the second will contribute $e^{S_0}k^2$. So if we take a very large value of $k$, there can be an interchange of dominance between the two.

We can now compute the entropy of the left brane degree of freedom, corresponding to the $i$-type index. We will refer to its density matrix as $\sigma$. For small $k$, a completely disconnected topology dominates, and one finds
\be
\frac{\Tr(\sigma^n)}{\Tr(\sigma)^n}={k|\tilde Z_1|^{2n}\over k^n |\tilde Z_1|^{2n}}={1\over k^{n-1}}.
\ee
The von Neumann entropy will be $\log(k)$. In the opposite limit $k\gg e^{S_0}$, the fully connected geometry dominates, both for the computation of $\tr(\sigma^n)$ and for $\tr(\sigma)^n$. We find
\be
\frac{\Tr(\sigma^n)}{\Tr(\sigma)^n}= {k^{2n} \tilde Z_{2n}\over k^{2n} |\tilde Z_2|^{n}}={ \tilde Z_{2n}\over |\tilde Z_2|^{n}};~~~~~~\tilde Z_{2n}= e^{S_0}\int ds\rho(s)(2|\Gamma(-{1\over 2}+is)|^2)^{2n}.
\ee
The $s$ integral just gives an order one coefficient, so the answer is roughly $e^{(1-n)S_0}$ and the von Neumann entropy is approximately $S_0$.

So there is a ``Page-like" transition in our simple de Sitter model.  As we increase the number of EOW brane states, the von Neumann entropy of one of the branes has a transition from the naive result $\log k$ to the de Sitter entropy $S_0$. Although the EOW brane setup is rather artificial, this does give a hint at a microscopic role for the de Sitter entropy \cite{PhysRevD.15.2738} (see also \cite{Freivogel:2006xu,Dong:2018cuv,Lewkowycz:2019xse,Geng:2019bnn}).

It would be interesting if a similar effect could be seen in more physical setup in which the large number of states $k$ emerges naturally from some bulk computation, rather than being put in by hand as we did here. A starting point could be the ``centaur'' geometry \cite{Anninos_2018} or possibly the large number of states produced by quantum field theory evolution over many e-folds, which was previously considered in the context of the de Sitter entropy in \cite{Arkani_Hamed_2007,Dubovsky_2009}.\footnote{We than Juan Maldacena for pointing this out to us.}

\section{Wormholes in non-averaged systems}\label{sec:discussion}
The arguments in this paper use spacetime wormhole geometries in an essential way.    But the results in section  \ref{sec:factorization and averaging}  for the overlap of individual black hole microstates $|\psi_i\rangle$ computed using such wormholes seem  only to be  consistent if they are interpreted as ensemble averages.  In this case it seems natural to interpret the wormholes as part of an effective description, not a microscopic one.   They do not know about the exponentially large amount of microscopic data contained in the fluctuating phases  
$R_{ij}$ in the non-averaged matrix elements $\langle \psi_{i}|\psi_j\rangle = \delta_{ij} + e^{-S_0/2}R_{ij}$.         We should stress that the inconsistency in overlaps that is resolved by an averaged description is not limited to the simple model discussed in section \ref{sec:asimplemodel}.    As discussed in section \ref{sec:factorizationProblem}, we expect a similar situation for the radiating black hole.

The question we want to address here is what role wormholes play in systems \emph{without} averaging.

We would like to describe a somewhat analogous situation in semiclassical quantum chaos, which may provide some guidance \cite{Saad:2019lba}.   Consider a few body quantum chaotic system, like a quantum billiard.   Semiclassically, matrix elements in the position basis can be written as sums over classical trajectories connecting the bra point to the ket point.    This should allow an analysis of the overlap puzzle.      A simpler situation that illustrates  many of the  same ideas is to consider the Minkowski signature partition function $\tr e^{-iHt}$ which can be written semiclassically as a sum over periodic orbits $a$.  The product of two such partition functions, the spectral form factor, can be rewritten as  a double sum over  periodic orbits, which we can express schematically as follows:
\be
K(t) = \tr e^{-iHt} ~ \tr e^{iHt} = \sum_{ab} e^{i (S_a -S_b)} .
\ee
Here $S_a$ is the classical action of orbit $a$.  

The spectral form factor obviously factorizes into the product of partition functions --  this requires a double sum over orbits.
On averaging,\footnote{One could average over a small time intervals, for example, or over the shape of the billiard table.   In the following discussion we imagine the times are long, but well before the plateau time, of order $e^{S}$.}   only the   ``diagonal" terms corresponding to the same orbit in each sum (up to a time shift) survive. This gives the ramp.\footnote{This ``diagonal" approximation is due to Berry \cite{Berry1985SemiclassicalTO}.}   After the pairing connection is made factorization is lost, as expected for an averaged system. In this picture the diagonal pairing pattern is an effective, coarse-grained description of the exponentially large number of long,  diagonally paired orbits, multiplied by their exponentially small one-loop determinant.  

Let's now try to make an analogy to quantum gravity. We view the quantum Hamiltonian of the billiard, $H$ as the ``boundary" description.  The sum over orbits would be the microscopic  ``dual  bulk"  description. The diagonal pairing pattern in the coarse-grained sum over long orbits we take to be the analog of the wormhole geometry in the gravitational context.\footnote{Random tensor networks \cite{Hayden:2016cfa} give another example of the formation of such effective wormhole connections after averaging.   The Ising domain walls discussed in \cite{Hayden:2016cfa} describe the structure of these effective connections. A closer analogy to the microstate overlaps discussed in section \ref{sec:factorization and averaging} would be to compute the averages of individual density matrix elements, not purities.}

It is well-known that wormholes conflict with the factorization of e.g.~partition functions of decoupled systems \cite{Maldacena:2004rf}.  In a non-averaged situation where factorization must hold, what is one supposed to do with the wormholes? The solution the periodic orbit analogy suggests is related to one already offered in \cite{Maldacena:2004rf}.    To restore factorization in the non-averaged theory, one doesn't eliminate the paired diagonal terms corresponding to the wormhole.  Instead one adds back in all the other unpaired off-diagonal terms.   So to have a gravitational  bulk understanding of non-averaged theories we need a gravitational  bulk  understanding of these off-diagonal terms. These might well not have a simple geometrical description, even an effective one.   

The periodic orbits are defined in the microscopic phase space that semiclassically determines all the microstates of the quantum system.   So a variant of the issue at hand is to have a gravitational bulk understanding, geometrical or not,  of all the microstates of the system.  This is related to  the fuzzball program.\footnote{For reviews see \cite{Mathur:2008nj,Bena:2013dka}. For a critique of this program see \cite{Raju:2018xue}.}   

On occasion another idea has been suggested: wormholes in non-averaged decoupled systems could be ruled out because they are not actual saddle points. However, in JT gravity, the wormhole describing the ramp \emph{is} a saddle point in the microcanonical ensemble \cite{Saad:2018bqo,Saad:2019pqd}.  We see no obstruction to the existence of such microcanonically stable wormholes in more complicated higher dimensional gravitational theories dual to non-averaged systems \cite{Saad:2018bqo}.  But they would give the wrong answer in a non-averaged theory.\footnote{We thank Phil Saad for discussions on this point.}  They do not describe the erratic fluctuations due to the fine-grained structure of energy levels  that we expect, as illustrated for example in figure 10 of \cite{Cotler:2016fpe}.  These erratic fluctuations are not a small effect -- their magnitude is of the order of the signal itself. So the existence of a saddle point is not a sharp criterion for including wormholes.\footnote{It would be interesting to find some internal signature of this failure within the geometric bulk theory.   In the presence of bulk matter there is a UV ``Hagedorn" type divergence at small wormhole diameter \cite{Saad:2019lba} that may have some bearing on this issue.}  Again, we stress that the periodic orbit analogy tells us that  the off-diagonal contributions are responsible for the erratic behavior.  Any complete gravitational bulk description \emph{must} contain a description of the analog of these  off-diagonal terms.  

In the periodic orbit example we proposed that the wormhole geometry is  analogous  to the diagonal pairing pattern, an effective, coarse-grained description of which orbits contribute.   This raises the question of whether bulk geometry in general is only an effective description of some different, more fundamental degrees of freedom -- the analog of the periodic orbits.   Another possibility is that the fundamental bulk description contains ``geometric" degrees of freedom, like perturbative strings, in addition to other non-geometric ones -- complicated configurations of large numbers of branes, for example.  It is even conceivable that geometry could actually describe everything, in some subtle way as yet not understood.\footnote{An impressive example of microstate information being completely  described by geometry is the elegant calculation of \cite{Dabholkar:2014ema}  of the exact entropy of certain supersymmetric black holes.  The exact integer degeneracies are recovered from a sum over an infinite number of instanton corrections computed using supersymmetric localization.    What would be required in the present context is an analogous result that would explain the much more complicated pattern of energy levels present in these non BPS chaotic systems as a sum over gravitational configurations.}   
 
 The current situation, though,  is that there is \emph{no} known  bulk description of a gravitational theory that is  rich enough to include the microscopic information necessary to explain, for example,  the erratic behavior in the spectral form factor.   The nature, or perhaps even the existence, of such a bulk description remains one of the deepest mysteries in quantum gravity.

We now turn to a more pragmatic question:  if  spacetime wormholes are only an effective description, are such  configurations useful in non-averaged bulk theories?  We believe the answer is clearly ``yes."      For example, the entropies computed in  section \ref{sec:asimplemodel} are ``self-averaging."  This means that they  have small variance in an averaged theory,  basically because they are sums of large numbers of fluctuating terms.  This variance can be computed from the bulk by considering additional wormholes linking the two copies used to compute the variance.    Roughly speaking, 
self-averaging quantities are those  where the ``off-diagonal" terms make a small contribution compared to the diagonal ones.\footnote{In the periodic orbit situation, the spectral form factor averaged over a long time interval is an example of a self-averaged quantity in a non-disordered situation. Clearly the off-diagonal terms are suppressed here.} We expect that a wormhole calculation of these self-averaging quantities will be quite accurate even in a single realization taken from  the ensemble of theories, and hence in a non-averaged system.    But we emphasize that wormholes will not give the exact answer.  Worse, without understanding the bulk origin of the off-diagonal microscopic effects, there is no clear procedure to systematically  improve the calculation into an arbitrarily accurate one.   

Systems with a direct coupling between subsystems like those  discussed in \cite{Maldacena:2018lmt} are another example where self-averaging behavior occurs.  Here again a wormhole calculation will be  useful, even in a non-averaged theory.  But again, a precise calculation would require the microscopic information.

In the absence of a complete microscopic description how is one to decide whether to trust a wormhole calculation in a non-averaged theory?   An empirical test might be the following.   Pretend the theory is part of an ensemble and compute the variance of the quantity of interest, again using wormholes.  If it is small, the wormhole calculation should be trustworthy.  If it is of order of the signal, beware!

\subsection*{Acknowledgements}
We are grateful to Ahmed Almheiri, Raphael Bousso, Tom Hartman, Jorrit Kruthoff, Juan Maldacena, Don Marolf, Henry Maxfield, Mehrdad Mirbabayi, Onkar Parrikar, Xiaoliang Qi, Phil Saad, and Eva Silverstein for helpful discussions. SS is supported in part by NSF grant PHY-1720397, ZY is supported in part by the Simons Foundation.  GP is supported in part by AFOSR award FA9550-16-1- 0082 and DOE award {DE-SC0019380}.

\appendix

\section{Details on the computation of $Z_n$}\label{app:zn}

In this appendix, we will discuss details about the calculation of $I_n$ and $Z_n$. Two important formulas will be:
\be
8\int_{-\infty}^\infty \mathrm{d}\ell\, K_{2is}(4 e^{-\frac{\ell}{2}})K_{2is'}(4 e^{-\frac{\ell}{2}}) = \frac{\delta(s-s')}{\rho(s)}, \hspace{20pt}
\int_{-\infty}^{\infty}d\ell e^{({1\over 2}-\mu)\ell}K_{2is}(4 e^{-{\ell\over 2}})={|\Gamma(\mu-{1\over 2}+is)|^2\over 2^{2\mu}}.\label{eq:appendixA}
\ee
Using the boundary particle formalism \cite{Kitaev:2018wpr,Yang:2018gdb}, the integral measure in general is the following ($x_i,z_i$ is the location of the insertion of brane):
\be
{1\over 2}\int_{x_1<x_2...x_n} {dx_1dx_2...dx_n dz_1dz_2...dz_n\over z_1^2 z_2^2...z_n^2 \text{V(SL(2,R))}}
\ee
where $z$ is rescaled such that the boundary is located near $z=1$.
The measure of the geodesic lengths can be determined from the above measure by first gauge fixing the SL(2,R) symmetry \cite{Yang:2018gdb} and then a change of variable. 
For computation of $Z_1$, we need to consider $n=2$. By gauge fixing $z_1=z_2=1,x_1=0$, we get ($x_2=e^{\ell\over 2}$):
\be
\begin{split}
	Z_1&=\int_0^{\infty}dx_2 x_2^{1-2\mu}\varphi_{\beta}(2\log x_2)
	= 2e^{S_0}\int_0^{\infty} ds\rho(s)e^{-{\beta s^2\over 2}}\int_{-\infty}^{\infty} d\ell e^{({1\over 2}-\mu)\ell}K_{2is}(4e^{-\ell\over 2})\\
	&=e^{S_0}\int_0^{\infty} ds\rho(s)e^{-\beta {s^2\over 2}}2^{1-2\mu}|\Gamma(\mu-{1\over 2}+is)|^2.
\end{split}
\ee
For computation of $Z_{n>1}$, we can first gauge fix $x_1=0,x_2=1,x_3=2$, the Faddeev-Popov determinant gives:
\be
\int_{2<x_4<..x_n} {dz_1dz_2dz_3dx_4 dz_4...dx_n dz_n\over z_1^2z_2^2z_3^2 z_4^2...z_n^2}
\ee
By defining the regularized length $\ell_{ij}=2\log(|x_i-x_j|)-\log z_i-\log z_j$, we can introduce a new basis $\lbrace \ell_{12},\ell_{13},\ell_{23}...\ell_{1n},\ell_{n-1,n}\rbrace$, which forms a triangulation of a hyperbolic polygon:
\be
 \includegraphics[scale=0.25,valign = c]{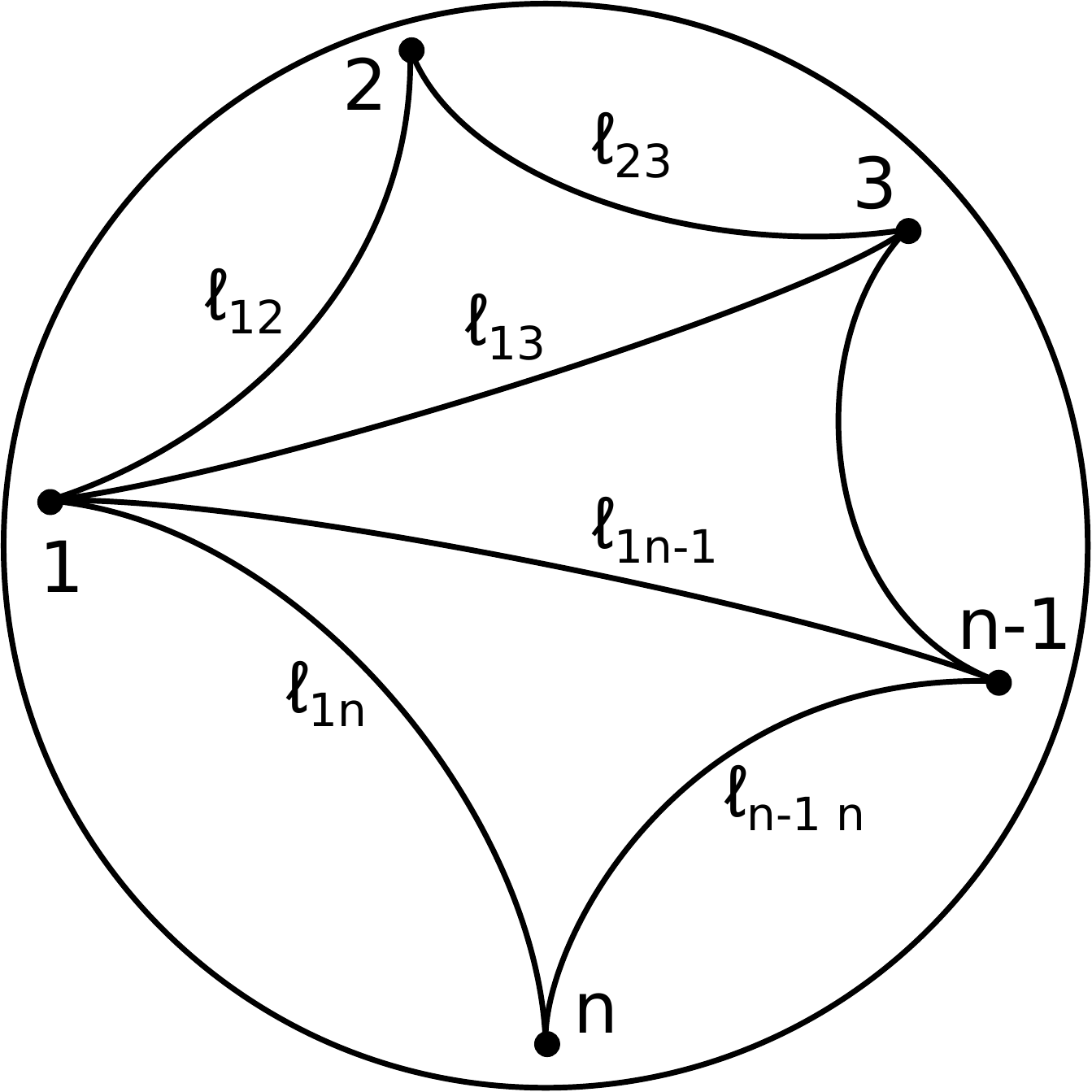}.
 \ee
Its Jacobian matrix can be shown to be a lower triangular matrix:
\be
\bold{J}=\left[
\begin{array}{c|c}
A & 0 \\
\hline
\# & S
\end{array}
\right];
~~~A=\left[\begin{matrix}
 -{1\over z_1}&-{1\over z_2}&0\\
  -{1\over z_1}&0&-{1\over z_3}\\
 0&-{1\over z_2}&-{1\over z_3}
 \end{matrix}\right];~~~
S=\left[\begin{matrix}
{2\over x_{41}}& -{1\over z_4}&0&0&...\\
{2\over x_{43}}&-{1\over z_4}&0&0&...\\
0&0&{2\over x_{51}}&-{1\over z_5}&...\\
0&0&{2\over x_{54}}&-{1\over z_5}&...\\
0&0&0&0&...&	
\end{matrix}
\right],
\ee
whose determinant is:
\be
|\det \bold{J}|=|\det A \det S|=|{2\over z_1 z_2 z_3} {2 x_{31}\over z_4 x_{43}x_{41}}...{2 x_{n-1,1}\over z_n x_{n,n-1}x_{n1}}|.
\ee
This gives the measure:
\be
{1\over 2^{n-1}}\int_{-\infty}^{\infty} d\ell_{12}d\ell_{13}d\ell_{23}... e^{\ell_{12}/2+\ell_{23}/2+\ell_{34}/2+...+\ell_{n1}/2}
\ee
Notice that only the exterior boundaries of the hyperbolic polygon has nontrivial measure.
To get the expression of $I_{n}$, we can glue multiple copies of $I_3$ along the inner boundaries. Using the formula of $I_3$ in \cite{Yang:2018gdb}, we get the expression of $I_n$:
\be
I_n=(\ell_{12},\ell_{23},...\ell_{n1})=2^n\int_0^{\infty}ds\rho(s)K_{2is}(4 e^{-\ell_{12}\over 2})...K_{2is}(4 e^{-\ell_{n1}\over 2}),
\ee
with integral measure (\ref{finalZn}):
\be
\int_{-\infty}^{\infty} d\ell_{12}d\ell_{23}...d\ell_{n1} e^{\ell_{12}/2+\ell_{23}/2+\ell_{34}/2+...+\ell_{n1}/2}.
\ee

\section{General two dimensional dilaton gravity}\label{app:dilatongravity}
Near-extremal black holes are universally described by Jackiw-Teitelboim gravity with negative cosmological constant, and therefore they also have replica wormhole configurations where the area of the transverse direction is the dilaton field (together with $S_0$). 
It is interesting to ask whether replica wormholes exist for  more general classes of black holes, especially those far from extremality.

Assuming spherical symmetry, general gravitational systems can be reduced to  two dimensional dilaton gravity \cite{Strominger:1994tn}, so we are led to  consider  JT gravity with a more general dilaton potential in our simple model:
\be
I=-{S_0\over 2\pi}\left[{1\over 2}\int_{M}\sqrt{g}R+\int_{\partial M}\sqrt{h}K\right]-\left[{1\over 2}\int_M \phi R-U(\phi)+\int_{\partial M}\sqrt{h}\phi K\right ]+\mu \int_{brane}ds '
\ee
Again we assume the brane follows a geodesic.
And again we  entangle the brane states with the radiation,  and consider the behavior of the entropy of the radiation when we  vary $k$. Note that flat space black holes (Rindler space) can be described in this framework using the special case of a constant dilaton potential.

While in general understanding  euclidean wormholes requires a good understanding of  off-shell geometries, in the planar limit only knowledge of the path integral on the disk topology $Z_n$ is required.   The derivation of the Schwinger-Dyson equation still applies since it only depends on the topological action and $k$.  Moreover in the heavy brane case, the branes become local near the boundary and $Z_n$ becomes proportional to the disk partition function of boundary length $n\beta$.\footnote{We consider the case where there exists an asymptotic boundary.} Since the effects of the heavy branes are local, they cancel out in the ratio of ${Z_n\over Z_1^n}$ and the result reduces to just the ratio of the disk partition functions
\be
{Z_n\over Z_1^n}={Z(n\beta)\over Z(\beta)^n} .
\ee

Writing the partition function in terms of the energy density $Z(\beta)=\int dE \tilde \rho(E)e^{-\beta E}$, we can again sum over $n$ in the Schwinger-Dyson equation (\ref{eqn:SD-eqn}) and get the resolvent equation:
\be
\lambda R=k+\int dE \tilde \rho(E){\omega (E) R\over k-\omega(E) R};~~~\omega(E)={e^{-\beta E}\over Z(\beta)}.
\ee

In general the exact form of $\tilde\rho(E)$ is not known and one can only solve the equation in the classical limit.
Using the classical thermodynamic relation for general dilaton gravity (Appendix E.3 in \cite{Maldacena:2019cbz}):
\be
E=-W(\phi_h);~~~S=S_0+4\pi \phi_h
\ee
where $W(\phi)=\int^{\phi}d\phi' U(\phi)$ is called the prepotential, we can write down the general semiclassical resolvent equation:
\be\label{eqn: resolvent general dilaton gravity}
\lambda R=k+\int d\phi_h e^{S_0+4\pi \phi_h}{e^{\beta W(\phi_h)} R\over kZ(\beta)- e^{\beta W(\phi_h)}R};~~~Z(\beta)=\int d\phi_h e^{S_0+4\pi \phi_h+\beta W(\phi_h)}.
\ee
In the microcanonical  case, we will still get Page's result.
In the canonical ensemble case, one needs to analyze the equation based on $W(\phi_h)$ and we expect that most of our analysis will still apply.

\section{General entanglement-wedge reconstruction using Petz}\label{app:ew}

In this appendix, we will indicate how the argument in sections \ref{sec:gravPetz} and \ref{sec:ftpetz} extends to the case of general entanglement-wedge reconstruction. We start with a code space of bulk field theory excitations $|a\rangle_{\sf{FT}}$ around some particular background, and an operator $\mathcal{O}_{\sf FT}$ acting in this space. $|\Psi_a\rangle$ is the boundary CFT state corresponding to $|a\rangle$, and $\mathcal{O}$ is the CFT operator corresponding to $\mathcal{O}_{\sf FT}$.

If $A$ is a subregion of the boundary theory, then the Petz map gives a guess for the reconstruction of $\mathcal{O}$ on region $A$:
\be
\mathcal{O}_A = \sigma_A^{-1/2}\tr_{\overline A}(\mathcal{O})\sigma_A^{-1/2}, \hspace{20pt}\sigma_A = \sum_{a = 1}^{d_{\sf{code}}}\tr_{\overline A}|\Psi_a\rangle\langle \Psi_a|.
\ee
Defining a replica version as in (\ref{sigmaR}), we have
\begin{align}
\langle \Psi_a|\mathcal{O}_A^{(n)}|\Psi_b\rangle &= \sum_{a'b'=1}^{d_{\sf code}}\tr_A\left[{\bf M}(a,b)\widehat{\bf M}^n {\bf M}(b',a')\widehat{\bf M}^n\right]\mathcal{O}_{a'b'}\label{ywu}
\end{align}
where 
\be
{\bf M}(a,b) = \tr_{\overline A}|\Psi_a\rangle\langle \Psi_b|, \hspace{20pt} \widehat{\bf M} = \sum_{a = 1}^{d_{\sf code}}{\bf M}(a,a).
\ee
In the bulk dual, the RHS of (\ref{ywu}) is a gravitational path integral with operator insertions to prepare the different state $|\Psi_a\rangle$ in the code subspace. At order $G_N^{-1}$, the answer doesn't depend on these operator insertions, and it reduces to the gravitational path integral for the Renyi $(2n+2)$-entropy, which we will refer to as $Z^{\text{grav}}_{2n+2}$.

At order $G_N^0$, we have a field theory computation on this fixed background, which we can write
\begin{align}
\langle \Psi_a|\mathcal{O}_A^{(n)}|\Psi_b\rangle &=\frac{Z^{\text{grav}}_{2n+2}}{(Z^{\text{grav}}_1)^{2n+2}}\sum_{a'b'=1}^{d_{\sf code}}\tr_{\mathcal{A}_n}\left[{M}(a,b;\theta)\widehat{ M}(\theta)^n { M}(b',a';\theta)\widehat{ M}(\theta)^n\right]\mathcal{O}_{a'b'} + O(G_N).
\end{align}
This formula requires some explanation. The replica-symmetric geometry $\mathcal{M}_{2n+2}$ that dominates $Z^{\text{grav}}_{2n+2}$ has a codimension-two surface $\Sigma$ that is fixed by the cyclic replica symmetry. We divide $\mathcal{M}_{2n+2}$ into $2n+2$ equal pieces by cutting along codimension-one surfaces $\mathcal{A}_n$ that connect the $2n+2$ copies of region $A$ on the boundary to $\Sigma$. Each surface $\mathcal{A}_n$ can be understood as a backreacted Renyi version of the Cauchy slice of the entanglement wedge of $A$. They intersect $\Sigma$ with equally spaced angles $\theta = 2\pi/(2n+2)$. The operator $M(a,b;\theta)$ is defined as the path integral on the geometry between two of these cuts, with boundary conditions that include the operator insertions for the states $|\Psi_a\rangle$ and $|\Psi_b\rangle$. In the limit where we take the total number of replicas to one, we find the simple formula
\be
M(a,b;2\pi) = \tr_{\overline{\mathcal{A}}}|a\rangle\langle b|_{\sf{FT}}.
\ee

As in section \ref{sec:ftpetz}, one can now take the $n\to -1/2$ limit, and find
\be
\langle \Psi_a|\mathcal{O}_A|\Psi_b\rangle = \langle a|\mathcal{O}_{\mathcal{A}}|b\rangle_{\sf{FT}} + O(G_N)\label{RHSOF}
\ee
where the RHS is an auxiliary Petz map computation, defined purely in the bulk field theory. $\mathcal{A}$ is the Cauchy slice of the entanglement wedge of $A$, and 
\be
\mathcal{O}_{\mathcal{A}} = \sigma_{\mathcal{A}}^{-1/2} \tr_{\overline{\mathcal{A}}}(\mathcal{O}_{\sf{FT}})\sigma_{\mathcal{A}}^{-1/2}, \hspace{20pt} \sigma_{\mathcal{A}} = \sum_{a = 1}^{d_{\sf code}}\tr_{\overline{\mathcal{A}}}|a\rangle\langle a|_{\sf{FT}}.
\ee
is the Petz map for the channel corresponding to erasure of the complement of the entanglement wedge. In particular, for the case where $\mathcal{O}_{\sf FT}$ acts within the entanglement wedge $\mathcal{A}$ to a good approximation, then $\mathcal{O}_{\mathcal{A}} = \mathcal{O}_{\sf FT}$, and reconstruction succeeds.

We also note that there is a more general version of the Petz reconstruction map, defined using a fixed, but arbitrary, state $\rho$ that is not necessarily maximally mixed. In this case, we define
\begin{align}
\mathcal{O}_A^{(\rho)} =\rho_A^{-1/2}\tr_{\overline A}(\rho^{1/2} \mathcal{O} \rho^{1/2})\rho_A^{-1/2}.
\end{align}
It is easy to see, by analogous arguments to those above, that this reconstruction reduces in the bulk to the field theory Petz reconstruction
\begin{align}
\mathcal{O}_{\mathcal{A}}^{(\rho)} = \rho_{\mathcal{A}}^{-1/2} \tr_{\overline{\mathcal{A}}}(\rho^{1/2}\mathcal{O}_{\sf{FT}}\rho^{1/2})\rho_{\mathcal{A}}^{-1/2}.
\end{align}
An advantage of this more general construction is that it can be made well-defined even for infinite-dimensional code spaces, where the maximally mixed state does not exist. However, one has to be somewhat careful here: if the code space includes degrees of freedom outside the entanglement wedge (i.e. we have a \emph{subsystem} rather than a \emph{subspace} code), the field theory Petz map, constructed using an arbitrary, non-maximally mixed state $\rho$, will not necessarily recover the original operator. Even if we use the twirled Petz map, the reconstruction is only guaranteed to work if the code space state $\rho$ has no entanglement between the inside and outside of the entanglement wedge \cite{Cotler:2017erl}.

In an infinite-dimensional code space (such as the entire field theory Hilbert space), such product states do not necessarily exist. However, by using a thermal state at very high temperature, we can make the entanglement arbitrarily short range. We should then expect that the field theory Petz map will recover the original operator with high accuracy so long as the inverse temperature $\beta$ is much smaller than the distance from the original operator to the edge of the entanglement wedge. This can be verified in simple cases where the action of the field theory modular flow (and hence the field theory Petz map) is known explicitly.

Finally, we emphasize that the definition of the Petz map reconstructions relies on being able to create arbitrary states in the code space using gravitational path integrals. For most situations of interest, this is not a problem. For example, if our code space is the state of a diary that was thrown into a black hole, we can easily construct arbitrary states in the code space simply by changing the state of the diary, before it was thrown into the black hole.

If we want to reconstruct the interior partners of Hawking quanta -- to understand, for example, whether there is a firewall \cite{Almheiri:2012rt} -- the situation is somewhat more complicated. We cannot directly manipulate the interior modes because they become trans-Planckian when evolved back in time. Instead, they are always produced, together with the Hawking quanta themselves, in a fixed, entangled state $|\psi_0\rangle$. However, we can use the gravitational path integral to manipulate the state of the Hawking quanta. Because the state $|\psi_0\rangle$ has maximal rank, using these manipulations, we can produce an overcomplete basis of states for the Hawking quanta and interior partners. 

By taking linear superpositions of such path integrals, we can therefore construct arbitrary states in code spaces that include interior partner modes. We can use these path integrals to construct Petz map reconstructions of interior modes, after the Page time, that act only on the Hawking radiation. 

This construction assumes that the interior modes were initially in the state $|\psi_0\rangle$, i.e. that there wasn't a firewall. Indeed, it has always been the case that gravitational calculations implied the absense of a firewall. The new result that one can see using the Petz map is that gravitational calculations also imply that the interior modes can be reconstructed on the Hawking radiation, i.e. ER=EPR \cite{Bousso:2012as,Nomura:2012sw,Verlinde:2012cy,Papadodimas:2012aq,Maldacena:2013xja}.

\section{A ensemble boundary dual of the simple model}\label{app:ensemble}
In this appendix, we show that the simple model, with pure JT gravity plus EOW branes, is dual to a boundary ensemble of theories, just like pure JT gravity \cite{Saad:2019lba}. This ensemble of theories is defined by a randomly chosen Hamiltonian $H$, and a set of randomly chosen special states $|i \rangle$ (the brane states). The dual bulk theory is valid to all order in $e^{-S_0}$. It includes nontrivial bulk topologies, including topologies with handles (suppressed by powers of $e^{-S_0}$). However it doesn't include any contributions of EOW brane loops. All brane world lines have to begin and end on the boundary. This provides some justification for the fact that we ignored the possibility of end-of-the-world brane loops in all our bulk gravity calculations in this paper.

We now give a precise definition of the ensemble of boundary theories. First, the Hamiltonian $H$ is chosen from the usual JT gravity ensemble of Hamiltonians, as in \cite{Saad:2019lba}. Let the eigenstates of this Hamiltonian be labelled $| E_a \rangle$. Then the brane states are chosen to be
\begin{align}\label{ensState}
|\psi_i(\beta) \rangle = \sum_{a} 2^{1/2- \mu} \Gamma(\mu-{1\over 2}+i\sqrt{2 E_a}) e^{-\beta E_a/2}C_{i,a} | E_a \rangle
\end{align}
where the coefficients $C_{i,a}$ are i.i.d. complex Gaussian random variables.

Let us see why this works. Our aim is to show that expressions of the form
\begin{align}
\mathbb{E}_{H,\{C\}}\left(\prod_{m=1}^p \langle \psi_{i_m}(\beta_m) | \psi_{j_m}(\beta_m) \rangle \prod_{n=1}^q \Tr(e^{- \tilde{\beta}_n H})\right),\label{exp1}
\end{align}
where the  expectation is over the ensemble of Hamiltonians and states (\ref{ensState}), matches a bulk computation in JT gravity with the following boundary conditions. We have $p$ asymptotic boundaries that are intervals of renormalized lengths $\{\beta_m\}$ bounded by EOW branes, and we have $q$ standard $S^1$ boundaries with renormalized lengths $\{\tilde{\beta}_n\}$.

We now simply integrate out the Gaussian random variables $C_{i,a}$. We find that (\ref{exp1}) equals
\begin{align} \label{eq:branepartition}
\sum_\pi \left[\prod_{m=1}^p \delta_{i_{\pi(m)},j_m} \mathbb{E}_H \left(\prod_{\gamma \in c(\pi)} \Tr\left( \prod_{m \in \gamma} \left[e^{- \beta_m H} \sum_a \frac{|\Gamma(\mu-{1\over 2}+i\sqrt{2 E_a})|^2}{2^{2\mu-1}} |E_a \rangle\langle E_a|\right]\right) \prod_{n=1}^q \Tr(e^{- \tilde{\beta}_n H}) \right)\right]
\end{align}
Here, we are summing over permutations $\pi$ on $p$ elements that take $m$ to $\pi(m)$ and the subsets $\gamma \subseteq \{ 1, 2, \dots, p\}$ are elements of the set $c(\pi)$ of cycles of the permutation $\pi$. This formula no longer involves the brane states $|i\rangle$. We can therefore hope to make contact with the arguments from \cite{Saad:2019lba} for evaluating products of partition functions, in this ensemble of Hamiltonians, using bulk JT gravity path integrals. 

Of course, \eqref{eq:branepartition} isn't quite in the right form that would allow us to do this, since it includes insertions of $|\Gamma(\mu-{1\over 2}+i\sqrt{2 E_a})|^2 |E_a \rangle\langle E_a|$. We therefore use the inverse Laplace transform $f (\beta') =\text{IL}[2^{1 - 2\mu} |\Gamma(\mu-{1\over 2}+i\sqrt{2 E_a})|^2](\beta')$ to write
\begin{align}
\sum_a 2^{1-2 \mu} |\Gamma(\mu-{1\over 2}+i\sqrt{2 E_a})|^2 |E_a \rangle\langle E_a| = \int d \beta'  \,f(\beta') \sum_a e^{-\beta' E_a}   |E_a \rangle\langle E_a|.
\end{align}
This turns the evaluation of \eqref{eq:branepartition}  into an integral over products of partition functions, with `inverse temperatures' that depend on the auxiliary variables $\beta'$. Explicitly, we have
\begin{align} \label{eq:branepartition2}
\sum_\pi \left[\int \prod_{m} d\beta_m' \prod_{m}[ f(\beta'_m) \delta_{i_{\pi(m)},j_m}]  \mathbb{E}_H \left(\prod_{\gamma \in c(\pi)} \Tr\left(e^{- \sum_{m \in \gamma} (\beta_m + \beta_m') H}\right) \prod_{n=1}^q \Tr(e^{- \tilde{\beta}_n H}) \right)\right].
\end{align}
The expectation values in this formula can be evaluated using bulk JT gravity by considering all topologies (including topologies with handles) of JT gravity with asymptotic boundaries of renormalised length $\sum_{m \in \gamma} (\beta_m + \beta_m')$ for each cycle $\gamma$ in the permutation $\pi$, as well as additional asymptotic boundaries of renormalised length $\tilde{\beta}_n$ for each $n$.

Once we include the sum over permutations $\pi$, this is exactly the set of topologies that appear in the simple model. The different permutations correspond to the different ways of pairing up brane start and end points. The factors of $\delta_{i_{\pi(m)},j_m}$ ensure that we only get nonzero contributions when the state of each brane is the same at both ends; it tells us that the branes have no dynamics, as desired. 

We are still not quite done however. In this formula, we consider geometry of fixed asymptotic length, and integrate over different lengths. In contrast, in JT gravity with branes, the topologies are partially bounded by geodesics, which contribute the EOW brane action $\mu l$. Fortunately, as argued in \cite{Yang:2018gdb,Saad:2019pqd}, any boundary segment of length $\beta'_m$ can be replaced in the JT gravity calculations by the insertion of the Wheeler-de Witt wavefunction at the homologous, non-selfintersecting bulk geodesic $\chi$. In the $l$ basis, this Wheeler-de Witt wavefunction is given by
\begin{align}
\psi_{\beta_m'} (l) = 4 e^{-l/2} \int_0^\infty ds \rho(s) e^{- \beta' s^2/2} K_{2 i s} (4e^{-l/2}),
\end{align}
as in \eqref{intrepphi}. When we rewrite the JT gravity calculation in this way, the only dependence of the expectation value in  \eqref{eq:branepartition2} on the auxiliary variables $\beta'_m$ comes in the choice of wavefunction that we insert at the geodesic.

We can now do the integrals over $\beta'_m$ explicitly by simply taking a superposition over these Wheeler-de Witt wavefunctions. After doing this superposition, we find that the full wavefunction on each geodesic is given by
\begin{align}
\psi (l) = 2^{3 - 2 \mu} e^{-l/2} \int_0^\infty ds \rho(s) |\Gamma(\mu-{1\over 2}+is)|^2 K_{2 is } (4 e^{-l/2}).
\end{align}
Using \eqref{eq:appendixA} and Eqn. 25 of \cite{yakubovich2013use},
\begin{align}
8 \int_0^\infty ds\rho(s) K_{2is}(4e^{-l/2}) K_{2is}(4e^{-l'/2}) = \delta(l - l'),
\end{align}
 which is the orthogonality condition for the $l$-basis of wavefunctions, we find
\begin{align}
\psi (l) = e^{- \mu l }.
\end{align}
This is just the insertion of a brane wavefunction on the geodesic $\chi$. We have therefore recovered the bulk theory of JT gravity with EOW branes.

We can now use the random state formula (\ref{ensState}) to write a random matrix formula for the density matrix $\rho_{\sf{R}}$ in the state (\ref{StateWholeSystem}):
\be
\rho_{\sf{R}} = C F(H)C^\dagger, \hspace{20pt} F(x) = |\Gamma(\mu-\frac{1}{2}+i\sqrt{2x})|^2e^{-\beta x}.
\ee
The matrix $H$ is drawn from the double-scaled matrix integral dual to ordinary JT gravity. The matrix $C$ is a rectangular $k\times \infty$ complex matrix with Gaussian random entries, rescaled so that the density matrix is correctly normalized.

\section{Fixed Area States and Random Tensor Networks}
In this appendix, we show that the non-perturbative corrections to the Ryu-Takayanagi formula from additional extremal surfaces are identical in a) ``fixed-area'' states in AdS/CFT \cite{Akers:2018fow,Dong:2018seb} and b) random tensor networks \cite{Hayden:2016cfa}. This is consistent with the intuition from  \cite{Akers:2018fow,Dong:2018seb} that fixed-area states are analogous to tensor network states, since both have flat entanglement spectra at leading order. However, our calculations suggest that the connection to \emph{random} tensor networks specifically (as opposed, for example, to perfect tensor networks \cite{Pastawski:2015qua}) goes much deeper than might previously suspected, with non-perturbative instanton corrections that are the same in both models. {\bf Note: } this appendix was added in v2; similar results were derived independently in upcoming work by Akers, Faulkner, Lin and Rath \cite{AFLR}.

We first calculate the von Neumann entropy, including non-perturbative corrections, for the simplest possible random tensor network: a bipartite pure state with subsystem Hilbert space dimensions $d_B$ and $d_{\bar B}$. (We assume for notational convenience that $d_B < d_{\bar B}$.) This is, of course, just the famous Page calculation \cite{Page:1993df}. Let us review how the answer can be found using the replica trick.

We first evaluate $\langle \tr (\rho_B^n) \rangle$, where the expectation is over the Haar measure for the random state. Using the standard result
\begin{align}
 \left\langle  |\psi \rangle\langle \psi|^{\otimes n}\right \rangle \propto \sum_\pi \pi_B \pi_{\bar B},
\end{align}
where the sum is over permutations $\pi$ on the $n$ copies of the system and $\pi_B$ and $\pi_{\bar B}$ are operators that permute the $n$ copies of their respective subsystem, we find
\begin{align}\label{eq:permeq}
\left\langle \tr (\rho_B^n)\right \rangle =\left \langle \frac{\tr (\rho_B^n)}{\langle\psi | \psi\rangle^n} \right\rangle = \frac{\sum_\pi \tr(\tau_B \,\pi_B) \tr(\pi_{\bar B})}{\sum_\pi \tr(\pi_B) \tr(\pi_{\bar B})} = \frac{\sum_\pi d_B^{\,C(\tau \circ \pi)} d_{\bar B}^{\,C(\pi)}}{\sum_\pi d_B^{\,C(\pi)} d_{\bar B}^{\,C(\pi)}},
\end{align}
where $\tau$ is any fixed cyclic permutation and $C(\Pi)$ is the number of cycles in the permutation $\Pi$. Here we focus on  the limit where $d_B$ and $d_{\bar B}$ are both large but their ratio is arbitrary. In this limit  \ref{eq:permeq} simplifies to
\begin{align} \label{eq:trrhoAn}
\left\langle \tr (\rho_B^n)\right \rangle = \sum_{\pi \in G} d_B^{\,C(\tau \circ \pi) - n} d_{\bar B}^{\,C(\pi) - n},
\end{align}
where the sum is now over permutations in the set $G$ of permutations such that $C(\pi) + C(\tau \circ \pi)$ takes its maximal value, which is $n +1$. These are the permutations that lie on a geodesic (shortest paths in permutation space, where each step is a transposition) between the identity permutation and the cyclic permutation $\tau$. Such permutations are in one-to-one correspondence with the set of non-crossing partitions. We note that, in the limit of large $d_B, d_{\bar{B}}$, fluctuations are suppressed and so the expectation over states is unnecessary. 

The number of non-crossing permutations such that $C(\tau\circ\pi) = k$ is equal to the Narayana number, $N(n,k)$. Substituting in the explicit expression for $N(n,k)$, we have
\be
\left\langle \tr (\rho_B^n)\right \rangle = \frac{1}{d_B^{n-1}}\sum_{k = 1}^n N(n,k) \left(\frac{d_B}{d_{\bar B}}\right)^{k-1} = \frac{1}{d_B^{n-1}}\sum_{k = 1}^n \frac{1}{n}\binom{n}{k}\binom{n}{k-1} \left(\frac{d_B}{d_{\bar B}}\right)^{k-1}.
\ee
To find the von Neumann entropy, we would like to analytically continue this expression in $n$ and then expand the answer near $n = 1$. This looks awkward, because the range of the sum itself depends on $n$. However, we can use the fact that the binomial coefficients vanish for all $k>n$ to extend the sum from $k = 1$ to infinity. The expression is then easy to continue in $n$, and we find
\begin{align}
\tr(\rho_B^n) = \frac{1}{d_B^{n-1}} \left[ 1 + \frac{n(n-1)}{2}\frac{d_B}{d_{\bar B}} + O\left((n-1)^2 \frac{d_B^2}{d_{\bar B}^2}\right)\right].
\end{align}
The first term comes from the identity permutation ($k = 1$) and the second comes from the permutations consisting of a single transposition ($k = 2$), which give the leading correction in the limit $d_{\bar B} \gg d_B$. We therefore find
\begin{align}
S = - \tr(\rho_B \log \rho_B) = - \partial_n \tr(\rho_B^n)|_{n=1} = \log d_B - \frac{d_B}{2 d_{\bar B}}.
\end{align} 
Naively, one would expect additional subleading corrections from permutations that are suppressed by higher powers of $d_B/d_{\bar B}$, as there were for $\tr(\rho_B^n)$. However, one can check that for $k>2$, the continuation of the Narayana number $N(n,k)$ near $n =1$ is proportional $(n-1)^2$, which does not contribute to the von Neumann entropy. Morally speaking, the von Neumann entropy is ``one-instanton exact".

We note that much of the above discussion extends to more complicated random tensor networks. For example, when evaluating $\tr(\rho_B^2)$, one ends up with the partition function of a classical Ising model. However, this will not be necessary for our present purposes. See \cite{Hayden:2016cfa} for more details.

Instead, we shall move on to discussing fixed-area states in quantum gravity. For concreteness, we shall mostly focus on the example of two disjoint boundary regions  (which we collectively denote by $B$) in AdS$_3$, although the story is completely general. In this case, there are two extremal surfaces, one homotopic to the region $B$ (with area $A_1$) and the other homotopic to the complementary region $\bar B$ (with area $A_2$). 

We consider states where the area of both surfaces has been measured, and so saddle points of the gravitational path integral can have conical singularities at the extremal surfaces, so long as the deficit angle is constant everywhere on the surface. As discussed in \cite{Akers:2018fow,Dong:2018seb} and section \ref{sec:petzlite}, this means that the replicated geometries used to calculate $\tr(\rho_A^n)$ are very simple: they are just $n$ copies of the original unbackreacted geometry, with different sheets glued together around the extremal surfaces.
\begin{figure}[t]
\begin{center}
\includegraphics[width = .4\textwidth]{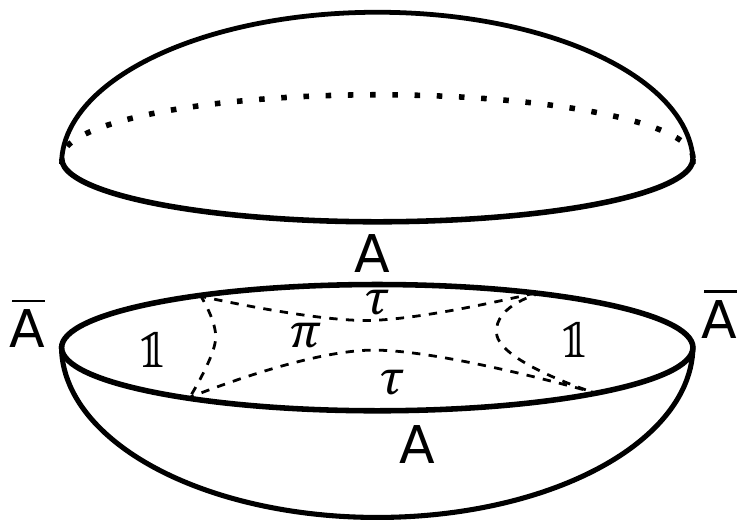}
\caption{{\small An example of a replicated geometry for two intervals in AdS, cut open.}}\label{fig:twointerval}
\end{center}
\end{figure}

More specifically, as shown in figure \ref{fig:twointerval}, the boundary conditions used to compute $\tr(\rho_B^n)$ mean that the bottom half (the 'ket') of the bulk geometry near region $B$ needs to be glued to the top half of the bulk geometry on the neighbouring sheet. Passing upwards through this region applies a cyclic permutation $\tau$ to the different sheets. In contrast, when upwards through the bulk geometry near region $\bar B$, one must remain on the same sheet. What about the central region, between the two extremal surfaces? This is not fixed by the boundary conditions, and so there should be saddles associated to each possible permutation $\pi$ of the sheets.

What is the action of each of these saddle points? We first note that, when the state is correctly normalised (dividing through by $\tr(\rho_B)^n$) the contribution to the gravitational action from everything except the conical singularities cancels between numerator and dominator because the geometry is unbackreacted. The conical singularites themselves give a contribution to the action of $(\hat{\phi} - 2 \pi) A/8 \pi G_N$, where $\hat{\phi}$ is the angle around the conical singularity. If the extremal surfaces had conical singularities with angles $\phi_1$ and $\phi_2$ in the unreplicated geometry, the contribution to the action in the replicated geometry is
$$
\left[n \phi_1 - 2 \pi C(\tau \circ \pi)\right] \frac{A_1}{8 \pi G_N} + \left[n \phi_2 - 2 \pi C(\pi)\right] \frac{A_2}{8 \pi G_N}.
$$

After normalisation, the dependence on $\phi_1$ and $\phi_2$ vanishes and we are left with the final result
\begin{align}
\tr(\rho_A^n) = \sum_\pi \exp\left(\left[C(\tau \circ \pi) - n\right] \frac{A_1}{4 G_N} + \left[C(\pi) - n\right] \frac{A_2}{4 G_N}\right).
\end{align}
Note that, since $A_1$ and $A_2$ are individually divergent, all permutations $\pi$ that do not minimise $(C(\tau \circ \pi) + C(\pi)$ are infinitely suppressed, even at finite $G_N$.

This is exactly the result that we found for the bipartite random pure state, except with $d_B$ and $d_{\bar B}$ replaced by $\exp(A_1/4G_N)$ and $\exp(A_2/4G_N)$ respectively. It therefore follows that the non-perturbative corrections to the von Neumann entropy for the fixed-area state are identical to the corrections that we found above for the bipartite random pure state.
\begin{figure}[t]
\begin{center}
\includegraphics[width = .3\textwidth]{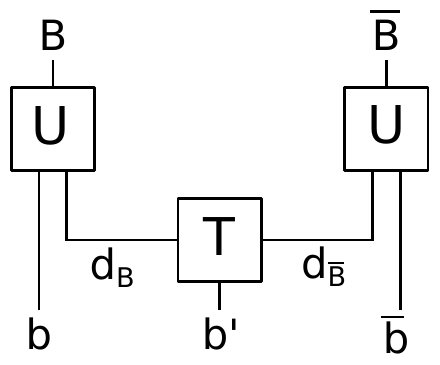}
\caption{{\small A simple random tensor network, with a single random tensor $|T\rangle$ and three bulk legs. One bulk leg flows into $| T\rangle$, while the other two are mapped directly into the boundary Hilbert spaces $B$ and $\bar B$.}}\label{fig:bulklegsTN}
\end{center}
\end{figure}

Finally, we comment briefly on what happens when bulk degrees of freedom are added in each case. For a much more careful analysis of perturbative correction in fixed-area states see \cite{Dong:2019piw}. If we treat the gravitational path integral semiclassically, the Euclidean field theory path integral in the bottom-half of the geometry prepares a particular state $|\psi\rangle_{b b' \bar{b}}$ of the bulk quantum fields. Here $b$ is the bulk region near $B$, $\bar b$ is the bulk region near $\bar B$ and $b'$ is the central region between the extremal surfaces. The field theory partition function on the full replicated geometry then evaluates to
\begin{align} \label{eq:bulklegs}
\langle\psi|^{\otimes n}\, \tau_b\, \pi_{b'}\, |\psi\rangle^{\otimes n},
\end{align}
where the operators $\tau_b$ and $\pi_{b'}$ permute the subsystems in regions $b$ and $b'$ respectively.

What about if we add bulk legs to the random tensor network, as shown in figure \ref{fig:bulklegsTN}? The answer is exactly the same thing. If we input some bulk state $|\psi\rangle_{b b' \bar b}$ and then evaluate $\tr(\rho_B^n)$ by replacing the random tripartite states $| T \rangle$ by a sum over permutations $\pi$, as in \eqref{eq:trrhoAn}, we find that the term associated to each permutation has an additional factor that is exactly \eqref{eq:bulklegs}. The analogy between fixed-area states and random tensor networks continues to hold.

\section{The Page transition in the simple model}\label{app:pagephases}
The aim of this appendix is to comprehensively analyse the full Page transition in the simple model. Despite the simplicity of this model, the structure of the transition turns out to be somewhat complicated. Different features have transitions at different times, creating a number of distinct ``sub-phases''. 

As a simple example, the transition happens earlier for larger $n$ R\'{e}nyi entropies, creating an infinite number of different Page transitions. We will ignore the R\'{e}nyi entropies, and instead focus only on changes in either
\begin{enumerate}[(a)]
\item qualitative features of the shape of the entanglement spectrum, or 
\item in the von Neumann entropy. 
\end{enumerate}
Nevertheless, we still find seven distinct phases within the transition, as the entanglement spectrum slowly switches from a flat spectrum, with all eigenvalues equal to $1/k$, to a thermal spectrum, with entropy given by the Bekenstein-Hawking entropy of the black hole. Because the details of the calculations in this appendix are fairly technical, we first summarise the main conclusions about each phase, and then proceed to analysing each phase in detail. 

\subsubsection*{Assumptions and Notation}
In this appendix, we shall always assume that we are in the semiclassical limit $\beta \ll 1$.\footnote{Recall that, in our units, this corresponds to the limit $G_N \to 0$.} At certain points, we shall also assume, for convenience, that the brane mass $\mu \gg 1/\beta$.

We first formalise some notation. Recall that
\begin{align}
Z_n = e^{S_0} \int ds \rho(s) y (s)^n.
\end{align}
We use notation where the saddle point value of $s$ for $Z_n$ is denoted by $s^{(n)}$.  In the limit $\mu \gg 1/\beta$, we have
\begin{align}
\rho(s)\, y(s)^n \sim \frac{s}{2 \pi^2}\, y(0)^n\, e^{2 \pi s - n\beta s^2/2}.
\end{align}
Hence
\begin{align}
s^{(n)} = \frac{2 \pi}{n\beta} + O(1).
\end{align}
There is one more saddle point that will be relevant for our calculations. This is the saddle point for
\begin{align}
e^{S_0} \int ds \rho(s)^2 y (s),
\end{align}
which we denote by $\sprime$. For $\mu \gg 1/\beta$, $\sprime = 4 \pi /\beta$. Note that we have $\sprime > s^{(1)}$ and $s^{(n)} > s^{(m)}$ for $n < m$.

For a particular choice $k$, there are two other values of $s$ which will be important. The first, denoted by $\smax$, is the value of $s$ corresponding to the $k$th  largest thermal eigenstate, as defined in \eqref{eq:defskmaintext}. In the semiclassical limit, this is defined by 
\begin{align} \label{eq:defsk}
\frac{e^{S_0} \rho(\smax)}{2 \pi} = k.
\end{align} 
The second, which we denote by $\stilde$ labels the thermal eigenstates with eigenvalue $1/k$. This is defined by 
\begin{align}\label{eq:defstildek}
\frac{y(\stilde)}{ Z_1} = \frac{1}{k}.
\end{align}
Note that the normalisation of the thermal state ensures that we always have $\stilde < \smax$. 

Finally, an important value of $k$, which we shall denote by $k_{3 \to 4}$, is defined to be the smallest value of $k$ for which
\begin{align} \label{eq:k3to4}
e^{S_0} \rho(\tilde{s}_{k_{3 \to 4}}) = \frac{k_{3 \to 4}^3 \, y(s_{k_{3\to 4}})^2}{Z_1^2}.
\end{align}
For $\mu \gg 1/\beta$, we have $\stilde = \sqrt{2 \pi (2 \smax/\beta -  2 \pi/\beta^2)} + o(1/\beta)$ and hence $s_{k_{3\to 4}} = (4 - 2\sqrt{2}) \pi/ \beta + o(1/\beta)$. More generally, we expect that $s^{(2)} < s_{k_{3\to 4}} < s^{(1)}$. Note that for $k \ll k_{3 \to 4}$, the left hand side of \eqref{eq:k3to4} is much smaller than the right hand side.

\subsection*{Summary of the phase structure}

At small $k$, specifically $k \ll Z_2/y(0)^2$ (phase I), the shape of the entanglement spectrum is a narrow semicircle distribution, of width $4 \sqrt{ Z_2 / k Z_1^2},$ 
centred on $1/k$. The von Neumann entropy is approximately equal to $\log k$, with a small correction 
\begin{align}
\delta S = - k Z_2 /2 Z_1^2
\end{align}
 that comes from the finite width of the semicircle. This is exponentially small with respect to $(S_{BH} - \log k)$.

For $Z_2/y^2(0) \ll k \ll e^{S_0} \rho( s^{(2)})$ (phase II), the entanglement spectrum is dominated by the same narrow semicircle distribution. The width of this semicircle continues to give the leading correction to the von Neumann entropy. However, there is now a small tail of much larger eigenvalues, that emerge out of the semicircle as $k$ increases. This tail looks like the thermal spectrum, except all the eigenvalues are increase by a constant shift of $\delta \lambda = 1/k$.

When $e^{S_0} \rho( s^{(2)}) \ll k \ll k_{3 \to 4}$ (phase III), the original semicircle disappears. The spectrum looks like a thermal spectrum, shifted by $\delta \lambda = 1/k$, which is then cutoff after $O(k)$ eigenvalues. We do not have good analytic control over the cutoff region itself. The von Neumann entropy is dominated by eigenvalues near the cutoff, and is still approximately equal to $\log k$. The largest correction to the entropy comes from the width of the cutoff region and has magnitude $O(k^2 y(\smax)^2/Z_1^2)$.

When $k_{3 \to 4} \ll k \lesssim e^{S_0} \rho( s^{(1)})e^{-O(1/\beta)}$ (phase IV), the leading correction instead comes from the eigenvalues in the shifted thermal spectrum, and has size $O(e^{S_0} \rho(\stilde)/k)$. Unlike in phase III, this correction is well controlled analytic; its exact size is given by a simple integral.

The actual Page transition (phase V) for the von Neumann entropy happens at $k \sim e^{S_0} \rho(s^{(1)})$. At this point, the shift in the thermal tail of eigenvalues starts decreasing in order to preserve the normalisation of the state. The shift is now $(1- p_k)/k$, where $p_k$ is the probability of the thermal ensemble being in one of its $k$ largest eigenvalues. The von Neumann entropy is well approximated, up to a correction that peaks at $O(\beta)$ size near the transition, by the entropy of a shifted thermal spectrum, with a hard cutoff after exactly $k$ eigenvalues. This gives a large $O(1/\sqrt{\beta})$ correction compared to the naive Page curve
\begin{align} \label{eq:naivepagesummary}
S = \min\left( \log k, S_{BH}\right),
\end{align}
where $S_{BH}$ is the black hole entropy in the canonical ensemble. This is parametrically larger than the correction for the microcanonical ensemble, because of the fluctuations in the energy.

As $k$ continues to grow, the shift in the thermal part of the spectrum decays, until it becomes neglible for all eigenvalues where the spectrum actually looks thermal. The entanglement spectrum is now the ordinary unshifted thermal spectrum, except that it is smoothly cutoff after $k$ eigenvalues. The von Neumann entropy is approximately equal to the black hole entropy $S_{BH}$.

Initially, the largest correction to that entropy still comes from the existence of the cutoff (phase VI) and has size 
\begin{align}
O(1- p_k) = O\left(\frac{\rho(\smax) y(\smax)}{Z_1}\right). 
\end{align}
However, when $k \gg e^{S_0} \rho(\sprime)$ (phase VII), this correction is smaller than the effect of small corrections to the thermal spectrum at $s \sim \sprime$. These corrections give a correction to the entropy of size 
\begin{align}
\delta S = - O\left(\frac{e^{2S_0} \rho(\sprime)^2 y(\sprime)}{k Z_1}\right).
\end{align}
 In the limit $k \to \infty$, we also regain some analytic control over the shape of the spectral cutoff.

Throughout the transition, the von Neumann entropy is well approximated by assuming that the density of states is a thermal spectrum, with a hard cutoff after $k$ eigenvalues, and with the eigenvalues increased by a uniform shift to ensure the correct normalisation. This approximation is worst at the Page transition, when it gives an $O(\beta)$ error. 

It is important to note, however, that this shifted thermal spectrum only accurately computes the leading correction to the naive Page curve in phases IV and V. Outside this range, the shifted thermal spectrum entropy is no more accurate than the naive Page curve. Nonetheless,  the shifted thermal spectrum is still very accurate everywhere else, since the naive Page curve is itself a very good approximation away from the transition.

\begin{figure}[t]
\includegraphics[width = 0.7\linewidth]{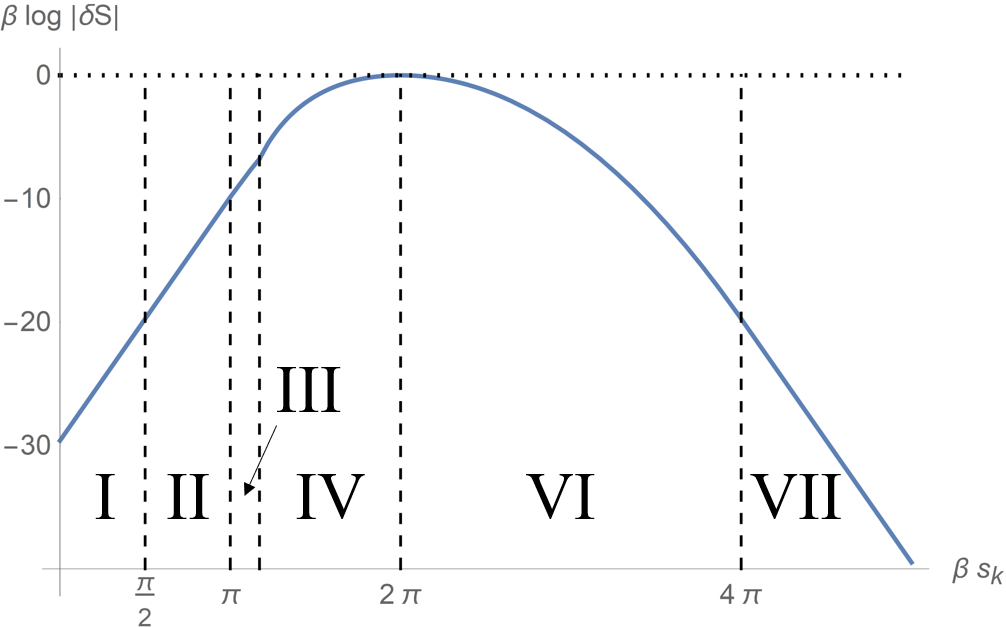}
\centering
\caption{A plot of $\beta \log |\delta S|$ against $\beta\, \smax$ in the limit where $\mu \gg 1/\beta$ and $\beta \to 0$. Near the transition, the entropy correction decays as $\delta S = -O(\exp(-\beta(\smax - s^{(1)})^2/2))$ because the largest correction comes from energy fluctuations causing the transition to happen earlier or later than expected. Far from the transition, the entropy decays as $\delta S = -\exp(-O(|\smax - s^{(1)}|))$, as for the microcanonical ensemble. Note that some important features of the correction, such as the maximal size $\delta S = -O(\sqrt{1/\beta})$, are not visible in this plot.}
\label{fig:deltaS}
\end{figure}

In Figure \ref{fig:deltaS}, we plot the correction $\delta S$ to the naive Page curve entropy over the course of the Page transition.

\subsection*{Phase I: $k \ll Z_2/y(0)^2$}
When $k$ is sufficiently small, we can solve the resolvent equation
\begin{align}
\lambda R(\lambda)  = k + e^{S_0}\int_0^\infty ds \rho(s) \frac{y(s) R}{k Z_1 - y(s) R},
\end{align}
by assuming that the resolvent $R$ satsifies $|y(s) R(\lambda)| \ll k Z_1$  for all $\lambda$ and  $s$. We justify this assumption by checking that the resulting solution for $R$ is self-consistent. We then have
\begin{align} \label{eq:taylor}
 \lambda R(\lambda) & = k + \int_0^\infty ds \rho(s) \left[\frac{y(s) R}{k Z_1} + \frac{y(s)^2 R^2}{k^2 Z_1^2} + O(\frac{y(s^3) R^3}{k^3 Z_1^3})\right],
\\& = k + \frac{R}{k} + \frac{R^2 Z_2}{k^2 Z_1^2}.
\end{align}
This is a quadratic equation in $R$ with solution
\begin{align} \label{eq:semicircle}
R(\lambda) = \frac{- (1/k - \lambda) + \sqrt{(\lambda - 1/k)^2 - 4 Z_2 / k Z_1^2 }}{2 Z_2 / k^2 Z_1^2},
\end{align}
where we chose the correct solution by demanding that $R \to 0$ as $\lambda \to \infty$. The density of states $D(\lambda)$ has support for $1/k - \sqrt{4 Z_2 / k Z_1^2} \leq \lambda \leq 1/k + \sqrt{4 Z_2 / k Z_1^2}$, where it has the form
\begin{align}
D(\lambda) = \frac{k^2 Z_1^2}{2 \pi  Z_2 } \sqrt{ 4 \frac{Z_2 }{ k Z_1^2} - ( \lambda - 1/k)^2 }.
\end{align}
This is a semicircle distribution that is sharply peaked around $1/k$.

We now need to check that our assumptions were self-consistent. We have $| R( \lambda) | \lesssim k^{3/2} Z_1 Z_2^{-1/2}$. Hence
\begin{align}
| R(\lambda) y(s) | \lesssim  k^{3/2} Z_1 Z_2^{-1/2} y(0),
\end{align}
for all $s$ and $\lambda$. Our assumption that $| R(\lambda) y(s) | \ll k Z_1$ is therefore valid so long as $k \ll Z_2 / y(0)^2$, as claimed.

What about the von Neumann entropy in this phase? At leading order, it is clearly given by $\log k$. To calculate the leading correction to this, we expand $\lambda$ about $1/k$ to second order. We find
\begin{align} \label{eq:vonNeumannsemicircle}
S = \int d \lambda D(\lambda) (1/k + (\lambda - 1/k)) \left( \log k - k (\lambda - 1/k)  + \frac{k^2 (\lambda - 1/k)^2}{2} + O((\lambda - 1/k)^3)\right).
\end{align}
Using
\begin{align} \label{eq:trace}
\int d \lambda D(\lambda) = k,
\end{align}
\begin{align} \label{eq:trace2}
\int d \lambda D(\lambda) (\lambda  - 1/k)= 0,
\end{align}
and
\begin{align} \label{eq:2ndmoment}
\int d \lambda D(\lambda) (\lambda -1/k)^2 = \frac{Z_2 }{ Z_1^2},
\end{align}
we therefore obtain
\begin{align} \label{eq:vn1}
\delta S = - \frac{k Z_2}{2 Z_1^2}.
\end{align}
This is very similar to the Page result, except that the correction is enhanced by a factor of $Z_2 / (Z_1 y(s^{(1)}))$, because the largest correction comes from values of s near $s^{(2)}$ rather than $s^{(1)}$. 

It is also worth noting that \eqref{eq:trace} and \eqref{eq:trace2} must also be true in the exact solution for $D(\lambda)$, since we know $\mathrm{Tr}(\rho^0) = k$ and  $\mathrm{Tr}(\rho) = 1$. This ensures that higher-order perturbative corrections to $D(\lambda)$ give corrections to the von Neumann entropy $S$ that are suppressed compared to \eqref{eq:vn1}.

A simpler way to reach the same result, without going through the full density of states calculation, is to look at the leading correction to the disconnected geometry in the calculation of the purities $\mathrm{Tr}(\rho^n)$ when $k$ is small. This leading correction comes from geometries where two replicas are connected, with the rest disconnected. There are $n(n-1)/2$ such geometries (from all the ways of pairing replicas), and each geometry is suppressed compared to the disconnected geometry by a factor of $Z_2 k / Z_1^2$. This gives a correction to the R\'{e}nyi entropy 
\begin{align}
S_n = -\frac{1}{n-1} \log \mathrm{Tr}(\rho^n)
\end{align}
of size $\delta S_n = -n Z_2 k / 2 Z_1^2$. Taking the limit $n \to 1$, reproduces \eqref{eq:vn1}.

\subsection*{Phase II: $Z_2/y(0)^2 \ll k \ll e^{S_0} \rho( s^{(2)})$}
In this regime, the approximation from the previous subsection breaks down at sufficiently small values of $s$. We need a new approach. Our strategy is to split the integral over $s$ into two pieces: $s < \stran$ and $s >s_\text{tran}$ as
\begin{align}\label{eq:splitintegral}
\lambda R(\lambda)  = k + e^{S_0}\int_0^{\stran} ds \rho(s) \frac{y(s) R}{k Z_1 - y(s) R} +  e^{S_0}\int_{\stran}^\infty ds \rho(s) \frac{y(s) R}{k Z_1 - y(s) R}.
\end{align}
We treat the integral for $s > \stran$ in the same way as before. This assumes that
\begin{align}\label{eq:assume1stran}
|y(\stran) R(\lambda)| \ll k Z_1
\end{align}
for all $\lambda$. We then treat the integral for $s < \stran$ as a small perturbation that can be ignored to leading order. For this second approximation to be justified, we require
\begin{align} \label{eq:assume0stran}
\left|e^{S_0}\int_0^{\stran} ds \rho(s) \frac{y(s) R}{k Z_1 - y(s) R}\right| \lesssim e^{S_0}\int_0^{\stran} ds \rho(s) \lesssim e^{S_0} \rho(\stran) \ll k.
\end{align}
In the first approximate inequality we used the fact that $|y(s) R| \lesssim |k Z_1 - y(s) R|$, except for a small neighbourhood near the pole at $k Z_1 = y(s) R$. This neighbourhood only gives a small contribution to the integral.

Can we choose $\stran$ to simultaneously satisfy the required conditions on both parts of the integral? The answer is yes. 
To satisfy our assumption \eqref{eq:assume0stran}, we require $\smax - \stran \gg 1$. In this phase, we have $\smax \ll s^{(2)} \ll s^{(1)}$ and so
\begin{align}
e^{S_0} \int_{\stran}^\infty ds  \rho(s) y(s)^2 \approx Z_2, \,\,\,\,\,\,\,\,\, e^{S_0} \int_{\stran}^\infty ds  \rho(s) y(s) \approx Z_1.
\end{align}
Treating the second term on the RHS of \eqref{eq:splitintegral} as a small perturbation, we find that the unperturbed solution for the resolvent $R$ is again given by \eqref{eq:semicircle}, as in phase I. 

Is this result consistent with our assumption \eqref{eq:assume1stran}? We have
\begin{align}
|y(\smax) R(\lambda)| \lesssim k^{3/2} Z_1 Z_2^{-1/2} y(\smax) = k Z_1 \frac{e^{S_0 /2} \rho({\smax})^{1/2} y(\smax)}{Z_2^{1/2}} \ll k Z_1,
\end{align}
since $\rho(\smax) y(\smax)^2 \ll \rho(s^{(2)}) y(s^{(2)})^2$. We can therefore consistently choose $\stran = \smax - \kappa$ for some large, but $O(1)$, constant $\kappa$, and simultaneously satisfy \eqref{eq:assume1stran} and \eqref{eq:assume0stran}.

Having justified our assumptions, we now treat the integral
\begin{align} \label{eq:phase2perturbation}
e^{S_0}\int_0^{\stran} ds \rho(s) \frac{y(s) R}{k Z_1 - y(s) R}
\end{align}
as a small perturbation. In the limit $\lambda - 1/k \gg \sqrt{Z_2 / k Z_1^2}$, the unperturbed solution \eqref{eq:semicircle} reduces to
\begin{align}
R_0 (\lambda) = \frac{k}{\lambda - 1/k}.
\end{align}
Including the perturbation \eqref{eq:phase2perturbation}, we find the first order correction
\begin{align} \label{eq:R1lambda}
R_1(\lambda) = \frac{ e^{S_0}}{\lambda - 1/k }\int_0^{\stran} ds \rho(s) \frac{y(s) R_0}{k Z_1 - y(s) R_0} = \frac{e^{S_0}}{\lambda - 1/k}\int_0^{\stran} ds \rho(s) \frac{y(s) /Z_1}{\lambda - 1/k - y(s)/ Z_1}.
\end{align}
Since we now have $y(0)/Z_1 \gg \sqrt{Z_2 / k Z_1^2}$, the pole at $\lambda - 1/k = y(s)/ Z_1$, for sufficiently small $s$, is in the regime where our approximations are valid. This gives a nonzero density of states, at far larger eigenvalues than anywhere in the semicircle. 

Specifically, for $\lambda - 1/k \gg \sqrt{Z_2 / k Z_1^2}$, we have
\begin{align} \label{eq:shiftedthermaltail}
D(\lambda) = e^{S_0} \int_0^{\stran} ds \rho(s) \delta( \lambda - 1/k - y(s)/ Z_1).
\end{align}
This looks like a thermal spectrum, with all the eigenvalues shifted upwards by $1/k$. At $\lambda - 1/k \approx 2 \sqrt{Z_2 / k Z_1^2}$, this shifted thermal spectrum merges into the main semicircle. 

The full spectrum is therefore a semicircle, centred on $1/k$ and with width $4 \sqrt{ Z_2 / k Z_1^2}$, plus a very small tail of larger eigenvalues that look like the largest eigenvalues of the thermal spectrum, shifted upwards by $\delta \lambda = 1/k$. Note that there is no tail of small eigenvalues, outside of the semicircle, because in this regime the unperturbed solution is negative, and the perturbed solution does not include any poles.

To understand the exact transition between the shifted thermal tail and the semicircle, we would need to consider the first order perturbation of the full unperturbed solution \eqref{eq:semicircle}. However, we will not do so here, since it is unimportant for our purposes. 

We note that the exact solution must satisfy
\begin{align} \label{eq:tracecorrect}
\int_{1/k - \sqrt{4Z_2 / k Z_1^2} }^{1/k - \sqrt{4Z_2 / k Z_1^2} } d \lambda\, D(\lambda) = k - \int_{1/k + \sqrt{4Z_2 / k Z_1^2} }^1 d \lambda\, D(\lambda),
\end{align}
and
\begin{align} \label{eq:tracecorrect2}
\int_{1/k - \sqrt{4Z_2 / k Z_1^2} }^{1/k - \sqrt{4Z_2 / k Z_1^2} } d \lambda\, D(\lambda) (\lambda - 1/k) = - \int_{1/k + \sqrt{4Z_2 / k Z_1^2} }^1 d \lambda\, D(\lambda) (\lambda - 1/k),
\end{align}
because of the constraints $\tr(\rho^0) = k$ and $\tr(\rho) = 1$. These results will be crucially to controlling the size of the correction to the von Neumann entropy.

What is the von Neumann entropy? Again, to leading order, it is given by $\log k$. There are two possible sources for the dominant correction. The first is the nonzero width of the semicircle, which gives the same correction that was found in \eqref{eq:vn1}. The second comes from the existence of the small tail of large eigenvalues. The existence of this tail gives a correction to the entropy of size
\begin{align}
\delta S_2 = \int_{1/k +\sqrt{4Z_2 / k Z_1^2}}^1 d \lambda D(\lambda) \left[-\lambda \log(\lambda) - \frac{\log k}{k} -(\log k - 1) (\lambda - 1/k)\right].
\end{align}
The first term here  is the direct contribution to the entropy from eigenvalues in the small tail, while the second and third terms come from corrections to the contribution from eigenvalues in the semicircle (calculated by the formula in \eqref{eq:vonNeumannsemicircle}) given \eqref{eq:tracecorrect} and \eqref{eq:tracecorrect2} respectively. We therefore find
\begin{align} \label{eq:thermalcorrection}
\delta S_2 &= e^{S_0}\int_{1/k +\sqrt{4Z_2 / k Z_1^2}}^1 d \lambda \int_0^{\stran} ds \,\rho(s)\, \delta( \lambda - 1/k - y(s)/ Z_1) (-\lambda\log \lambda - 1/k - \lambda (\log k - 1))
\\& = e^{S_0} \int_0^{s^*} ds \rho(s) \left[\left(\frac{1}{k} + \frac{y(s)}{Z_1}\right) \left(- \log \left(1 + \frac{k y(s)}{Z_1}\right) + 1\right) - \frac{1}{k}\right],
\\& = \frac{e^{S_0}}{k} \int_0^{s^*} ds \rho(s) \left[\frac{ky(s)}{Z_1} - \left(1 + \frac{ky(s)}{Z_1}\right) \log \left(1 + \frac{k y(s)}{Z_1}\right)\right],
\end{align}
where $y(s^*) = 2\sqrt{Z_2 / k}$ (and hence $s^* \ll \stran$). Since $0 \geq x - (1+x) \log(1 + x) \geq - x^2$ for $x > 0$, we have
\begin{align}
|\delta S_2| \lesssim \frac{e^{S_0} k \rho(s^*) y(s^*)^2}{k Z_1^2} \ll  \frac{k Z_2}{Z_1^2}.
\end{align}
We conclude that \eqref{eq:vn1} is still the leading correction to the von Neumann entropy.

We note that the largest contribution to \eqref{eq:thermalcorrection} comes from values of $\lambda \sim 1/k + O(\sqrt{Z_2 / k Z_1^2})$. In this regime, the approximation $R_0 (\lambda) = k/(\lambda - 1/k)$ is not very accurate. Hence the precise size of the correction \eqref{eq:thermalcorrection} should not be trusted. 

This is fine: since \eqref{eq:trace} and \eqref{eq:trace2} are fixed by the constraints, the leading effect on the von Neumann entropy of \emph{any} small correction to $D(\lambda)$, with $\lambda - 1/k = O(\sqrt{Z_2 / k Z_1^2})$, will be subleading compared to \eqref{eq:2ndmoment}. It will therefore be small compared to the correction in \eqref{eq:vn1}. The point of the calculation in \eqref{eq:thermalcorrection} is to bound the correction from the small density of states with $\lambda - 1/k  \gg \sqrt{Z_2 / k Z_1^2}$, where a small correction to $D(\lambda)$ could, in principle, have given a larger correction to the von Neumann entropy than \eqref{eq:vn1}.

\subsection*{Phase III: $e^{S_0} \rho( s^{(2)}) \ll k \ll k_{3 \to 4}$}
In this phase, we will have somewhat less control over the shape of the peak of the density of states $D(\lambda)$. We will focus on understanding the density of states \emph{away} from this peak. Fortunately this approach will still give us good control over the von Neumann entropy. Our strategy will be similar to our strategy for phase II. However we will find that our approximation for the resolvent $R(\lambda)$ will now only be valid outside of a small range of $\lambda$ (which includes the peak of the spectrum). 

Explicitly, we choose some $\stran$ as for $Z_2/y(0)^2 \ll k \ll e^{S_0} \rho( s^{(2)})$, and approximate the resolvent equation by
\begin{align}
\lambda R(\lambda) = k  + e^{S_0}\int_0^{\stran} ds \rho(s) \frac{y(s) R}{k Z_1 - y(s) R} +  e^{S_0}\int_{\stran}^\infty ds \rho(s) \frac{y(s) R}{k Z_1} ,
\end{align}
where the second term on the right hand side is treated as a small perturbation. As before, for this approximation to be valid, we require $\smax - \stran \gg 1$. as well as $|y(\stran) R(\lambda)| \ll k Z_1$, for the values of $\lambda$ where we want to be able to trust our approximation.

Unlike for phase II, we do not include a term
\begin{align} \label{eq:quadraticterm}
e^{S_0}\int_{\stran}^\infty ds \rho(s) \frac{y(s)^2 R^2}{k^2 Z_1^2},
\end{align}
in our approximation for the integral over $s > \stran$. In this phase we will have $\stran \gg s^{(2)}$, causing \eqref{eq:quadraticterm} to be dominated by $s \sim \stran$. It will therefore be much smaller than $k$, and can be safely ignored, whenever our approximations are themselves valid.

Ignoring the small perturbation, we obtain the initial unperturbed solution
\begin{align} \label{eq:R0lambda3}
R_0 (\lambda) = \frac{k}{\lambda - 1}.
\end{align}
Here, we have used the fact that $\stran < \smax \ll s^{(1)}$, to see that
\begin{align}
e^{S_0} \int_{\stran}^\infty ds \rho(s) y(s)  \approx Z_1.
\end{align}
Our unperturbed solution $R_0(\lambda)$ has a pole at $\lambda = 1/k$, so it is clearly not consistent with our assumption that $|y(\stran) R(\lambda)| \ll k Z_1$ for all values of $\lambda$. 

What values of $\lambda$ allow us to choose $\stran$ so that our solution is self-consistent? We have  $|y(\smax) R(\lambda)| \ll k Z_1$ so long as
\begin{align} \label{eq:lambdanottoobig}
\left|\lambda - 1/k\right| \gg y(\smax)/Z_1.
\end{align}
Since $\smax = O(1/\beta)$ in the semiclassical limit, we have $y(s) \propto e^{O(s)}$  for $s \approx \smax$. Both assumptions are therefore valid so long as \eqref{eq:lambdanottoobig} holds and we choose $\stran =\smax - \kappa$, for some large, but $O(1)$, constant $\kappa$. As for phase II, the first order correction to \eqref{eq:R0lambda3} is given by \eqref{eq:R1lambda}. 

We find that the density of states $D(\lambda) = 0$ for sufficiently small $\lambda$, satisfying $1/k - \lambda \gg y(\smax)/Z_1$. For large values of $\lambda$, with $\lambda - 1/k \gg y(\smax)/Z_1$, we have
\begin{align}
D(\lambda) = e^{S_0} \int ds \rho(s) \delta( \lambda - 1/k - y(s)/ Z_1),
\end{align}
as in \eqref{eq:shiftedthermaltail}. This region only contains a small fraction of the $k$ total eigenvalues in the entanglement spectrum -- by definition, the $k$th largest eigenvalue in the shifted thermal spectrum is given by $ \lambda = 1/k + y(\smax)/Z_1$. The remaining eigenvalues must lie in the range $1/k - O(y(\smax)/Z_1) \lesssim \lambda \lesssim 1/k + O(y(\smax)/Z_1)$. Since we know that $R_0 (\lambda)$ is not a good approximate solution for any values of $\lambda$ within this range, we conclude that the density of states $D(\lambda)$ is large throughout this region, rather than being concentrated parametrically closer to $1/k$. 

In summary, we conclude the following. The density of states looks like a thermal spectrum, shifted by $\delta \lambda = 1/k$, cutoff by some unknown function of width $O(y(\smax)/Z_1)$ at $\lambda = 1/k$. We do not know any compelling way to get analytic traction on this cutoff function, although it can of course be calculated numerically, by solving either the entire resolvent equation numerically or an appropriate approximation to it. 

We note that when $k \sim e^{S_0} \rho(s^{(2)})$, the width of the semicircle that we found in phase II is equal to the width of this new uncontrolled cutoff function. There is a smooth transition between the two.

We now move on to calculating the von Neumann entropy. To leading order we still have $S = \log k$. As for phase II, there are two potential sources for the leading correction. The first is a correction from the width of the cutoff, as in \eqref{eq:vonNeumannsemicircle}. This is equal to
\begin{align}  \label{eq:phase3cutoffcorrection}
\delta S_1 = - \frac{k}{2}\int_{1/k - O(y(\smax)/Z_1)}^{1/k + O(y(\smax)/Z_1)}  d \lambda D(\lambda) (\lambda - \frac{1}{k})^2 = O(k^2 \frac{y(\smax)^2}{Z_1^2}).
\end{align}
To calculate the exact size of this correction, we would need to use a numerical approximation for the density of states in the cutoff region. The second possible correction comes from the existence of the thermal tail of large eigenvalues. As in \eqref{eq:thermalcorrection}, this is given by
\begin{align} \label{eq:phaseIVcorrection}
\delta S_2 = \frac{e^{S_0}}{k} \int_0^{\smax - O(1)} ds \rho(s) \left[\frac{ky(s)}{Z_1} - \left(1 +\frac{k y(s)}{Z_1}\right) \log \left(1 + \frac{k y(s)}{Z_1}\right)\right].
\end{align}
Since we have $s^{(2)} \ll \smax \ll s^{(1)}$, this integral is dominated by values of $s$ with $k y(s)\sim Z_1$. Unlike for $k \ll e^{S_0} \rho(s^{(2)})$, this means that the dominant contribution comes from eigenvalues where the approximation can be trusted. Its size is therefore $O(e^{S_0} \rho(\stilde)/k)$, where $\stilde$ is defined in \eqref{eq:defstildek}.

Comparing the sizes of the two corrections, we have
\begin{align}
\frac{\delta S_2}{\delta S_1} = O\left(\frac{e^{S_0} \rho(\stilde) Z_1^2}{k^3 y(\smax)^2}\right) \ll 1.
\end{align}
since we are assuming that $k \ll k_{3 \to 4}$, for $k_{3 \to 4}$ defined in \eqref{eq:k3to4}.

We therefore conclude that the dominant correction in this phase comes from the width of the cutoff rather than the thermal tail.

\subsection*{Phase IV: $k_{3 \to 4} \ll k \lesssim e^{S_0} \rho( s^{(1)})e^{-O(1/\beta)}$ }
This phase can be dealt with using exactly the same strategy as phase III. The only change is that we now find that the source of the leading correction to the von Neumann entropy (which is still given by $\log k$ at leading order) has changed. Specifically, the leading correction $\delta S_2 = O(e^{S_0} \rho(\stilde)/k)$ now comes from the existence of the thermal tail, rather than the width of the cutoff region. Its exact size is given \eqref{eq:phaseIVcorrection}, with high accuracy in the semiclassical limit.

One might worry about whether we can still assume that
\begin{align}
e^{S_0} \int_{\stran}^\infty ds \rho(s) y(s) \approx Z_1.
\end{align}
As we shall see in the next section, a more accurate treatment of this integral would decrease the shift in the thermal tail by $p_k/ k$, where $p_k$ is the probability that the thermal ensemble is in one of its $k$ largest eigenvalues. When $k/e^{S_0} \rho( s^{(1)}) = e^{-O(1/\beta)}$, we have $p_k/k = O( y(\smax)/Z_1)$, so this shift can be ignored everywhere that the shifted thermal spectrum can be trusted.

The leading \emph{uncontrolled} correction to the von Neumann entropy still comes from the width of the cutoff region, and has size $\delta S_1 =  O(k^2 y(\smax)^2/Z_1^2)$, as in \eqref{eq:phase3cutoffcorrection}.

\subsection*{Phase V: $e^{S_0} \rho( s^{(1)})e^{-O(1/\beta))} \ll k \ll e^{S_0} \rho( s^{(1)}) e^{O(1/\beta)}$}
We have now finally reached the actual Page transition for the von Neumann entropy! In this phase, we can still use the same approximations that we used in phases III and IV to find the shape of the density of states away from its peak. Again, we want to choose $\stran = \smax - \kappa$, for a large, but $O(1)$, constant $\kappa$. However, we can no longer assume that
\begin{align}
e^{S_0} \int_{\stran}^\infty ds \rho(s) y(s) = Z_1.
\end{align}
Instead we have
\begin{align}
e^{S_0} \int_{\stran}^\infty ds \rho(s) y(s) \approx e^{S_0} \int_{\smax}^\infty ds \rho(s) y(s) = (1 - p_k) Z_1,
\end{align}
where $p_k$ is the probability of the thermal density matrix being in one of its $k$ largest eigenvalues. We therefore find that the density of states $D(\lambda)$ is given by
\begin{align}
D(\lambda) = \begin{cases} 
      e^{S_0} \int ds \rho(s) \delta( \lambda - \frac{1-p_k}{k} - \frac{y(s)}{ Z_1}) & \lambda -\frac{1-p_k}{k} \gg \frac{y(\smax)}{Z_1}, \\
      \text{large and uncontrolled} & \lambda -\frac{1-p_k}{k} = O( \frac{y(\smax)}{Z_1}) ,\\
      0 &  \frac{1-p_k}{k} - \lambda \gg \frac{y(\smax)}{Z_1},
   \end{cases}
\end{align}
The region $\lambda - (1-p_k)/k = O( y(\smax)/Z_1)$, which we do not have good analytic control over, contains most of the eigenvalues. 

We see that the shift in the thermal part of the spectrum decreases as we move through the Page transition. This can be easily understood as the eigenvalues adjusting to match the twin constraints that $\Tr(\rho) = 1$ and $\Tr(\rho^0) = k$; if we had $O(k)$ thermal eigenvalues, all shifted by $\delta \lambda = 1/k$, we would have $\Tr(\rho) > 1$.

Near $s^{(1)}$, the function $\rho(s) y(s)$ is well approximated by a Gaussian of width $O(\sqrt{1/\beta})$. More precisely, in the limit where the brane mass $\mu$ is large, we have
\begin{align}
\rho(s)y(s) \propto e^{- \frac{\beta}{2} (s -s^{(1)})^2}.
\end{align}
We therefore have $Z_1 = O(\sqrt{1/\beta}\, \rho(s^{(1)}) y(s^{(1)}))$. Also, for $\smax - s^{(1)} \gg \sqrt{1/\beta}$, we have
\begin{align}
1 - p_k = O\left( \frac{\rho(\smax) y(\smax)}{\sqrt{\beta}(\smax - s^{(1)})\rho(s^{(1)}) y(s^{(1)})}\right).
\end{align}
It follows that
\begin{align}
\frac{y(\smax)/Z_1}{(1-p_k)/ k}  = O\left(\frac{\sqrt{\beta}\rho(\smax) y(\smax)}{(1-p_k) \rho(s^{(1)}) y(s^{(1)})}\right) \ll 1,
\end{align}
since in this phase we have $\smax - s^{(1)} \ll 1/\beta$. The width of the uncontrolled region is therefore always small compared to the size of a typical eigenvalue in the uncontrolled region.

Let us try to calculate the von Neumann entropy. As a first approximation, we can consider the von Neumann entropy of a thermal spectrum, with a hard cutoff after $k$ eigenvalues, and with each eigenvalue increased by $(1- p_k)/k$. This agrees with the actual spectrum everywhere that we have control, and obeys the constraints $\Tr \rho^0 = k$ and $\Tr \rho = 1$. The von Neumann entropy of such a state would be given by
\begin{align}\label{eq:shiftedthermal}
S = - e^{S_0} \int_0^{\smax} ds \rho(s) \left(\frac{y(s)}{Z_1} + \frac{1 - p_k}{k} \right) \log \left(\frac{y(s)}{Z_1} + \frac{1 - p_k}{k} \right).
\end{align}
What is the error from this approximation? We have significantly changed the spectrum in the small region where $\lambda -\frac{1-p_k}{k} = O( \frac{y(\smax)}{Z_1})$. To leading order, this part of the spectrum gives a contribution to the von Neumann entropy of $(1 - p_k) \log((1-p_k)/k)$. Since the zeroth and first moments are fixed by the rest of the spectrum and the constraints $\Tr(\rho) = 1$ and $\Tr(\rho^0) = k$, the leading difference between the shifted thermal spectrum and the actual spectrum will be controlled by
\begin{align} \label{eq:blaaaaah}
\frac{k}{(1-p_k)}\int_{(1-p_k)/k - O(y(\smax)/Z_1)}^{(1-p_k)/k + O(y(\smax)/Z_1)} d \lambda\, \Delta D(\lambda) \left(\lambda - \frac{(1 - p_k)}{k}\right)^2  = O\left(\frac{k^2}{(1-p_k)} \frac{y(\smax)^2}{Z_1^2}\right),
\end{align}
where $\Delta D(\lambda)$ is the diference between the two spectral densities. To find this formula, we expanded $\lambda$ to second order about $(1 - p_k)/k$. In estimating its size, we used the fact that most of the $k$ eigenvalues are in this region of the spectrum. 

How large can this error get? It is largest when $\smax \sim s^{(1)} + O(\sqrt{1/\beta})$. In this case, we have $(1-p_k) = O(1)$, so the size of the correction is $O(\beta)$.

A cruder approximation is to replace $ \log \left(y(s)/Z_1 +(1 - p_k)/k \right)$ in \eqref{eq:shiftedthermal} by $\log \max(y(s)/Z_1, (1 - p_k)/k)$. The largest error from this simplification comes from values of $s$ where $y(s)/Z_1 \sim (1 - p_k)/k$. From such eigenvalues, we have an $O(1)$ error in our estimate of the logarithm. When $\smax \sim s^{(1)}$, this gives a total error of $O(\rho(s)y(s)/Z_1) = O(\sqrt{\beta})$, which is parametrically worse than the shifted thermal spectrum approximation.

How large is the difference between the shifted thermal entropy \eqref{eq:shiftedthermal} and a naive Page curve given by $S =  \min(\log k , S_{BH})$? Assuming $\mu \gg 1/\beta$,
\begin{align}
S_{BH} = \frac{e^{S_0}}{Z_1} \int ds \rho(s) y(s) \log(Z_1/ y(s)) = S_0 + \frac{4 \pi^2}{\beta} + \frac{1}{2} \log \left(\frac{2 \pi}{\beta}\right) + O(1).
\end{align}
Let $k = Z_1/y(s^{(1)})$, i.e $\stilde = s^{(1)}$. The naive Page curve entropy would be $\log k$. Instead, using the crude approximation discussed above, we find
\begin{align}
\begin{split}
S & = \frac{e^{S_0}}{Z_1}\int_0^{s^{(1)}} ds \rho(s) y(s) \left[\frac{1}{2} \beta s^2 +  \log \frac{Z_1}{y(0)}\right] +  \frac{1}{2} \log k + O(1) \\&= \log k  - \sqrt{2 \pi \beta} \int_0^{s^{(1)}} ds \,e^{-\frac{\beta}{2} (s - s^{(1)})^2}  (s^{(1)} - s) + O(1) \\&= \log k - \sqrt{\frac{2 \pi}{ \beta}} + O(1).
\end{split}
\end{align}
In the first line, we used the fact that $p_k = 1/2 + o(1)$. In the second line, we used the fact that $\log(Z_1/y(s^{(1)}) = \log k$ and assumed $\mu \gg 1/\beta$.

The leading correction is therefore $O(1/\sqrt{\beta})$; it becomes very large in the semiclassical limit. This is in sharp contrast to the Page curve for Haar random states, or the microcanonical ensemble, where the correction is never larger than $O(1)$. The energy fluctuations in a thermal state parametrically increase the size of the corrections to the naive Page curve.

\subsection*{Phase VI: $e^{S_0} \rho( s^{(1)})e^{O(1/\beta)} \lesssim k \ll e^{S_0} \rho(\sprime)$}
In this phase, we have $(1-p_k)/k = O(y(\smax)/Z_1)$. Hence the approximations made in phase V no longer give us good control over the bottom of the spectrum. Instead, we treat the regimes $\lambda \gg y(\smax)/Z_1$ and $\lambda \ll y(\smax)/Z_1$ separately.

For the former case, we can treat the entire second term in
\begin{align} \label{eq:yetagain}
R(\lambda) = \frac{k}{\lambda}  +  \frac{e^{S_0}}{\lambda}\int_{0}^\infty ds \rho(s) \frac{y(s) R}{k Z_1 - y(s) R}
\end{align}
as a small perturbation. Our initial solution is $R_0 (\lambda) = k/\lambda$. Hence we have $R_0\, y(\smax) \ll k Z_1$ and our treatment of the second term \eqref{eq:yetagain} as a small perturbation is self-consistent. We therefore have the first order perturbative correction
\begin{align} \label{eq:R1lambda5}
R_1(\lambda) = \frac{e^{S_0}}{\lambda}\int_{0}^\infty ds \rho(s) \frac{y(s)/Z_1}{\lambda - y(s)/Z_1},
\end{align}
which gives
\begin{align}
D(\lambda) = e^{S_0} \int ds \rho(s) \delta( \lambda - y(s)/ Z_1).
\end{align}
The thermal spectrum is essentially unshifted, everywhere that it is under good analytic control.

For $\lambda \ll y(\smax)/Z_1$, it is easiest to consider $\lambda$ as a function of negative, real $R$. Rewriting \eqref{eq:yetagain}, we find
\begin{align}
\lambda = \frac{k}{R} -  \frac{e^{S_0}}{R} \int_{0}^\infty ds \rho(s) \frac{y(s) }{y(s) - k Z_1 / R}.
\end{align}
For small negative $R$, the first term dominates, and so $\lambda$ is negative. For very large negative $R$, the second term dominates, and so $\lambda$ is positive, although it approaches zero as $R \to - \infty$. Note that for negative $R$, the contribution from the integral is positive for all values of $s$. At some intermediate value of $R$, $\lambda$ should have a maximum. This value will be the bottom of the eigenvalue spectrum.

Suppose we have $R = - \kappa k Z_1 / y(\smax)$ for some large, but $O(1)$, constant $\kappa$ as usual. In this case, the second term dominates and we have
\begin{align}
\lambda \gg - \frac{k}{R} = \frac{y(\smax)}{\kappa Z_1}.
\end{align}
We can therefore be confident that $D(\lambda) = 0$ for $\lambda \ll y(\smax)/Z_1$. As for  phases III-V, we do not have good control over the shape of the spectrum in the cutoff region, when $\lambda = O(y(\smax)/Z_1)$.

The largest correction to the von Neumann entropy comes from the cutoff region. For eigenvalues in this region, there is an $O(1)$ multiplicative uncertainty in the eigenvalue, which corresponds to an $O(1)$ uncertainty in $\log \lambda$. The total probability of an eigenvalue being in this region is $O(1-p_k)$. Hence we have
\begin{align}
S = S_{BH} - O(1-p_k) = S_{BH} - O\left(\frac{e^{S_0} \rho(\smax)y(\smax)}{Z_1}\right),
\end{align}
where $S_{BH}$ is the entropy of the canonical ensemble. Note that this correction is the same order of magnitude as the difference between $S_{BH}$ and the entropy of the shifted thermal spectrum \eqref{eq:shiftedthermal}. However, we have no good reason to think that the two corrections are the same. The size will be affected by the details of the cutoff, over which we have very little control.

\subsection*{Phase VII: $k \gg e^{S_0} \rho(\sprime)$}
In the limit $k \to \infty$, the correction to the thermal entropy from the cutoff region, discussed above, decays as
\begin{align}
O(\frac{\rho(\smax)y(\smax)}{Z_1}) = O(e^{-\frac{\beta}{2}(\smax - s^{(1)})^2}).
\end{align}
This decays much faster than the largest corrections to the Renyi entropies, which come from planar diagrams that consist of two discs and are suppressed by a factor of $O(1/k)$
compared to the leading connected topology. One can also make general arguments that it is inconsistent with entanglement wedge reconstruction for the correction to decay faster than $O(1/k)$ in this limit \cite{Hayden:2018khn}.

It follows that there should exist some other correction, which cannot come from the cutoff region, that becomes dominant at sufficiently large $k$. The source of that correction is the second-order perturbative correction to the resolven. As for phase VI, the unperturbed solution is given by $R_0(\lambda) = k/ \lambda$ and the first order perturbative correction $R_1(\lambda)$ is given by \eqref{eq:R1lambda5}. The second order correction is
\begin{align}
R_2 (\lambda) =\frac{e^{2S_0} Z_1}{k} \int ds_1 ds_2 \frac{\rho(s_1) \rho(s_2) y(s_1) y(s_2)}{(Z_1 \lambda - y(s_1))^2 (Z_1 \lambda - y(s_2))}.
\end{align}
The second order perturbative correction to the density of states is therefore
\begin{align}
D_2 (\lambda) = \frac{e^{2S_0}}{k Z_1} \int ds_1 ds_2 \frac{\rho(s_1) \rho(s_2) y(s_1) y(s_2)}{y(s_2) - y(s_1)} \delta'(\lambda - y(s_1) / Z_1).
\end{align}
The pole at $s_1 = s_2$ is dealt with by taking the principal value. The contributions proportional to $\delta(\lambda - y(s_1) / Z_1)$ and $\delta(\lambda - y(s_2) / Z_1)$, which would otherwise exist, cancel under the relabelling $s_1 \leftrightarrow s_2$. We therefore get a correction to the von Neumann entropy of
\begin{align}
\delta S &=  - \int^1_{O(y(\smax)/Z_1} d \lambda\, D_2(\lambda) \,\lambda \log(\lambda),
\\& = \frac{e^{2S_0}}{k Z_1} \int_0^{\smax - O(1)} ds_1 \int_0^\infty ds_2 \frac{\rho(s_1) \rho(s_2) y(s_1) y(s_2)}{y(s_2) - y(s_1)} [\log(y(s_1)/Z_1) + 1],
\\& = -\frac{e^{2S_0}}{2 k Z_1} \int_0^{\smax - O(1)} ds_1 \int_0^{\smax - O(1)} ds_2 \frac{\rho(s_1) \rho(s_2) y(s_1) y(s_2)}{y(s_1) - y(s_2)} \log\frac{y(s_1)}{y(s_2)}. \label{eq:vNcorrectionlargeK}
\end{align}
In the last step, we dropped the integral over $s_2 > \smax - O(1)$, which is highly suppressed, and then symmetrised the integrand with respect to the relabelling $s_1 \leftrightarrow s_2$. This integral is dominated by values where $s_1 - s_2 = O(1)$. Its magnitude is therefore
\begin{align}
\delta S = - O\left(\frac{e^{2S_0}}{k Z_1} \int_0^{\smax - O(1)} ds \rho(s)^2 y(s)\right) = -O\left(\frac{e^{2S_0} \rho(\sprime)^2 y(\sprime)}{k Z_1}\right),
\end{align}
where $\sprime$ is the saddle point for $\rho(s)^2 y(s)$. This is proportional to $1/k$, as expected. We emphasize that this final result depends crucially on the fact that $k \gg e^{S_0}\rho(\sprime)$. If we still had $k \ll e^{S_0}\rho(\sprime)$, the integral would be dominated by values of $s$ close to $\smax$ and we would have 
\begin{align}
\delta S =  -O\left(\frac{e^{S_0} \rho(\smax) y(\smax)}{ Z_1}\right).
\end{align}
 This is the same order of magnitude as the correction we found in phase VI. We emphasize that the exact size of the correction still cannot be computed, since the perturbative approximation is poor for $s = \smax- O(1)$.

As with the correction to the entropy in phases I and II, there is a much simpler way to find this entropy correction, by analytically continuing the leading correction to the Renyi entropies. The leading correction to the purities $\Tr(\rho^n)$ in the limit $k\to \infty$ comes from planar diagrams consisting of two connected components. These give a contribution to the purity of
\begin{align}
\delta \,\Tr(\rho^n) = \frac{n}{2 k Z_1^n} \sum_{p=1}^{n-1} Z_p Z_{n-p} =  \frac{n e^{2S_0}}{2 k Z_1^n} \int ds_1 ds_2 \rho(s_1) \rho(s_2) \frac{y(s_1)^n y(s_2) - y(s_1) y(s_2)^n}{y(s_1) - y(s_2)}.
\end{align}
They therefore give a correction to the R\'{e}nyi entropies $S_n = 1/(1-n) \log(Tr(\rho^n))$ of
\begin{align}
\delta S_n = - \frac{n  e^{2S_0}}{(n-1) 2 k Z_n}  \int ds_1 ds_2 \rho(s_1) \rho(s_2) \frac{y(s_1)^n y(s_2) - y(s_1) y(s_2)^n}{y(s_1) - y(s_2)},
\end{align}
In the limit $n \to 1$, this gives the correction to the von Neumann entropy
\begin{align}
\delta S = - \frac{ e^{2S_0}}{2 k Z_1}  \int ds_1 ds_2 \rho(s_1) \rho(s_2) \frac{y(s_1)y(s_2)}{y(s_1) - y(s_2)} \log \frac{y(s_1)}{y(s_2)},
\end{align}
which is exactly what we found before.

The disadvantage of this approach is that it does not know \emph{when} this calculation breaks down. If we only looked at the two connected component topologies, we would expect that the corrections to the Page curve would become large when $k =  O(e^{2S_0} \rho(\sprime)^2 y(\sprime)/Z_1)$, which is much larger than the value of $k$ at the Page transition. In contrast, our more careful approach using the resolvent can see that this large correction doesn't exist when $k \ll e^{S_0} \rho(\sprime)$, because the integral in \eqref{eq:vNcorrectionlargeK} is cutoff at $\smax$.

Finally, in the limit of very large $k$, we actually regain some increased analytic control over the cutoff region. Specifically, we need $\smax \gg 1/\beta$. We assume for this section that the brane mass $\mu \gg 1/\beta$ and so we have $y(s) = y(0)e^{-\beta s^2/2}$. Suppose we rewrite \eqref{eq:yetagain} as
\begin{align} \label{eq:yetyetagain}
\lambda R = k - e^{S_0} \int ds \frac{\rho(s)}{- \frac{k Z_1}{y(s) R} + 1}.
\end{align}
For $s \gg O(1/\beta)$, $y(s)$ is changing much faster than $\rho(s)$. Hence, for sufficiently large, negative, real $R$, we have
\begin{align}
e^{S_0} \int ds \frac{\rho(s)}{- \frac{k Z_1}{y(s) R} + 1} \approx e^{S_0} \int_0^{s_R} ds \rho(s),
\end{align}
where $y(s_R) = - k Z_1 / R$ or equivalently 
\begin{align}\label{eq:sR}
s_R = \sqrt{(2/\beta)  \log (- Ry(0)/k Z_1)}.
\end{align} 
For self-consistency of our assumptions, we will need $\log (- R \,y(0)/k Z_1) \gg 1$. We know that in the cutoff region we have $|R| = O(kZ_1/y(\smax))$, which implies $s_R \approx \smax$. Our approximation is therefore well controlled so long as $\smax \gg O(1/\beta)$, as claimed. 

So far, we have only considered negative, real $R$. For imaginary or positive $R$, the integral is seemingly not so simple. However, we can simply deform the integration contour in the complex plane, without passing through any poles, to absorb the phase of $R$ into a phase of $y(s_R)$. Hence \eqref{eq:sR} is valid when $R$ has arbitrary phase so long as we use complex logarithms. Writing $R = -k Z_1e^{r + i \theta}/y(0)$, we obtain
\begin{align}
s_R = \sqrt{2r/\beta } + \frac{i \theta}{\sqrt{2 \beta  r}},
\end{align}
where we used the fact that $r \gg 1/\beta$. We can also do the integral in \eqref{eq:yetyetagain} explicitly to get
\begin{align} \label{eq:realimag}
- Z_1 \lambda e^{r + i \theta} = 1 - \frac{e^{S_0}}{k} \frac{s_R}{4 \pi ^3} e^{2 \pi s_R}.
\end{align}
Comparing the imaginary parts of the left and right hand sides, we find 
\begin{align} \label{eq:imagpart}
Z_1 \lambda e^{r} \sin \theta = \frac{e^{S_0}}{4 \pi^3 k} \sqrt{\frac{2r}{\beta  }} e^{2 \pi\sqrt{2r/\beta }} \sin \frac{\theta}{\sqrt{2 \beta  r}}
\end{align}
Hence, for any fixed $0 < \theta < 2 \pi$, we have
\begin{align} \label{eq:nottoobig}
Z_1 \lambda e^{r} \ll \frac{e^{S_0}}{k} \sqrt{\frac{2r}{\beta  }} e^{2 \pi \sqrt{2r/\beta }}.
\end{align}
Comparing the real parts of \eqref{eq:realimag}, we have
\begin{align} \label{eq:realpart}
- Z_1 \lambda \,e^{r} \cos \theta = 1 - \frac{e^{S_0}}{4 \pi^3 k} \sqrt{\frac{2r}{\beta  }} e^{2 \pi \sqrt{2 r /\beta}},
\end{align}
which, combined with \eqref{eq:nottoobig} tells us that, for fixed $0 < \theta < 2 \pi$,
\begin{align} \label{eq:sRapproxsk}
1 - \frac{e^{S_0}}{4 \pi^3 k} \sqrt{\frac{2 r}{\beta }} e^{2 \pi \sqrt{2r/\beta }} \ll 1,
\end{align}
Hence $r - \beta \smax^2/2 = \Delta r \ll \beta\smax $. Substituting this back into \eqref{eq:imagpart}, and using the small angle approximation for $\sin (\theta/\sqrt{\beta  r})$, we find
\begin{align}
\frac{Z_1 \lambda \,e^{\Delta r}}{y(\smax)}  \sin \theta = \frac{2 \pi \,\theta}{\beta \smax}.
\end{align}
Hence
\begin{align}
\theta = \text{sinc}^{-1}\left(\frac{2 \pi \,y(\smax)}{ Z_1\, \lambda \,\beta\, \smax\, e^{\Delta r}} \right),
\end{align}
where the inverse-sinc function $\text{sinc}^{-1}: [0,1] \to [0,\pi]$ is single-valued on its domain. We now need to solve for $\Delta r$. We now turn to \eqref{eq:realpart}. Given \eqref{eq:sRapproxsk}, we can rewrite \eqref{eq:realpart} as
\begin{align}
\frac{Z_1 \lambda  e^{ \Delta r}}{y(\smax)} \cos \theta = \frac{2\pi \Delta r}{\beta \smax}.
\end{align}
Solutions for this exist with $\theta = 0$ for $\lambda < 2 \pi\, y(\smax)/ (e Z_1 \beta \smax) = \lambda_\text{min}$, so there are no eigenvalues below this cutoff. At $\lambda = \lambda_\text{min}$, we have $\Delta r = 1$. 

For $\lambda > \lambda_\text{min}$, we have
\begin{align}
\frac{ \lambda\,  e^{ \Delta r - 1}}{\lambda_\text{min}} \cos\text{sinc}^{-1}\left(\frac{\lambda_\text{min}}{ \lambda e^{\Delta r - 1}} \right) = \Delta r.
\end{align}
In principle, this equation can be solved to find the fixed function $\Delta r(\lambda/\lambda_\text{min})$, which does not depend on any parameters. For $\lambda \gg \lambda_\text{min}$, we have $ \cos(\text{sinc}^{-1}\left(\lambda_\text{min}e^{-\Delta r + 1}/ \lambda  \right) \approx 1$ and so $\Delta r = - W_0(\lambda/\lambda_\text{min})$ where $W_0$ is the Lambert $W$ function. The resolvent $R$ decays at large $\lambda$, as expected, and we reenter the regime where our perturbative approximations are well controlled.

Finally, we obtain the density of states
\begin{align} \label{eq:sinsinc}
D(\lambda) = \frac{Z_1 k e^r \sin \theta}{\pi} =  \frac{k Z_1 e^{\Delta r(\lambda/\lambda_\text{min})}}{\pi y(\smax)} \sin \text{sinc}^{-1}\left(\frac{2\pi \, y(\smax) }{e^{\Delta r(\lambda/\lambda_\text{min})} Z_1 \lambda \beta \smax} \right).
\end{align}

For $\lambda \gg y(\smax)/( Z_1 \beta \smax)$, we can use the small $x$ approximation $\sin \text{sinc}^{-1}(x) = \pi x$. This gives
\begin{align}
D(\lambda) = \frac{2 \pi \,k }{ \lambda\, \beta \smax}.
\end{align}
For the thermal spectrum near the cutoff, we also have
\begin{align}
D(\lambda) = e^{S_0} \int_0^{\smax} ds \rho(s) \delta(\lambda - y(s)/Z_1) \approx 2 \pi k \int_0^{\smax} ds  \,\delta(\lambda - y(s)/Z_1) \approx \frac{2 \pi \,k}{\lambda \beta \smax },
\end{align}
in the first step, we used $\rho(s) \approx \rho(\smax)$, while the second step used $y'(s) = \beta\, s\, y(s)$. The cutoff spectrum connects smoothly to the thermal spectrum. 

This completes our analysis of the entanglement spectrum of the simple model.

{\footnotesize
\bibliography{references}
}

\bibliographystyle{utphys}

\end{document}